\pdfoutput=1
\documentclass{article}
\usepackage{iclr2027_conference,times}

\usepackage{amsmath,amssymb,amsfonts,amsthm,bm}
\usepackage{booktabs}
\usepackage{graphicx}
\usepackage{url}
\usepackage{xcolor}
\usepackage{algorithm}
\usepackage{algpseudocode}
\usepackage{mathtools}
\usepackage{tabularx}
\usepackage{multirow}
\usepackage{enumitem}
\usepackage{makecell}
\usepackage[font=small,labelfont=bf,skip=4pt]{caption}
\usepackage[list=off]{subcaption}
\usepackage{microtype}
\usepackage{wrapfig}

\AtBeginDocument{%
  \setlength{\abovedisplayskip}{5pt plus 2pt minus 2pt}%
  \setlength{\belowdisplayskip}{5pt plus 2pt minus 2pt}%
  \setlength{\abovedisplayshortskip}{2pt plus 2pt}%
  \setlength{\belowdisplayshortskip}{3pt plus 2pt minus 2pt}%
}
\graphicspath{{figures/}}

\newtheorem{theorem}{Theorem}

\title{Differentiate the Solver, Not the Equation:\\
Reverse-Sweep Adjoints for Block Implicit Simulation}

\author{%
Lei Shu$^{1}$ \hspace{0.7em}
Ying Jiang$^{2}$ \hspace{0.7em}
Kui Wu$^{3}$ \hspace{0.7em}
Yin Yang$^{4}$ \hspace{0.7em}
Leonidas Guibas$^{1}$ \hspace{0.7em}
Chenfanfu Jiang$^{2,*}$\\[6pt]
$^{1}$Stanford University \quad
$^{2}$University of California, Los Angeles\\
$^{3}$LightSpeed Studios, Tencent America \quad
$^{4}$The University of Utah\\[4pt]
{\footnotesize\texttt{leishu@stanford.edu, anajymua@gmail.com, walker.kui.wu@gmail.com}}\\
{\footnotesize\texttt{yin.yang@utah.edu, guibas@cs.stanford.edu, chenfanfu.jiang@gmail.com}}\\[2pt]
{\footnotesize $^{*}$Corresponding author}
}

\iclrfinalcopy   

\begin{document}
\maketitle

\begin{abstract}
Differentiable simulation is a key component in learning, control, and inverse
problems, where gradients through nonlinear implicit solvers are required.
Existing approaches either rely on unrolled automatic differentiation, whose
memory grows with solver depth, or on equation-level implicit differentiation,
which assembles global Jacobians and solves large sparse adjoint systems,
discarding the locality of the forward solver -- and differentiating the
\emph{converged} equation rather than the finite computation that actually
ran. To overcome these limitations, we propose \emph{solver-level
differentiation}, which differentiates the executed solver itself. Our key
insight is that when a solver is composed of block implicit updates, its
discrete adjoint is obtained by applying the corresponding adjoint updates in
reverse order, yielding a reverse-sweep formulation whose backward pass
mirrors the forward solver. From an operator perspective, the forward pass
realizes an approximate inverse through ordered local solves, and the backward
applies its transpose through reverse local adjoint solves, constructing no
global system. We instantiate this idea on Vertex Block Descent
\citep{Chen2024VBD}, yielding a differentiable solver whose reverse colored
Gauss--Seidel sweeps are composed entirely of local $3{\times}3$ adjoint
solves. The backward matches automatic differentiation through the identical
executed forward to machine precision at every solver depth, where the
equation-level adjoint is off by $37\%$ after one sweep; in a controlled
same-codebase, same-GPU comparison it is $33\times$ faster and uses
$71\times$ less memory than unrolled automatic differentiation; and the same
construction is exact on projective dynamics and extended position-based
dynamics. We scale differentiable elastodynamics to $10^{6}$ contact-coupled
soft bodies ($8$M vertices) on one GPU. Overall, this work highlights solver
structure as a practical organizing principle for efficient differentiable
simulation.
\end{abstract}

\section{Introduction}
\label{sec:intro}

Implicit integrators are the workhorse of stiff physical simulation
\citep{Baraff1998LargeSteps}. Each step solves a nonlinear stationary problem
\begin{equation}
\nabla G(\bm{x}) = \bm{0},
\qquad
G(\bm{x}) = \tfrac{1}{2h^{2}}(\bm{x}-\bm{y})^{\!\top} M (\bm{x}-\bm{y}) + E(\bm{x}),
\label{eq:stationary}
\end{equation}
where $G$ is the incremental potential of the step,
$\bm{x}\in\mathbb{R}^{3|V|}$ stacks the positions of $|V|$ mesh vertices,
$M$ is the mass matrix, $h$ the time step, $\bm{y}$ the inertial prediction
from the previous step, and $E$ the elastic potential. Differentiating through
such a step is what makes simulation usable as a layer: for system
identification, control and trajectory optimization, and training with a
simulator in the loop \citep{Hu2020DiffTaichi,Hu2019ChainQueen,Degrave2019DifferentiablePhysics}.

The obstacle is memory. \textbf{Unrolled automatic differentiation (AD)}
\citep{Griewank2008EvaluatingDerivatives} traces every elementary operation of
the executed program, so the tape grows with solver depth $\times$ rollout
length $\times$ mesh size; backpropagation through time
\citep{Werbos1990BPTT} over long rollouts then needs checkpointing.
It is exact -- we use it as ground truth -- but the first to run out of
memory. \textbf{Equation-level implicit differentiation}
\citep{Giles2000AdjointApproach,Blondel2022ImplicitDiff} instead discards the
executed solver, asserts $\nabla G(\bm{x}^{\star})=\bm{0}$, and applies the
implicit function theorem, requiring assembly and solution of a global adjoint
system. Memory becomes $O(1)$ in solver depth, but a global linear-algebra
object reappears -- and, less often noted, the gradient it returns is the
derivative of a \emph{converged} solve, not of the computation that ran when
the solver executes a small fixed number of sweeps, as it does in practice for
speed. The question is \emph{which function the backward should
differentiate: the equation the solver approximates, or the finite solver that
ran?}

We take the executed solver seriously as the object to differentiate. Many
modern simulators are \emph{block-coordinate implicit solvers}: they minimize
Eq.~\eqref{eq:stationary} by sweeping over small blocks of variables, solving a
tiny implicit problem per block and applying each update immediately. Vertex
Block Descent \citep{Chen2024VBD} is the instance we build on: one
$3{\times}3$ Newton solve per vertex, with vertices graph-colored so that a
color class updates in parallel. For any such solver the Jacobian of the sweep
is an ordered product of block-update Jacobians, so its transpose is that
product in reverse. The adjoint is therefore \emph{the same sweep, backwards},
with each block update replaced by its local transpose: same coloring, same
constant block size, same memory layout, and no global system at any point.
We read this as a structural duality -- forward block sweep
$\Leftrightarrow$ reverse adjoint sweep -- with a thesis attached:
\emph{equation-level differentiation destroys the solver locality that
block-coordinate methods are built around; solver-level differentiation
preserves it}. The backward pass is a solver of the same shape as the forward,
and everything that made the forward fast makes the backward fast.

\paragraph{What we are and are not claiming.} We do not claim custom gradients
are novel: a custom vector--Jacobian product can always be handed to an AD
framework. But that product still has to \emph{compute} something, and an implicit
step admits two computations: (a)~assemble and solve a global adjoint system, as
every prior differentiable elastodynamics method does, or (b)~run a reverse
sweep of constant-size local solves and never form such an object. Route (b)
removes the memory wall; we contribute its construction, exactness conditions,
proof, and GPU implementation, plus the measurement that (a) and (b) do not
merely differ in cost: at finite solver depth they return \emph{different
gradients}, and only (b) differentiates what was executed.

Exactness at finite depth is the subtle part. Three terms are required, and
dropping any one yields gradients that look correct at convergence and are badly
wrong before it: the local adjoint solve, the Hessian-tangent term, and, easily
overlooked, the Jacobians of non-smooth safeguards (update saturation,
regularization floors, validity gates). We found all three in our own
implementation against an autograd reference, and report their error laws because
they are invisible to the usual finite-difference-at-converged-$K$ protocol.

\paragraph{Contributions.}
\begin{itemize}[leftmargin=1.4em,itemsep=1pt,topsep=1pt]
\item The exact discrete adjoint of an executed block-coordinate implicit sweep
(\S\ref{sec:method}): a reverse sweep of constant-size local adjoint solves
that forms no global linear-algebra object, with the structural conditions
under which it is exact (Theorem~\ref{thm:adjoint}).
\item An analysis of what exactness at \emph{finite} solver depth requires
(\S\ref{sec:exactness}), including non-smooth safeguard terms that prior
hand-derived adjoints omit, with the measured error law of each omission
(\S\ref{sec:exp-ablate}).
\item A controlled comparison (one codebase, one GPU, identical meshes, and
forward trajectories verified identical to $10^{-16}$) against unrolled AD
and against both assembled and matrix-free equation-level adjoints
(\S\ref{sec:exp-cost}); plus inverse-problem tasks, machine-precision
transfer to projective and extended position-based dynamics
(App.~\ref{app:generality}), and differentiable elastodynamics at $10^{6}$
contact-coupled bodies ($8$M vertices).
\end{itemize}

\paragraph{Scope.} We implement and evaluate a single solver family, with
Vertex Block Descent as the concrete instantiation.
Theorem~\ref{thm:adjoint} states the structural assumptions under which the
construction transfers; minimal instantiations on extended
position-based dynamics \citep{Macklin2016XPBD} and projective dynamics
\citep{Bouaziz2014ProjectiveDynamics} confirm the transfer, exact to machine
precision at every tested depth (App.~\ref{app:generality}).

\section{Background and notation}
\label{sec:background}

\begin{figure}[t]
\centering
\begin{minipage}[c]{0.30\linewidth}
\centering
\includegraphics[width=\linewidth,trim=15 14 14 14,clip]{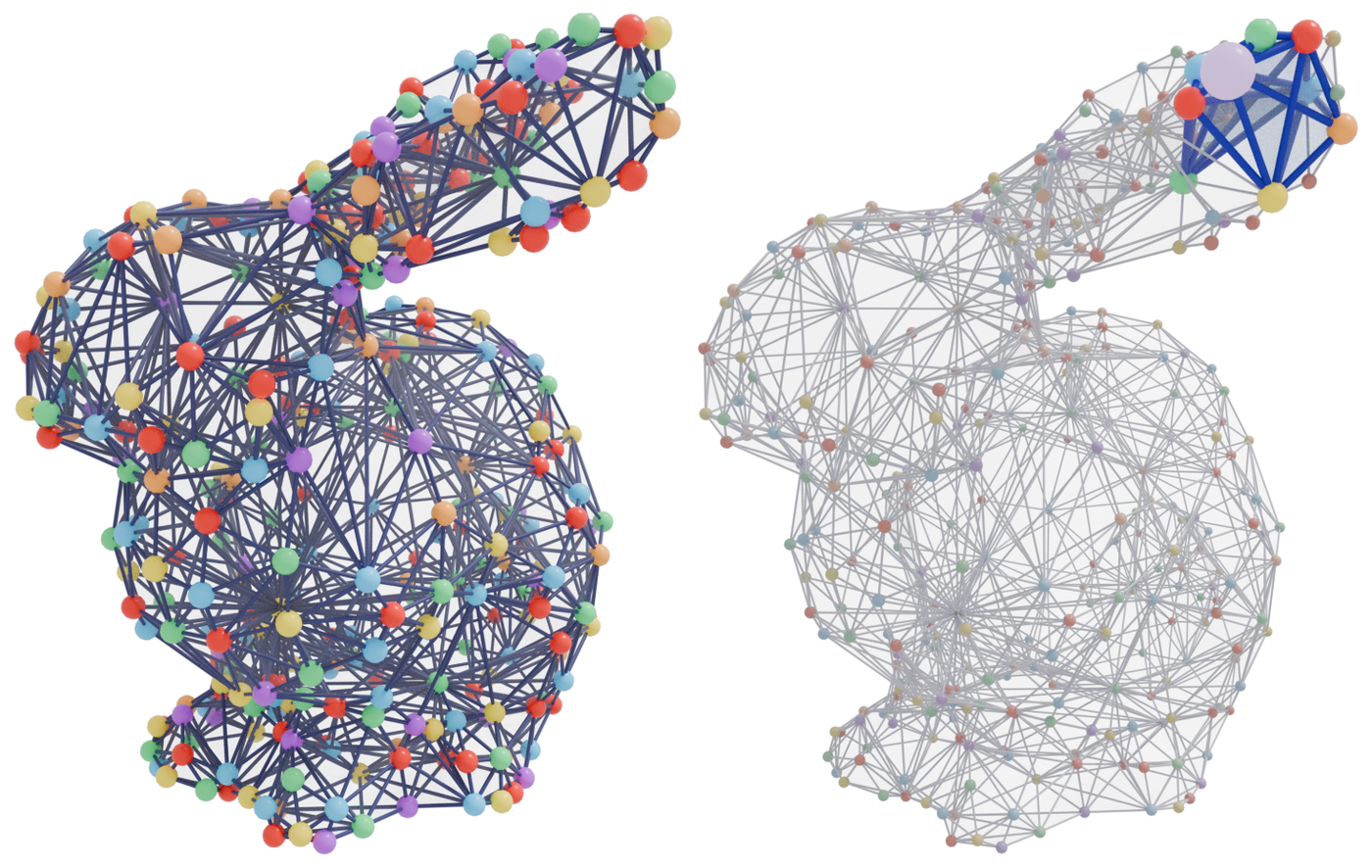}
\end{minipage}\hspace{0.035\linewidth}
\begin{minipage}[c]{0.60\linewidth}
\caption{Vertex Block Descent on a tetrahedral mesh. Left: distance-1 graph
coloring; vertices that share an element never share a color, so each color
class updates in parallel. Right: one local $3{\times}3$ Newton solve,
touching only the elements incident to the updated vertex.}
\label{fig:overview}
\end{minipage}
\vspace{-6pt}
\end{figure}

\paragraph{Notation.} Bold italic denotes vectors ($\bm{x}$, $\bm{v}$,
$\bm{y}$, $\bm{g}_i$, $\bm{p}_i$, $\Delta\bm{x}_i$); roman capitals denote
matrices ($M$, $H_i$, $I$); plain italic denotes scalars ($h$, $K$, $\mu$,
$\lambda$, $\epsilon$, and indices $i,j,t$); italic capitals denote operators on
state ($S_i$, $U_i$, $F_i$). An overbar denotes an \emph{adjoint}: for a
downstream scalar loss $L$, $\bar{\bm{x}} \coloneqq \partial L/\partial\bm{x}$.
We write $|V|$ for the vertex count, so a mesh has $3|V|$ degrees of freedom
(DoF), and $|T|$ for the tetrahedron count. \textbf{Solver depth} $K$ is the
number of inner sweeps executed per time step, where one sweep visits every
vertex once in color order; $K$ is a fixed user-chosen budget, not a
convergence tolerance, and small $K$ is the common regime because it is fast.

\paragraph{Vertex Block Descent.} \citet{Chen2024VBD} solve
Eq.~\eqref{eq:stationary} by block coordinate descent on the variational form of
implicit Euler, taking one vertex position as a block and sweeping the blocks in
Gauss--Seidel order. Each local update decreases the same global objective, so
the scheme is unconditionally stable and, unlike a truncated global Newton
solve, \emph{remains} stable when the iteration count is capped: the sweep count
is a compute budget rather than a tolerance -- so finite $K$ is the normal
operating point, and exactly the regime where an equation-level adjoint stops
describing what ran. Concretely, for vertex $i$ with lumped mass $m_i$ and incident
elements $\mathcal{E}(i)$,
\begin{equation}
\begin{aligned}
\bm{g}_i(\bm{x}) &= \tfrac{m_i}{h^{2}}(\bm{x}_i - \bm{y}_i)
  + \!\!\sum_{e \in \mathcal{E}(i)}\!\! \nabla_{\bm{x}_i} E_e(\bm{x}), \\
H_i(\bm{x}) &= \tfrac{m_i}{h^{2}} I_3
  + \!\!\sum_{e \in \mathcal{E}(i)}\!\! \nabla^{2}_{\bm{x}_i\bm{x}_i} E_e(\bm{x}),
\end{aligned}
\label{eq:local-gh}
\end{equation}
and the local Newton step is applied immediately (Gauss--Seidel):
\begin{equation}
H_i(\bm{x})\,\Delta\bm{x}_i = -\bm{g}_i(\bm{x}),
\qquad
\bm{x}_i \leftarrow \bm{x}_i + \Delta\bm{x}_i .
\label{eq:local-newton}
\end{equation}
Vertices are distance-1 graph-colored, so vertices sharing an element never
share a color and each color class updates in parallel
(Fig.~\ref{fig:overview}). Vertex Block Descent is thus a block implicit solver whose
fundamental operation is a constant-size $3{\times}3$ solve. Elastic energy is
stable Neo-Hookean \citep{Smith2018StableNeoHookean} throughout.

\section{The adjoint of the executed sweep}
\label{sec:method}

\subsection{Block updates and the solver-induced inverse}

Write the executed solver as a composition of block updates
$S = S_n \circ S_{n-1}\circ\cdots\circ S_1$, where $S_i$ modifies only block
$i$:
\begin{equation}
\bm{x}_i^{+} = U_i(\bm{x}) = \bm{x}_i + \Delta\bm{x}_i(\bm{x}),
\qquad
F_i(\Delta\bm{x}_i, \bm{x}) = \bm{0},
\label{eq:block-update}
\end{equation}
with $\Delta\bm{x}_i$ defined implicitly by $F_i$; for Vertex Block Descent,
$F_i(\Delta\bm{x}_i,\bm{x}) = H_i(\bm{x})\Delta\bm{x}_i + \bm{g}_i(\bm{x})$.
Iterating this composition realizes an \emph{approximate inverse} of the global
stationary system: $S$ maps the predictor $\bm{y}$ to a state of lower
$G$-energy, and finitely many sweeps reduce the residual. We call $S$ the
solver-induced inverse; it is block-structured by construction.

\subsection{Transpose action, and contrast with equation-level differentiation}

The Jacobian of the sweep is the ordered product of block-update Jacobians, so
\begin{equation}
(\partial S)^{\!\top} = (\partial S_1)^{\!\top}(\partial S_2)^{\!\top}\cdots(\partial S_n)^{\!\top}.
\label{eq:transpose}
\end{equation}
Two consequences. Reverse ordering is the computational form of the transpose
of a composition, not a design choice. And because each $\partial S_i$ touches
only one block, so does each $(\partial S_i)^{\!\top}$: the backward is the
transpose of the forward sweep block by block, with the same color partition in
reverse and the same constant block size.

This differentiates a different object from equation-level implicit
differentiation, which replaces the executed solver by the converged equation
and inverts $\nabla^2 G(\bm{x}^{\star})$:
\begin{equation}
\underbrace{(\partial S)^{\!\top}}_{\substack{\text{adjoint of the finite executed}\\\text{solver; block-local}}}
\qquad\text{vs.}\qquad
\underbrace{\bigl(\nabla^{2}G(\bm{x}^{\star})\bigr)^{-\!\top}}_{\substack{\text{adjoint of the converged}\\\text{system; global solve}}}
\label{eq:solver-vs-equation}
\end{equation}
The two coincide only in the convergence limit. \S\ref{sec:exp-exact} measures
the gap and shows it is large in the regime practitioners actually use.

\subsection{Local adjoint of one block update}
\label{sec:local-adjoint}

Differentiating $F_i = \bm{0}$ with respect to an arbitrary variable $z$ gives
$\frac{\partial F_i}{\partial \Delta\bm{x}_i}\frac{d\Delta\bm{x}_i}{dz} + \frac{\partial F_i}{\partial z} = \bm{0}$,
and since $\partial F_i/\partial\Delta\bm{x}_i = H_i$,
\begin{equation}
\frac{d\Delta\bm{x}_i}{dz}
= -H_i^{-1}\!\left(\frac{\partial \bm{g}_i}{\partial z}
+ \frac{\partial H_i}{\partial z}\,\Delta\bm{x}_i\right).
\label{eq:implicit-delta}
\end{equation}
We differentiate the local \emph{optimality condition}, not the implementation
of the local linear solver: the backward is organized by the implicit relation
itself rather than by an expanded reverse-mode graph.

\begin{figure}[t]
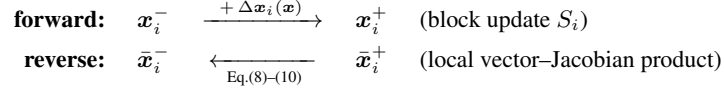

\centering
\footnotesize
\begin{tabular}{r c c c l}
\textbf{forward:} & $\bm{x}_i^{-}$ & $\xrightarrow{\;\;+\,\Delta\bm{x}_i(\bm{x})\;\;}$ & $\bm{x}_i^{+}$ & (block update $S_i$)\\[4pt]
\textbf{reverse:} & $\bar{\bm{x}}_i^{-}$ & $\xleftarrow[\text{\;\;Eq.\eqref{eq:identity-adjoint}--\eqref{eq:adjoint-update-general}\;\;}]{}$ & $\bar{\bm{x}}_i^{+}$ & (local vector--Jacobian product)
\end{tabular}
\caption{Data flow around one block update. The forward advances the block
state; the reverse consumes the \emph{incoming} adjoint $\bar{\bm{x}}_i^{+}$
(accumulated by reverse updates already executed) and produces the
\emph{outgoing} adjoint $\bar{\bm{x}}_i^{-}$. Superscripts denote position in
the sweep, not before/after in time.}
\label{fig:timeline}
\end{figure}

Let $\bar{\bm{x}}_i^{+}$ be the \emph{incoming} adjoint at block $i$, the
one accumulated by every reverse update that has already executed, and
$\bar{\bm{x}}_i^{-}$ the \emph{outgoing} adjoint passed further back
(Fig.~\ref{fig:timeline}). Because
$\bm{x}_i^{+} = \bm{x}_i^{-} + \Delta\bm{x}_i(\bm{x})$, the chain rule splits
into a direct identity branch and an implicit branch:
\begin{equation}
\bar{\bm{x}}_i^{-} \;=\; \bar{\bm{x}}_i^{+}
\;+\; \Bigl(\tfrac{d\Delta\bm{x}_i}{d\bm{x}_i}\Bigr)^{\!\top}\bar{\bm{x}}_i^{+}.
\label{eq:identity-adjoint}
\end{equation}
Introducing the local adjoint variable $\bm{p}_i$ defined by
\begin{equation}
H_i^{\!\top}\bm{p}_i = \bar{\bm{x}}_i^{+},
\label{eq:local-adjoint}
\end{equation}
and combining with Eq.~\eqref{eq:implicit-delta}, the implicit branch for
\emph{any} upstream variable $z$ becomes
\begin{equation}
\Bigl(\tfrac{d\Delta\bm{x}_i}{dz}\Bigr)^{\!\top}\bar{\bm{x}}_i^{+}
= -\Bigl(\tfrac{\partial \bm{g}_i}{\partial z}\Bigr)^{\!\top}\bm{p}_i
\;-\; \Bigl(\tfrac{\partial H_i}{\partial z}\Bigr)\!:\!\bigl(\bm{p}_i\,\Delta\bm{x}_i^{\!\top}\bigr).
\label{eq:adjoint-update-general}
\end{equation}
Here $A\!:\!B=\sum_{a,b}A_{ab}B_{ab}$, so the second term is the contraction of
the third-order tensor $\partial H_i/\partial z$ against the rank-one matrix
$\bm{p}_i\Delta\bm{x}_i^{\!\top}$,
\begin{equation}
\Bigl(\tfrac{\partial H_i}{\partial z}\Bigr)\!:\!\bigl(\bm{p}_i\Delta\bm{x}_i^{\!\top}\bigr)
\;=\;\sum_{a,b=1}^{3}\frac{\partial [H_i]_{ab}}{\partial z}\,[\Delta\bm{x}_i]_b\,[\bm{p}_i]_a .
\label{eq:hessian-tangent-index}
\end{equation}
No third-order tensor is ever formed: Eq.~\eqref{eq:hessian-tangent-index} is
evaluated as a directional derivative of $H_i$ along $\Delta\bm{x}_i$, tested
against $\bm{p}_i$, at the cost of a few $3{\times}3$ products.

Setting $z = \bm{x}_i$ recovers the implicit summand of
Eq.~\eqref{eq:identity-adjoint}; $z=\bm{x}_j$ accumulates the off-diagonal
contribution into $\bar{\bm{x}}_j^{-}$. For a material parameter
$\theta\in\{\mu,\lambda\}$, or any other upstream quantity the block depends on
(nodal forces, rest shape), the same expression accumulates
\begin{equation}
\bar{\theta} \;\mathrel{+}=\;
-\Bigl(\tfrac{\partial \bm{g}_i}{\partial \theta}\Bigr)^{\!\top}\bm{p}_i
\;-\;\Bigl(\tfrac{\partial H_i}{\partial \theta}\Bigr)\!:\!\bigl(\bm{p}_i\,\Delta\bm{x}_i^{\!\top}\bigr),
\label{eq:param-adjoint}
\end{equation}
summed over every block update in which $\theta$ participated. State and
parameter adjoints therefore share one local solve
$H_i^{\!\top}\bm{p}_i=\bar{\bm{x}}_i^{+}$; parameters cost only the extra
contractions, which is why adding material gradients to a rollout does not
change the asymptotic cost of the backward. Together
Eqs.~\eqref{eq:identity-adjoint}--\eqref{eq:adjoint-update-general} are the
complete local vector--Jacobian product of one block update. We call the second term of
Eq.~\eqref{eq:adjoint-update-general} the \textbf{Hessian tangent}
(``stage~2''), since it exists only because $H_i$ itself depends on state; the
first is the gradient tangent (``stage~1''). Every quantity is $3{\times}3$;
nothing global is formed.

\subsection{Reverse-sweep adjoint principle}
\label{sec:principle}

\begin{theorem}[Exact discrete adjoint of a block implicit solver]
\label{thm:adjoint}
Let $S = S_n\circ\cdots\circ S_1$ be a finite composition of block implicit
updates of the form \eqref{eq:block-update}. Assume each $F_i$ is $C^1$ in a
neighborhood of its executed linearization state and that
$\partial F_i/\partial\Delta\bm{x}_i$ is invertible there; for the affine
instance $F_i = H_i\Delta\bm{x}_i + \bm{g}_i$ it suffices that $H_i$ and
$\bm{g}_i$ are differentiable at that state with $H_i$ invertible. Then
$(\partial S)^{\!\top} = (\partial S_1)^{\!\top}\cdots(\partial S_n)^{\!\top}$,
with each $(\partial S_i)^{\!\top}$ realized by the local vector--Jacobian product of
Eqs.~\eqref{eq:identity-adjoint}--\eqref{eq:adjoint-update-general}: a direct
identity contribution, a local adjoint solve
$H_i^{\!\top}\bm{p}_i = \bar{\bm{x}}_i^{+}$, and a local scatter. Provided the
backward replays the same local linearization states as the forward, the
adjoint of one sweep is a reverse block sweep of constant-size local solves.
\end{theorem}

The proof is in Appendix~\ref{app:proof}. The qualifier that matters: this is
exact for the \emph{finite executed} $S$, not merely in the convergence limit.
It does not assert that the block factorization is unique or that it converges
to the global linearized inverse for any particular $K$.

\subsection{What exactness at finite depth actually requires}
\label{sec:exactness}

Theorem~\ref{thm:adjoint} assumes each $F_i$ is differentiated in full.
Production block solvers contain safeguards that are easy to treat as
invisible, and each omission produces gradients that are wrong at small $K$ and
self-correcting at large $K$, so they hide from convergence-regime testing.
We isolate three, all of which we found in our own implementation.

\textbf{(i) Update saturation.} Implementations commonly damp the raw step,
$\Delta\bm{x}\leftarrow\Delta\bm{x}\,c/(\|\Delta\bm{x}\|+c+\varepsilon)$
with damping constant $c$ (App.~\ref{app:impl} gives the values), to stabilize
early sweeps. Its Jacobian is a rank-one correction that must be applied to the
incoming adjoint \emph{before} the local solve:
\begin{equation}
J = s\,I - \tfrac{1}{c\,r}\,\Delta\bm{x}\,\Delta\bm{x}^{\!\top},
\qquad r = \|\Delta\bm{x}^{\text{raw}}\|,\quad s = \|\Delta\bm{x}\|/r .
\label{eq:sat-jac}
\end{equation}
Omitting it gives relative error $\propto \|\Delta\bm{x}\|/(\|\Delta\bm{x}\|+c)$,
measured up to $26\%$ at $K{=}1$ in Table~\ref{tab:ablation}, vanishing as
$\|\Delta\bm{x}\|\to0$.

\textbf{(ii) The Hessian tangent.} The second term of
Eq.~\eqref{eq:adjoint-update-general} is often dropped from the \emph{state}
adjoint and retained, if at all, only for parameter gradients. For the block
Hessian of Eq.~\eqref{eq:local-gh} the state dependence enters only through
$\bm{q}_i=\mathrm{cof}(F)\nabla N_i$, so the term reduces to two cross products
per (element, vertex) pair. Omitting it leaves error $\propto\|\Delta\bm{x}\|$.

\textbf{(iii) Validity gates.} Tangent terms are sometimes disabled on
compressed elements (e.g.\ applied only when $\det F > \tau$) while the forward
applies no such cutoff. This is the most damaging of the three: its effect is
non-local and \emph{grows with mesh refinement}: one gated element out of
$245$ moves the whole gradient by $8.2\times10^{-3}$, and the gated terms
contain no division by $\det F$, so the gate buys no safety.

None of the three is visible from finite-difference checks at converged $K$,
which is the standard validation protocol; all three are visible immediately
against an autograd reference at $K{=}1$ (\S\ref{sec:exp-ablate}).

\section{Differentiable Vertex Block Descent}
\label{sec:diffvbd}

We instantiate the principle on Vertex Block Descent. Each forward update records its local
displacement $\Delta\bm{x}_i$ and Hessian block; the backward traverses
iterations and colors in reverse, undoing the recorded displacement to
reconstruct the linearization point, applying Eq.~\eqref{eq:sat-jac}, passing
the direct adjoint contribution, solving
$H_i^{\!\top}\bm{p}_i=\bar{\bm{x}}_i^{+}$, and scattering the stage-1 and
stage-2 contributions to neighboring vertices and material parameters via
Eq.~\eqref{eq:adjoint-update-general}. The same backward yields gradients with
respect to initial state $(\bm{x}_0,\bm{v}_0)$, Lam\'e parameters
$(\mu,\lambda)$, time-varying external forces $\bm{f}_t$, and rest-shape
positions, in one reverse sweep with no global assembly and no graph rebuild
between parameter sets (closed forms: App.~\ref{app:closed-forms}). Forward and backward share the color partition,
neighborhood structure, memory layout, and thread schedule, so the backward
inherits the forward's parallelism directly (Algorithm~\ref{alg:diffvbd}).
Relative to unrolled AD the difference is what is \emph{not} stored: the tape
collapses to a per-sweep replay buffer of nine doubles per active vertex --
the applied update and the six entries of the symmetric block, $72$ bytes.

\begin{algorithm}[t]
\caption{Forward Vertex Block Descent and its reverse adjoint}
\label{alg:diffvbd}
\begin{algorithmic}[1]
\For{$k=0,\dots,K-1$}\Comment{forward}
  \For{color $c=1,\dots,C$}
    \ForAll{vertices $i$ in color $c$ \textbf{in parallel}}
      \State assemble $\bm{g}_i,H_i$; solve $H_i\Delta\bm{x}_i=-\bm{g}_i$; saturate
      \State record $\Delta\bm{x}_i,H_i$; \ $\bm{x}_i\leftarrow\bm{x}_i+\Delta\bm{x}_i$
    \EndFor
  \EndFor
\EndFor
\For{$k=K-1,\dots,0$}\Comment{reverse}
  \For{color $c=C,\dots,1$}
    \ForAll{vertices $i$ in color $c$ \textbf{in parallel}}
      \State undo $\Delta\bm{x}_i$; apply saturation Jacobian \eqref{eq:sat-jac} to $\bar{\bm{x}}_i^{+}$
      \State solve $H_i^{\!\top}\bm{p}_i=\bar{\bm{x}}_i^{+}$
      \State scatter stage-1 and stage-2 terms \eqref{eq:adjoint-update-general} to neighbors / parameters
    \EndFor
  \EndFor
\EndFor
\end{algorithmic}
\end{algorithm}

\section{Implementation}
\label{sec:impl}

We provide two GPU backends, NVIDIA Warp \citep{Macklin2022Warp} kernels and a
fused CUDA C++ path compiled on demand, differing by $7$--$22\times$ in runtime
through kernel fusion but agreeing to $10^{-16}$ on forward state and gradients.
Memory grows with replay information, not with an AD graph, and CUDA graph
capture removes launch overhead in outer loops.
The per-vertex block defaults to the positive-definite approximation
$H_i=\mu\|\nabla N_i\|^2 I+\lambda\bm{q}\bm{q}^{\!\top}$ (per element,
with $\nabla N_i$ the constant shape-function gradient) of
\citet{Chen2024VBD}; an exact $9{\times}9$-tangent variant covers the
indefinite stable Neo-Hookean Hessian. \paragraph{Contact.} Two primitives coexist, and both are differentiated the
same way as the elastic term, by the reverse sweep rather than by a separate
mechanism. The first is the incremental potential contact (IPC) log barrier
\citep{Li2020IPC} against the floor and against the vertex's own neighborhood,
$b(d) = -\kappa\,(d-\hat d)^2\log(d/\hat d)$ for gap $d<\hat d$, with lagged
dissipation applied only to downward motion. Because the barrier contributes
to the same per-vertex $\bm{g}_i$ and $H_i$ of Eq.~\eqref{eq:local-gh}, its
derivative enters the backward through the existing local solve, and no
contact-specific adjoint is added; the barrier's second derivative sits in
$H_i$, and its third derivative is carried by the Hessian-tangent term of
Eq.~\eqref{eq:adjoint-update-general}. Adjoints therefore flow through every
barrier evaluation of a rollout, not through a converged contact state. The
second primitive is a differentiable center-of-mass penalty between separate
bodies, with a uniform-grid broad phase (sorted cell keys, one binary search
per neighbor cell) that stays $O(N)$ at $10^{6}$ bodies, used when the object
count makes vertex-level self-collision impractical. Pair indices carry no
gradient, so the search algorithm never touches differentiability.

This is the substantive difference from DiffIPC \citep{Huang2024DiffIPC}, which
differentiates a converged incremental potential and therefore inherits a global
step Hessian on the backward; ours differentiates the executed barrier
evaluations and stays block-local, which is why the memory figures quoted in
\S\ref{sec:exp-runtime} differ by orders of magnitude. Symmetrically, where
the contact set is still changing, our gradient reflects the executed sweep,
not a converged contact configuration.
Both backends expose a tensor-native interface, so backpropagation through
time runs GPU-resident.
Appendix~\ref{app:impl} gives details.

\section{Experiments}
\label{sec:exp}

Optimization tasks use Adam \citep{Kingma2015Adam}.
\S\ref{sec:exp-exact}--\S\ref{sec:exp-inverse} run on a single RTX~4090 Laptop GPU
(16\,GB, CUDA 12.1, PyTorch 2.3), \S\ref{sec:exp-runtime}'s $K\times|V|$ grid
on an RTX~A6000, and \S\ref{sec:exp-scale} on one H100 (80\,GB). Every backward
route in \S\ref{sec:exp-exact}--\S\ref{sec:exp-cost} is measured on the
\emph{same} GPU within a single process, so no comparison below crosses
hardware.

\paragraph{Baselines and the reference gradient.} We compare four backward routes over an
\emph{identical} forward: (1)~\textbf{ours}; (2)~\textbf{unrolled AD}, PyTorch
autograd through the executed colored sweep; (3)~\textbf{global adjoint
(assembled)}, building the sparse $3|V|\times3|V|$ Hessian and solving the
adjoint system by preconditioned conjugate gradients (PCG), the route of
DiffPD, DiffCloth, DiffQN and DiffIPC; (4)~\textbf{global adjoint
(matrix-free)}, never forming the Hessian, applying it via double-backward
Hessian-vector products inside PCG. We include (4) deliberately: it is the
leanest possible opponent on memory, and is exactly the design point of
supplying a custom vector--Jacobian product for the linear solve inside an AD
framework. Routes (3)
and (4) agree to $2\times10^{-16}$, certifying they solve the same system.

Unrolled AD serves twice. As a competitor it shows the cost of a generic tape.
As a \emph{reference} it is exact by construction, since autograd through the
executed program is, to floating point, the derivative of what ran, so all
gradient errors below are relative errors against it. We verify that all
backends produce the same forward trajectory to $10^{-16}$ before comparing any
gradient; without that check a gradient comparison measures nothing.

\subsection{Exactness at finite solver depth}

\label{sec:exp-exact}

\begin{table}[t]
\caption{\textbf{Gradient error vs.\ the autograd reference} (relative, max over
$\bar{\bm{x}}_0,\bar{\bm{v}}_0$) as solver depth $K$ varies. Our reverse sweep
is exact at every depth. The equation-level adjoint differentiates a converged
solve, so it is accurate only once the executed state satisfies
$\nabla G=\bm{0}$; the residual column quantifies how badly that assumption
fails at small $K$. Mesh $|V|{=}115$ ($345$ DoF), backpropagation through
four rollout steps.}
\label{tab:exactness}
\centering\small
\renewcommand{\arraystretch}{1.12}
\begin{tabular}{lcccc}
\toprule
$K$ & $\|\nabla G(\bm{x}^{\star})\|$ & Ours & Global adj.\ (assembled) & Global adj.\ (matrix-free) \\
\midrule
$1$  & $1.6\times10^{-1}$  & $\bm{3.6\times10^{-15}}$ & $3.7\times10^{-1}$ & $3.7\times10^{-1}$ \\
$5$  & $4.7\times10^{-9}$  & $\bm{3.3\times10^{-15}}$ & $2.3\times10^{-5}$ & $2.3\times10^{-5}$ \\
$20$ & $1.0\times10^{-14}$ & $\bm{2.1\times10^{-15}}$ & $9.7\times10^{-10}$ & $1.6\times10^{-10}$ \\
\bottomrule
\end{tabular}
\end{table}

Table~\ref{tab:exactness} is the central measurement. At $K{=}1$ the executed
state violates the stationary equation by $1.6\times10^{-1}$ and the
equation-level gradient is wrong by $37\%$; ours is exact. By $K{=}20$ the
solver has converged and the routes agree, as Theorem~\ref{thm:adjoint}
predicts. This is the operational content of differentiating the solver rather
than the equation: the two are not two implementations of one gradient but
gradients of two different functions, and only one of them is the function the
forward pass evaluated. Exactness holds across every axis swept (meshes $|V|\in\{26,115,631\}$,
$K\in\{1,5,20\}$, one- and four-step rollouts, three decades of stiffness,
two step sizes, four decades of update magnitude): worst case
$3.6\times10^{-15}$, on both GPU backends (finite-difference validation:
App.~\ref{sec:exp-fd}).

\subsection{Ablation: the three exactness terms}
\label{sec:exp-ablate}

\begin{table}[t]
\caption{\textbf{Ablation of \S\ref{sec:exactness}}: relative gradient
error vs.\ the reference at $K{=}1$, across mesh refinement.}
\label{tab:ablation}
\centering\small
\renewcommand{\arraystretch}{1.12}
\begin{tabular}{lccc}
\toprule
Variant & $|V|{=}26$ & $|V|{=}115$ & $|V|{=}631$ \\
\midrule
Full adjoint (ours)              & $\bm{4.2\times10^{-16}}$ & $\bm{5.1\times10^{-16}}$ & $\bm{9.1\times10^{-16}}$ \\
w/o saturation Jacobian (i)      & $7.7\times10^{-2}$ & $1.6\times10^{-1}$ & $2.6\times10^{-1}$ \\
w/o Hessian tangent (ii)         & $2.9\times10^{-4}$ & $8.2\times10^{-3}$ & $1.1\times10^{-1}$ \\
with validity gate $\det F>0.5$ (iii) & $4.2\times10^{-16}$ & $8.2\times10^{-3}$ & $1.2\times10^{-1}$ \\
\bottomrule
\end{tabular}
\end{table}

Each omission in Table~\ref{tab:ablation} is invisible at convergence, severe
before it, and \emph{grows} under mesh refinement -- the opposite of what
large-scale use requires. The ablated variants are earlier revisions of our own
implementation, so each row is a defect that actually shipped. The gated
variant (iii) is exact on the coarse mesh only because no element there
approaches inversion: the gate never fires until refinement activates it, which
is precisely what makes this omission the easiest to miss.

\subsection{Cost: a controlled comparison}
\label{sec:exp-cost}

\begin{table}[t]
\caption{\textbf{Backward cost: same codebase, GPU, and mesh}, with forward
trajectories verified identical to $10^{-16}$, so no part of the gap is
attributable to a faster forward, a different mesh, or different hardware.
$|V|{=}115$, $K{=}20$, backpropagation through four rollout steps. ``PCG''
is the conjugate-gradient
iteration count. Peak memory is the caching-allocator peak
(\texttt{torch.cuda.max\_memory\_allocated}).}
\label{tab:cost}
\centering\small
\renewcommand{\arraystretch}{1.12}
\begin{tabular}{lcccc}
\toprule
Backward route & Time (ms) & Peak memory (MB) & PCG iters & Gradient error \\
\midrule
Ours                          & $\bm{28.5}$ & $\bm{0.9}$ & n/a & $2.1\times10^{-15}$ \\
Unrolled AD (reference)          & $939.9$ & $64.3$ & n/a & n/a \\
Global adjoint (assembled)    & $257.6$ & $19.8$ & $36$ & $9.7\times10^{-10}$ \\
Global adjoint (matrix-free)  & $89.4$  & $17.7$ & $40$ & $1.6\times10^{-10}$ \\
\bottomrule
\end{tabular}
\end{table}

Table~\ref{tab:cost} gives $33\times$ less time and $71\times$ less memory than
unrolled AD. Notably, the regimes separate cleanly. When the step is well
conditioned ($h{=}10^{-2}$, soft material) PCG converges in ${\sim}5$
iterations and the equation-level route is at its closest -- there, our
advantage is memory and the absence of assembly. As conditioning degrades the
routes diverge: at $\mu{=}\lambda{=}5\times10^{5}$ PCG needs $233$ iterations
and the equation-level backward rises to $247$--$389$\,ms while ours does not
move with stiffness: a reverse sweep never solves a global system whose
difficulty tracks the material.
The corresponding growth in problem size is Table~\ref{tab:scaling}
(App.~\ref{app:scaling}), where our backward stays within
$0.96$--$1.14\times$ the forward up to $|V|{=}201$k, a regime in which the
unrolled tape does not fit in device memory at all.

\subsection{Runtime and scaling}
\label{sec:exp-runtime}

On a bunny mesh ($339$ vertices, $897$ tetrahedra, $40$ steps) the Warp backend
runs $34.7$/$146.4$\,ms per forward/backward step; the fused CUDA C++ path
gives ${\sim}3$/$5$\,ms. The backward-to-forward ratio is \emph{not} a single
number: at small $|V|$ it is dominated by fixed per-color kernel-launch
overhead, and as $|V|$ grows per-vertex work overtakes that overhead and the
ratio approaches $1$, measuring $0.96$--$1.14$ across the full $K\times|V|$
grid up to $|V|{=}201$k (Appendix~\ref{app:scaling}). Memory is linear in
$K|V|$ at ${\approx}4$\,KB per vertex per sweep, holding to $\pm13\%$ across the measured $K\times|V|$ grid; $360\times$ more vertices give $370\times$ memory at
fixed $K$. For reference, DiffIPC \citep{Huang2024DiffIPC} reports
${\sim}100$\,GB at $|V|{=}9{,}939$ against our $1{,}074$\,MB at $|V|{=}27{,}169$, two orders of magnitude less at a larger problem, though we stress this
is a cross-paper comparison of published numbers, unlike
Table~\ref{tab:cost}.

Solver depth also affects what an optimizer achieves, not only gradient
accuracy, though on our small initial-state task the effect does not separate
from run-to-run spread; we report that null result in
App.~\ref{app:kdepth}.

\subsection{Inverse problems}
\label{sec:exp-inverse}

We verify the gradients are useful, not merely correct. In each task we state
what is given, what is optimized, and the success criterion (narratives:
App.~\ref{app:inverse-detail}; protocols: App.~\ref{app:tasks}; more tasks:
App.~\ref{app:extra-tasks}); Table~\ref{tab:inverse} reports the number of
scalar optimization variables (DoF), iterations, final loss with units, and a
task-interpretable outcome.

\begin{table}[t]
\caption{Inverse-problem results, all using the DiffVBD adjoint with Adam.
\textbf{\#vars} is the number of scalar optimization variables. \textbf{Final
loss} is the converged objective in task units: relative parameter-recovery
loss (dimensionless) for material ID; per-vertex squared position error
(m$^2$) for dress and silhouette; squared trajectory error (m$^2$) for wind;
squared center-of-mass-to-target distance (m$^2$) for the bounces.
\textbf{Outcome} is task-interpretable: relative material error; for wind,
$\|\Delta w\|$ is the $L_2$ deviation of the recovered piecewise-constant wind
from the reference, marked under-determined for the reason given in
\S\ref{sec:exp-inverse}; for the bounces, the final center-of-mass-to-target distance, against
trajectory envelopes of ${\approx}1.8$\,m (cube) and ${\approx}0.8$\,m (bunny),
so $0.013$\,m and $0.034$\,m are small relative to the motion.}
\label{tab:inverse}
\centering\small
\renewcommand{\arraystretch}{1.12}
\begin{tabular}{lcccl}
\toprule
Task & \#vars & Iters & Final loss & Outcome \\
\midrule
Beam $\mu$ ID              & $1$  & $76$  & ${<}10^{-4}$              & $\mu$: $0.1\%$ error \\
Beam $\mu,\lambda$ ID      & $2$  & $100$ & $2.1\times10^{-4}$        & $\mu,\lambda$: ${<}0.25\%$ \\
Dress $\mu,\lambda$ ID     & $2$  & $250{+}5$ & $4.8\times10^{-9}$ m$^2$ & $\mu$: $0.01\%$, $\lambda$: $0.03\%$ \\
Dress silhouette           & $2$  & $350{+}5$ & ${<}10^{-12}$ m$^2$   & $\mu$: ${<}0.01\%$, $\lambda$: ${<}0.01\%$ \\
Flag wind control          & $24$ & $199$ & $5.1\times10^{-3}$ m$^2$  & $\|\Delta w\|{=}15.3$ (under-det.) \\
Cube bounce                & $6$  & $151$ & $1.6\times10^{-2}$ m$^2$  & center-of-mass--target $0.30\to0.013$\,m \\
Bunny bounce               & $3$  & $100$ & $3.1\times10^{-2}$ m$^2$  & center-of-mass--target $0.035\to0.034$\,m \\
Cover the Spot             & $18$ & $600$ & n/a & uncovered area $0.0\%$ \\
\bottomrule
\end{tabular}
\end{table}

\subsection{Scale}
\label{sec:exp-scale}

We jointly optimize per-body $\bm{v}_0$ and $\mu$ for $10^{6}$
tetrahedralized cubes (edge $3$\,mm, ${\sim}8$M vertices, ${\sim}6$M tetrahedra
packed as one mega-mesh) so that at $t^{\star}{=}0.6$\,s each body's center of mass
reaches a target voxel; bodies continue ballistically past $t^{\star}$, so the
optimizer must place each on a trajectory whose apex crosses its target at
exactly that step. Rollouts are $240$ implicit steps ($h{=}5$\,ms, $K{=}5$)
with backpropagation through time, gradient-checkpointed
\citep{Chen2016SublinearMemory} every $5$ steps
under a Huber \citep{Huber1964Robust} objective. Bodies interact through the
differentiable center-of-mass contact penalty of \S\ref{sec:impl} (its
uniform-grid broad phase, over $10^{6}$ centers per step), and the backward differentiates
through every contact impulse. A contact-free stage first brings bodies near
their targets ($14.8$\,h, ${\sim}71$\,GB, mean error $0.291$\,m); $200$
Adam iterations under the contact-enabled dynamics then land at $0.311$\,m
($16.7$\,h, ${\sim}74$\,GB on one H100), the $0.02$\,m gap being the price
of forbidding overlap at the formation; overall the loss falls
$1.03\times10^{6}\to1.31\times10^{5}$ ($87\%$). The experiment is a stress test of differentiability
under simultaneous solver-depth, object-count, contact, and memory pressure,
not a visual benchmark.

\section{Related work}
\label{sec:related}

\paragraph{Differentiable simulators.} Differentiable physics is established
for soft bodies
\citep{Hu2020DiffTaichi,Hu2019ChainQueen,Murthy2021gradSim,Qiao2020ScalableDifferentiablePhysics},
cloth and frictional contact
\citep{Liang2019DifferentiableCloth,Geilinger2020ADD,Li2022DiffCloth},
articulated bodies \citep{Werling2021Nimble}, and Eulerian PDEs
\citep{Holl2020PhiFlowPDE}. The dominant implementation unrolls a
fixed-iteration stepper through reverse-mode AD, dissolving the block structure
of the forward by the time the gradient is produced.

\paragraph{Discrete adjoints of iterative solvers.} Differentiating the
stationary equation via the implicit function theorem
\citep{Giles2000AdjointApproach,Hinze2009PDEConstrainedOptimization,Blondel2022ImplicitDiff}
underlies Neural ODEs \citep{Chen2018NeuralODE}, deep equilibrium models
\citep{Bai2019DEQ}, optimization layers \citep{Amos2017OptNet}, and adjoint
multigrid for steady-state PDEs \citep{Giles2000RungeKuttaMultigridAdjoint}.
That literature classically distinguishes differentiating a converged
equation from a finite iterative process; we carry the second option down to
the granularity of a block-coordinate sweep, where it becomes matrix-free. In elastodynamics, DiffPD and DiffCloth
\citep{Du2022DiffPD,Li2022DiffCloth}, DiffQN \citep{Cai2025DiffQN}, and DiffIPC
\citep{Huang2024DiffIPC} all place a global linear-algebra object on the
backward path (Table~\ref{tab:methods}). DiffXPBD \citep{Stuyck2023DiffXPBD}
supplies analytic gradients of extended position-based-dynamics constraint projections, but its forward is
position-based rather than energy-minimizing. The nearest block-implicit neighbor outside graphics
likewise takes a global conjugate-gradient backward around a Gauss--Seidel
forward \citep{Chen2026GaussSeidelProjection}. Operator-level analytic adjoints
are long-standing in differentiable fluid control
\citep{McNamara2004FluidAdjoint,Tang2021HoneyShrunkDomain,Chen2024LaplacianFluidControl},
which supply hand-derived gradients for global operations such as pressure
projection in place of a generic AD trace; we share that principle and push
its granularity from a per-step global operator to the per-vertex
$3{\times}3$ block, so the backward inherits the forward's parallel
schedule.

\paragraph{Block-coordinate solvers.} Structured local updates run from
mass--spring solvers \citep{Liu2013MassSpring} through projective dynamics
\citep{Bouaziz2014ProjectiveDynamics}, position-based dynamics
\citep{Mueller2007PBD,Macklin2016XPBD}, to Vertex Block Descent
\citep{Chen2024VBD}. Their forwards are organized for cache locality and
color-parallel execution; our contribution is a backward that keeps that
organization.

\section{Discussion and limitations}
\label{sec:discussion}

Equation-level differentiation discards the locality that block implicit
solvers were designed to exploit; the solver-level alternative recovers it
without sacrificing correctness, closing an implementation gap that has held
differentiable elastodynamics near ${\sim}30$k DoF: a forward that scales to
millions of vertices becomes differentiable for one forward-equivalent
reverse pass and a per-sweep replay buffer, with no global linear-algebra
object.

\textbf{Limitations.} \emph{(i)}~The full evaluation covers one solver
family; projective and extended position-based dynamics are verified on
minimal spring systems (App.~\ref{app:generality}). \emph{(ii)}~Applications wanting the derivative of a fully
\emph{converged} solve are better served equation-level.
\emph{(iii)}~Backward replay uses atomic-add scatter. \emph{(iv)}~Coulomb
friction with active-set changes is future work.

\subsubsection*{Reproducibility statement}
All claims in \S\ref{sec:exp-exact}--\S\ref{sec:exp-cost} are produced by the
released harness, which constructs every scene
procedurally (no data dependencies), verifies forward-trajectory agreement
across backends before comparing gradients, and writes machine-readable
results. The exactness comparison, the ablation of \S\ref{sec:exactness}, and
the cost table are each a single command. Exact commands, hardware, and library
versions are in Appendix~\ref{app:repro}; Theorem~\ref{thm:adjoint} is proved
in Appendix~\ref{app:proof}.

\subsubsection*{Ethics statement}
This work concerns numerical methods for physical simulation and raises no
human-subjects, privacy, or fairness concerns known to us. Differentiable
simulation carries the dual-use potential common to physical modeling
generally; we identify no application-specific risk introduced here.

\bibliographystyle{iclr2027_conference}
\bibliography{refs}

\appendix
\section{Proof of Theorem~\ref{thm:adjoint}}\label{app:proof}
Equation~\eqref{eq:transpose} is the chain rule applied to
$S=S_n\circ\cdots\circ S_1$, so the adjoint visits blocks in reverse order. For
a single block, the $C^1$ hypothesis on $F_i$ together with invertibility of
$\partial F_i/\partial\Delta\bm{x}_i$ at the executed linearization state gives
Eq.~\eqref{eq:implicit-delta} by the implicit function theorem. Since
$\bm{x}_i^{+}=\bm{x}_i+\Delta\bm{x}_i(\bm{x})$, the local vector--Jacobian product splits into the
direct contribution of Eq.~\eqref{eq:identity-adjoint} and the implicit
contribution; introducing $\bm{p}_i$ through Eq.~\eqref{eq:local-adjoint}
rewrites the latter as Eq.~\eqref{eq:adjoint-update-general}. Finally, the
implementation updates all vertices of one color in parallel; distance-1
coloring makes same-color updates touch disjoint blocks through disjoint
stencils, so the parallel update equals their composition in any order and the
same holds for the adjoints. Composing in reverse yields the claim. $\square$

\section{The construction on other block solvers}\label{app:generality}

Theorem~\ref{thm:adjoint} asks only for a finite composition of differentiable
block updates with invertible local systems, so it should apply beyond Vertex
Block Descent. We verify this on minimal instantiations of two further solvers; both
implementations ship with the released code.

\emph{Extended position-based dynamics}: the block is one distance-constraint
projection (a closed-form update of one Lagrange multiplier and two
particles), swept sequentially, $K$ sweeps per step; the reverse sweep replays
the projections in reverse with a hand-derived local vector--Jacobian
product. \emph{Projective
dynamics}: one iteration is a per-spring local projection followed by a global
solve with a \emph{constant} prefactored symmetric positive-definite
matrix; the backward reuses the
same factorization (transpose solve) and hand-derived projection
vector--Jacobian products, so the
backward mirrors the forward's local--global structure exactly.

Both are validated the same way as the main experiments: against
\texttt{torch.autograd} through the identical executed forward, on spring and
constraint networks built over the \S\ref{sec:exp-exact} test meshes
($|V|{=}26$ and $115$), at $K\in\{1,5,20\}$ (XPBD: extended position-based
dynamics; PD: projective dynamics):

\begin{center}\small
\begin{tabular}{lcccc}
\toprule
Solver & $|V|$ & $K{=}1$ & $K{=}5$ & $K{=}20$ \\
\midrule
XPBD & $26$  & $3.9\times10^{-16}$ & $4.8\times10^{-16}$ & $7.0\times10^{-16}$ \\
XPBD & $115$ & $6.2\times10^{-16}$ & $6.0\times10^{-16}$ & $5.0\times10^{-16}$ \\
PD   & $26$  & $9.3\times10^{-16}$ & $9.7\times10^{-16}$ & $9.7\times10^{-16}$ \\
PD   & $115$ & $2.5\times10^{-15}$ & $5.9\times10^{-15}$ & $1.1\times10^{-14}$ \\
\bottomrule
\end{tabular}
\end{center}

Worst case $1.1\times10^{-14}$ (the projective-dynamics global solve adds a
factorization round-trip per iteration), i.e.\ the same finite-depth exactness
as the Vertex Block Descent instantiation. These are exactness demonstrations,
not performance claims: the implementations are deliberately plain NumPy.

\section{Closed forms for the local adjoint}\label{app:closed-forms}

Eq.~\eqref{eq:param-adjoint} gives the parameter contribution abstractly. For
the block Hessian actually used,
$H_i^{\text{elas}}=\sum_{e\ni i}w_e\bigl(\mu\|\nabla N_{i,e}\|^{2}I
+\lambda\,q_{i,e}q_{i,e}^{\!\top}\bigr)$, with $w_e=\mathrm{vol}_e h^{2}$ the
element weight in the $h^{2}$-scaled system and $\nabla N_{i,e}$ the constant
shape-function gradient, the Hessian-tangent contraction of
Eq.~\eqref{eq:hessian-tangent-index} collapses to scalars, with no tensor
assembled and no derivative of $q$ required:
\begin{align}
\Bigl(\tfrac{\partial H_{i}^{\text{elas}}}{\partial \mu}\Bigr)\!:\!\bigl(\bm{p}_{i}\Delta\bm{x}_{i}^{\!\top}\bigr)
&=\sum_{e\ni i}w_e\|\nabla N_{i,e}\|^{2}\,\bigl(\Delta\bm{x}_{i}\!\cdot\!\bm{p}_{i}\bigr),
\label{eq:stage2-mu}\\
\Bigl(\tfrac{\partial H_{i}^{\text{elas}}}{\partial \lambda}\Bigr)\!:\!\bigl(\bm{p}_{i}\Delta\bm{x}_{i}^{\!\top}\bigr)
&=\sum_{e\ni i}w_e\bigl(q_{i,e}^{\!\top}\Delta\bm{x}_{i}\bigr)\bigl(q_{i,e}^{\!\top}\bm{p}_{i}\bigr).
\label{eq:stage2-lam}
\end{align}
Combining these with the parameter derivatives of the stable Neo-Hookean
energy as implemented -- the reparameterized form
$\Psi=\tfrac{\mu}{2}(\|F\|_{F}^{2}-3)+\tfrac{\bar\lambda}{2}(J-\alpha)^{2}$
with $\bar\lambda=\mu+\lambda$, $\alpha=1+\mu/\bar\lambda$
\citep{Smith2018StableNeoHookean}, whose Piola derivatives are
$\partial P/\partial\mu=F+(J-2)\,\mathrm{cof}\,F$ and
$\partial P/\partial\lambda=(J-1)\,\mathrm{cof}\,F$
(equivalently $\partial V_{e}/\partial\mu=\tfrac{1}{2}(\|F_{e}\|_{F}^{2}-3)+\tfrac{1}{2}(J_{e}-2)^{2}$
and $\partial V_{e}/\partial\lambda=\tfrac{1}{2}(J_{e}-1)^{2}$, each up to a
state-independent constant) -- closes the material gradient: one dot
product and one quadratic form per incident element, reusing the $\bm{p}_i$
already computed for the state adjoint.

Two consequences are worth stating. First, the cost of material gradients is a
constant multiple of the state adjoint, which is why enabling them does not
change the asymptotic backward cost. Second,
Eqs.~\eqref{eq:stage2-mu}--\eqref{eq:stage2-lam} are precisely the terms that
defect (ii) of \S\ref{sec:exactness} drops: an implementation that keeps them
for parameters but not for the state adjoint produces material gradients that
look correct while the state gradient carries error $\propto\|\Delta\bm{x}\|$.

\section{Supporting figures and tables}\label{app:figs}

\begin{table}[h]
\caption{\textbf{Differentiation strategies for implicit solvers.}
\emph{Executed solver?}: does the backward operate on the same finite block
sweep the forward ran, rather than on a converged equation or an unrolled
implementation graph. \emph{Global object}: the global linear-algebra structure
the backward constructs, if any. \emph{Block-local?}: is the backward a
per-block solve in the forward's color partition. \emph{Memory vs.\ $K$}:
asymptotic in solver depth at fixed mesh size; our constant of ${\approx}4$
kilobytes per vertex per sweep is measured (App.~\ref{app:scaling}).
\emph{Scale}: largest configuration reported in each work.}
\label{tab:methods}
\centering\small
\setlength{\tabcolsep}{4pt}
\renewcommand{\arraystretch}{1.15}
\begin{tabular}{@{}lcccll@{}}
\toprule
Method & \makecell{Executed\\solver?} & \makecell{Global\\object} & \makecell{Block-\\local?}
       & Memory vs.\ $K$ & Scale \\
\midrule
Unrolled AD        & graph level    & none         & no      & linear in graph  & ${\sim}10$k DoF \\
Equation-level impl.\ diff. & no             & sparse adj.  & no      & $O(1)$           & varies \\
DiffPD / DiffCloth & no             & PD Cholesky  & no      & $O(1)$           & ${\sim}10$k DoF \\
DiffXPBD           & constr.\ proj. & none         & partial & linear in states & $66$k verts \\
DiffQN             & no             & quasi-Newton & no      & $O(1)$           & ${\sim}10$k DoF \\
DiffIPC            & no             & step Hessian & no      & $O(1)$           & ${\sim}20$k verts \\
\midrule
\textbf{Ours}      & \textbf{yes}   & \textbf{none} & \textbf{yes}
                   & \textbf{lin., ${\approx}4$\,KB/v/sw} & $\bm{8}$\textbf{M verts} \\
\bottomrule
\end{tabular}
\end{table}

This appendix collects the qualitative results and supporting tables
referenced from the main text.

\begin{figure}[h]
\centering
\includegraphics[width=0.58\linewidth]{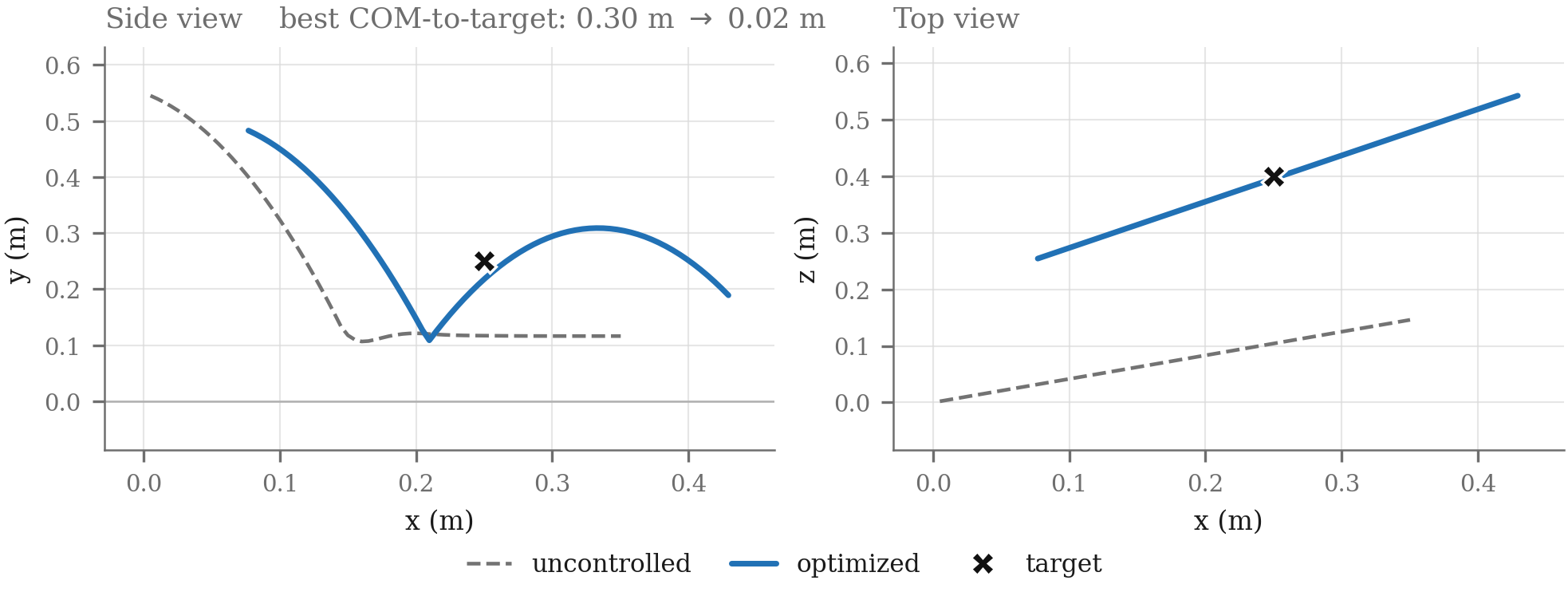}
\caption{Cube center of mass trajectory under the initial parameters versus the optimized
ones, side and top view. The optimizer, not the initialization, drives the
target hit.}
\label{fig:cube-traj}
\end{figure}

\begin{figure}[h]
\centering
\includegraphics[width=0.52\linewidth]{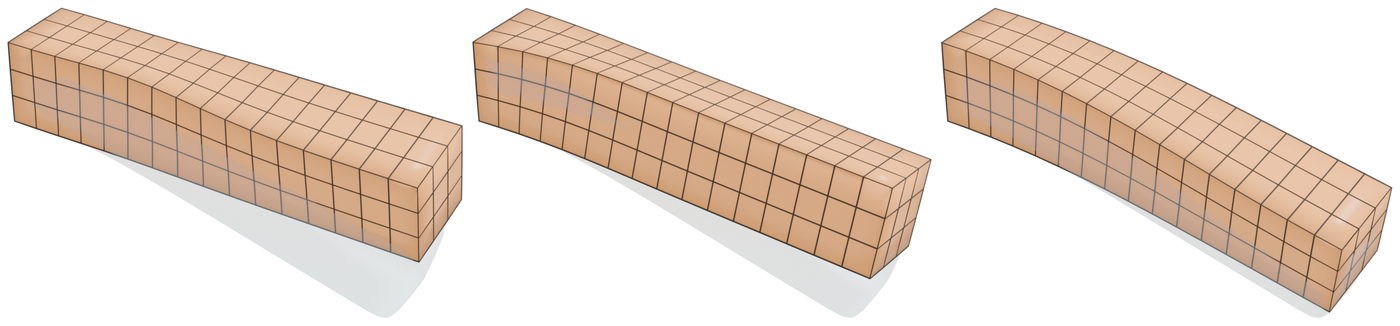}
\caption{Beam $\mu/\lambda$ identification, target deformation $(\mu,\lambda){=}(800,400)$ under a $60^\circ$ angular kick. Blue ghost: target equilibrium. Both parameters are recovered to $<0.3\%$.}
\label{fig:beam-mulam}
\end{figure}

\begin{figure}[h]
\centering
\includegraphics[width=0.55\linewidth]{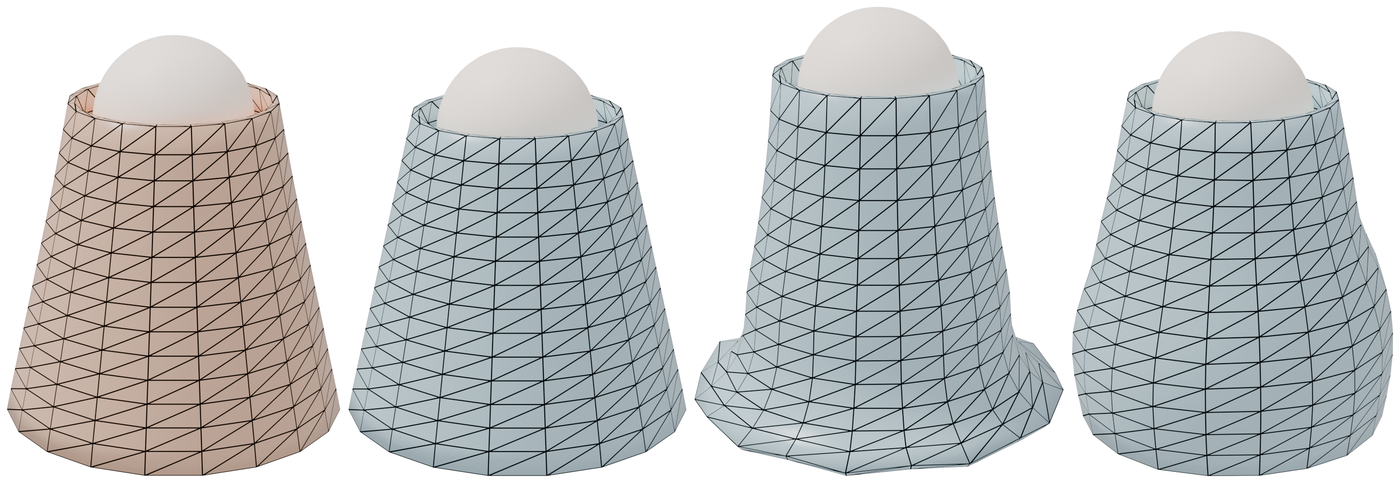}
\caption{Dress $\mu/\lambda$ identification. Left: initial stiff configuration ($\mu{=}5000$). Remaining panels: target draping progression ($\mu{=}1500$).}
\label{fig:applications}
\end{figure}

\begin{figure}[h]
\centering
\includegraphics[width=0.52\linewidth]{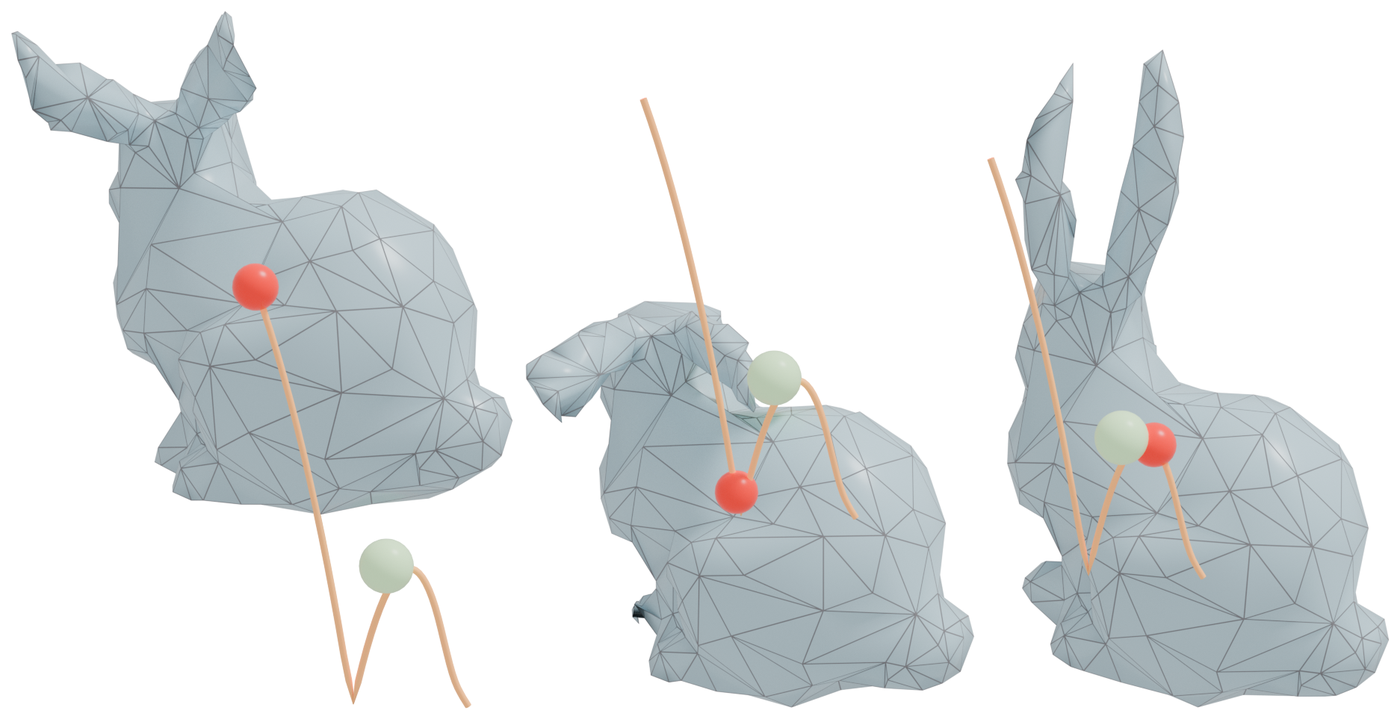}
\caption{Bunny initial-state optimization: start, mid-bounce, and at the target.}
\label{fig:bunny}
\end{figure}

\begin{figure}[h]
  \centering
  \includegraphics[width=0.70\linewidth]{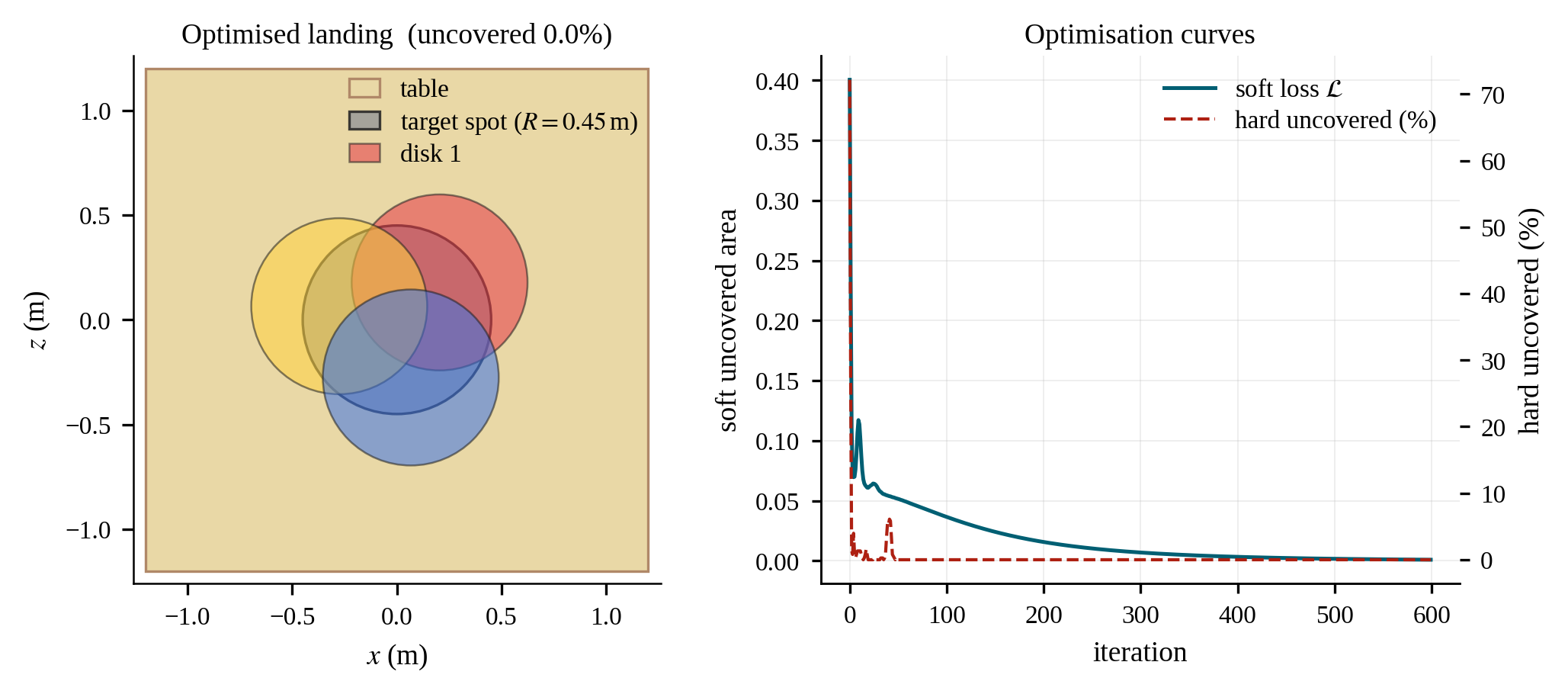}
  \\[4pt]
  \includegraphics[width=0.26\linewidth]{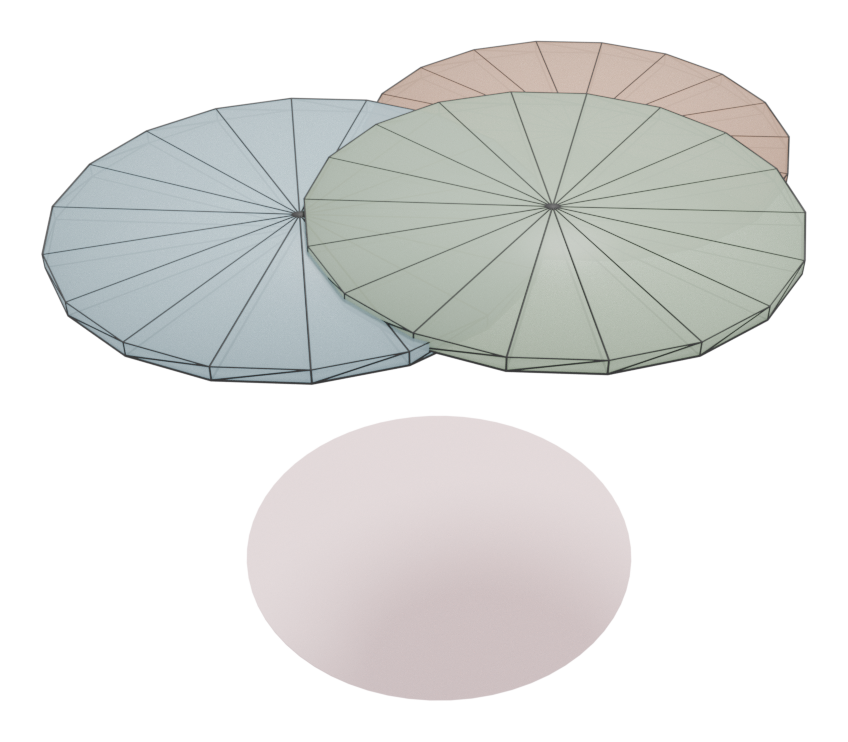}\hspace{0.01\linewidth}
  \includegraphics[width=0.26\linewidth]{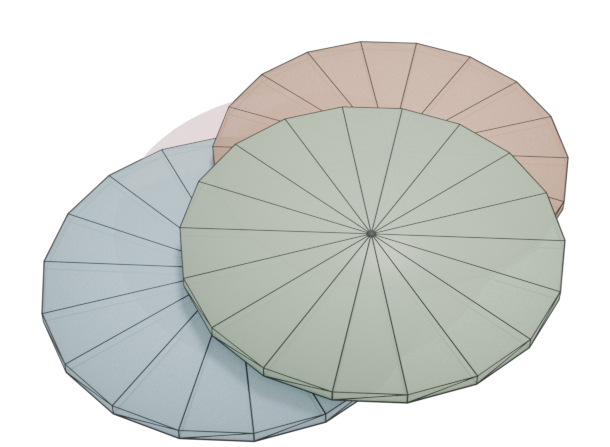}\hspace{0.01\linewidth}
  \includegraphics[width=0.26\linewidth]{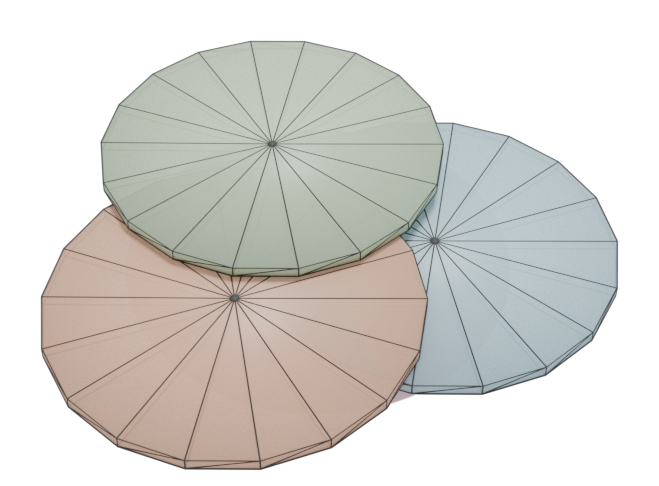}
  \caption{\textbf{Cover the Spot.} Three soft-body disks are jointly optimized by backpropagation through the full DiffVBD rollout to land such that their union covers a painted target spot. \emph{Top:} top-down view of the optimized landing (left) and optimization curves (right). \emph{Bottom:} 3D snapshots of the optimized throw at three moments: high mid-air, low mid-air, and settled. Final hard uncovered area: $0.0\%$ (full coverage).}
    \label{fig:cover-the-spot}
\end{figure}

\begin{table}[h]
\caption{Finite-difference (FD) validation. Maximum relative error between
analytic and FD gradients. The improvement with $K$ reflects the executed solver
approaching a stationary point, where the FD reference is best conditioned.}
\label{tab:fd}
\centering\small
\renewcommand{\arraystretch}{1.12}
\begin{tabular}{lcc}
\toprule
Configuration & Max relative error & Pass \\
\midrule
Baseline ($K{=}1$, CPU)   & $4.07\times10^{-10}$ & \checkmark \\
Baseline ($K{=}1$, GPU)   & $4.07\times10^{-10}$ & \checkmark \\
Multi-sweep $K{=}2$       & $1.64\times10^{-9}$  & \checkmark \\
Multi-sweep $K{=}3$       & $3.05\times10^{-10}$ & \checkmark \\
$\mu$-only ($\lambda{=}0$)   & $1.34\times10^{-9}$ & \checkmark \\
$\lambda$-only ($\mu{=}0$)   & $5.33\times10^{-10}$ & \checkmark \\
\bottomrule
\end{tabular}
\end{table}

\section{Additional inverse problems}\label{app:extra-tasks}

Three further inverse problems, each a scalar or low-dimensional recovery
by backpropagation through the full rollout, are reported here rather than in
\S\ref{sec:exp-inverse} for space. They are included because they probe a
regime the main-text tasks do not: the optimized quantity is not a material
constant or an initial state but a \emph{rest-shape} parameter (the dress
scale below) or a persistent external force (the cape wind), and in both cases
the gradient must survive pinning constraints and contact along the whole
trajectory. Table~\ref{tab:extra} summarizes them.

\begin{table}[h]
\caption{Additional inverse problems. All use the DiffVBD adjoint with Adam.
Loss is the task objective at convergence; error is relative to the known
ground truth used to generate the target.}
\label{tab:extra}
\centering\small
\renewcommand{\arraystretch}{1.12}
\begin{tabular}{lccll}
\toprule
Task & \#vars & Iters & Loss (start $\to$ end) & Outcome \\
\midrule
Cape $\mu$ (center sag)   & $1$ & $80$  & $1.67{\times}10^{-2}\to1.64{\times}10^{-7}$ & $\hat\mu{=}1996.7$\,Pa ($0.16\%$) \\
Dress size (hem height)   & $1$ & $80$  & $3.4{\times}10^{-2}\to1.8{\times}10^{-7}$   & $\hat s{=}0.910$ (${<}0.5\%$) \\
Cape wind (hem shape)     & $3$ & $120$ & $4{\times}10^{-2}\to2{\times}10^{-8}$       & ${<}0.3\%$ relative \\
\bottomrule
\end{tabular}
\end{table}

\begin{figure}[h]
  \centering
  \footnotesize\textit{(top) Init $\mu_0{=}8000$\,Pa (over-stiff, hammock barely sags);
                       (bottom) recovered $\hat\mu{=}1996.7$\,Pa (target $\mu^\star{=}2000$, err $0.16\%$)}\par\vspace{2pt}
  \includegraphics[width=0.163\linewidth]{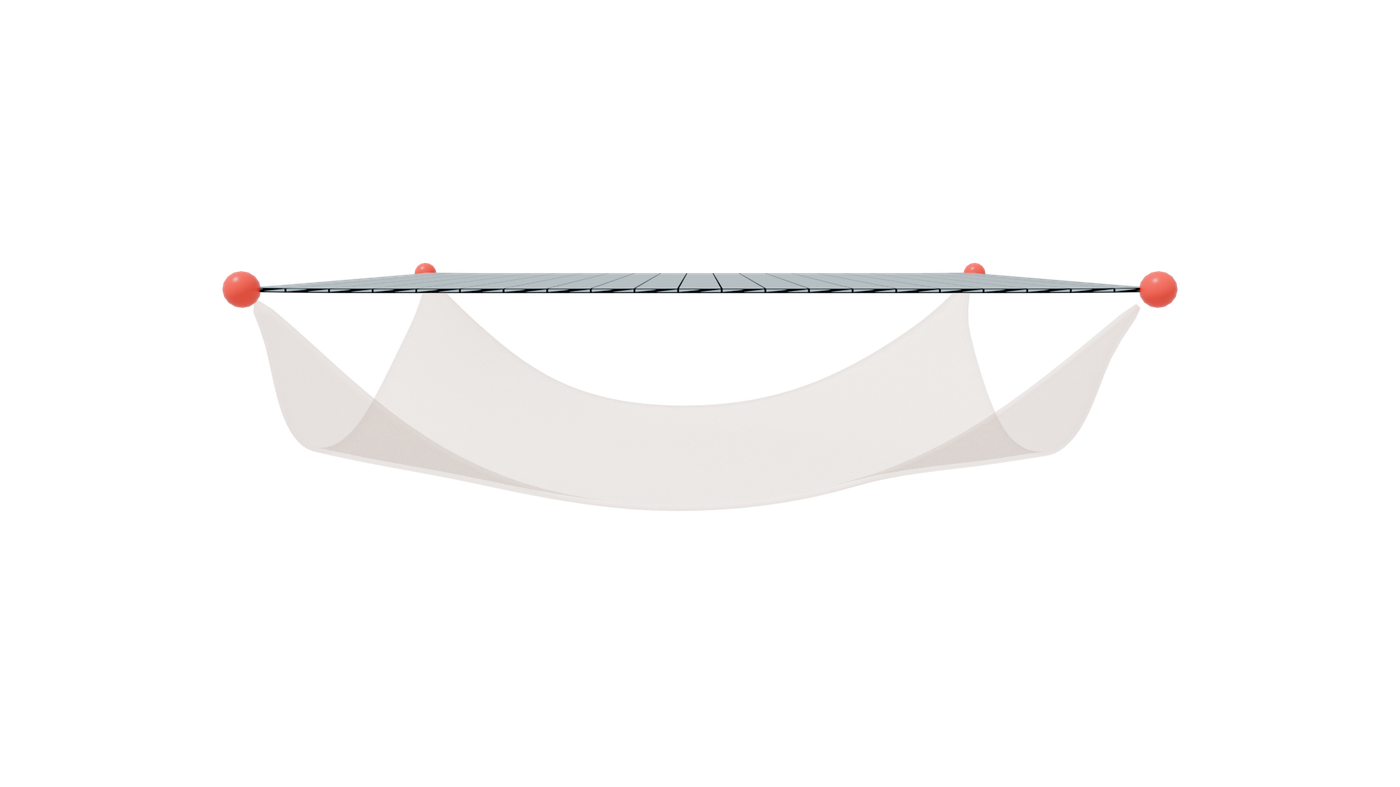}\hfill
  \includegraphics[width=0.163\linewidth]{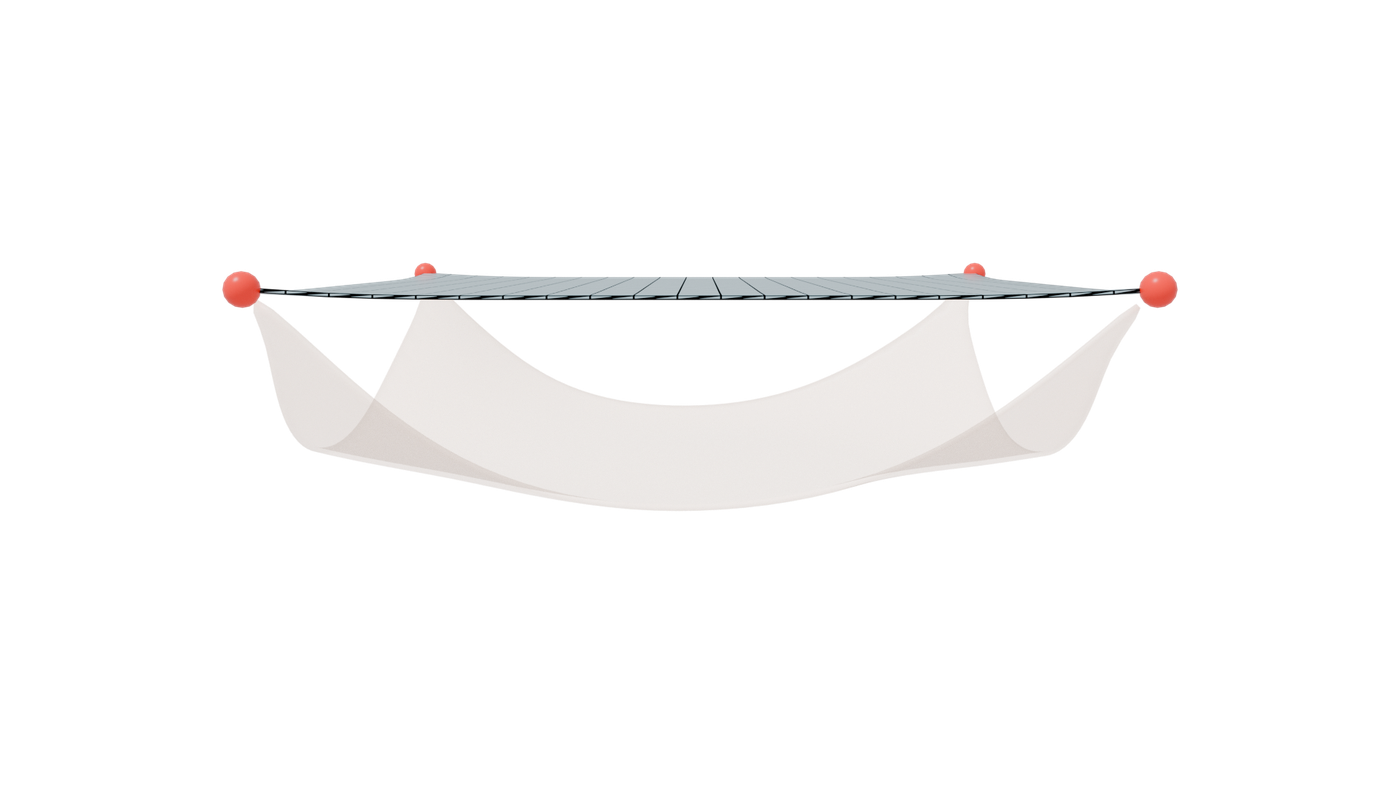}\hfill
  \includegraphics[width=0.163\linewidth]{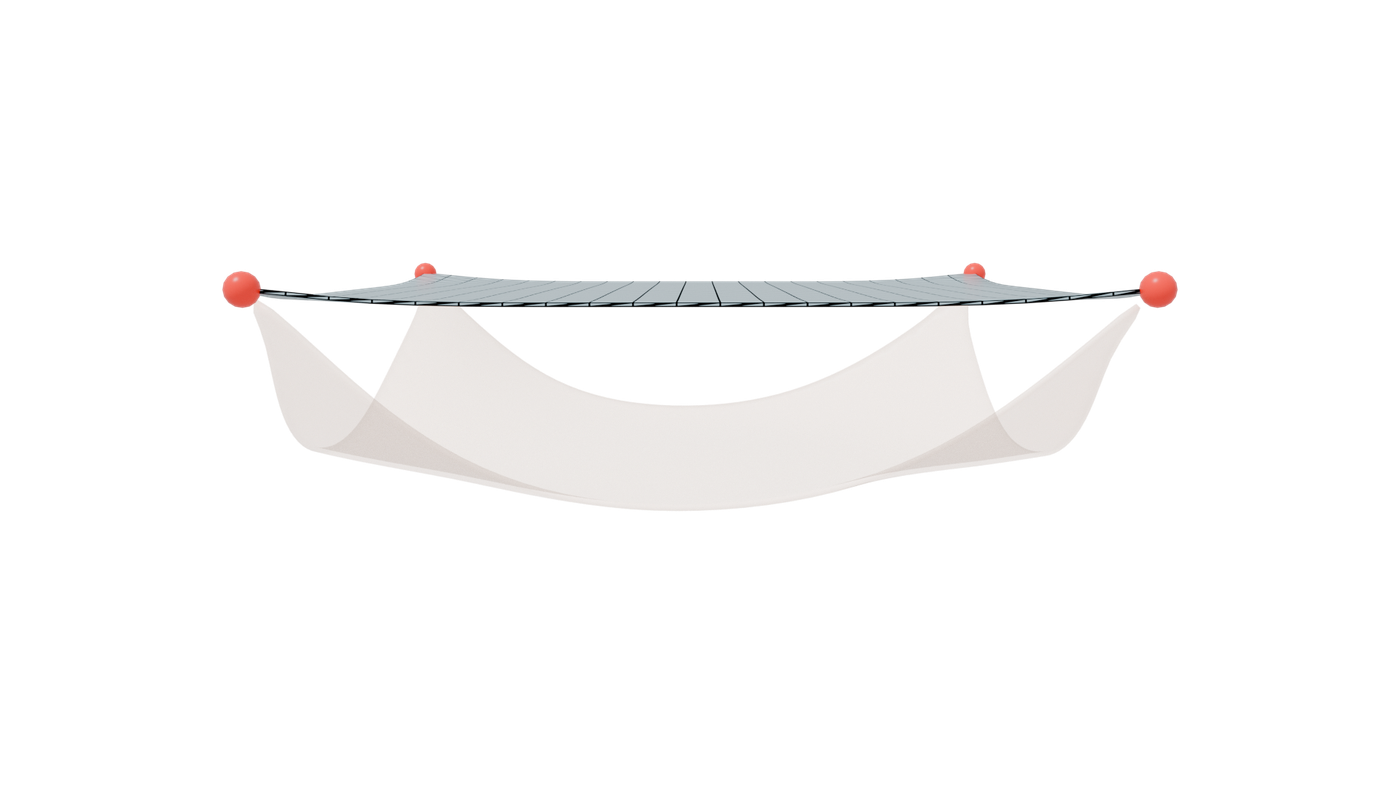}\hfill
  \includegraphics[width=0.163\linewidth]{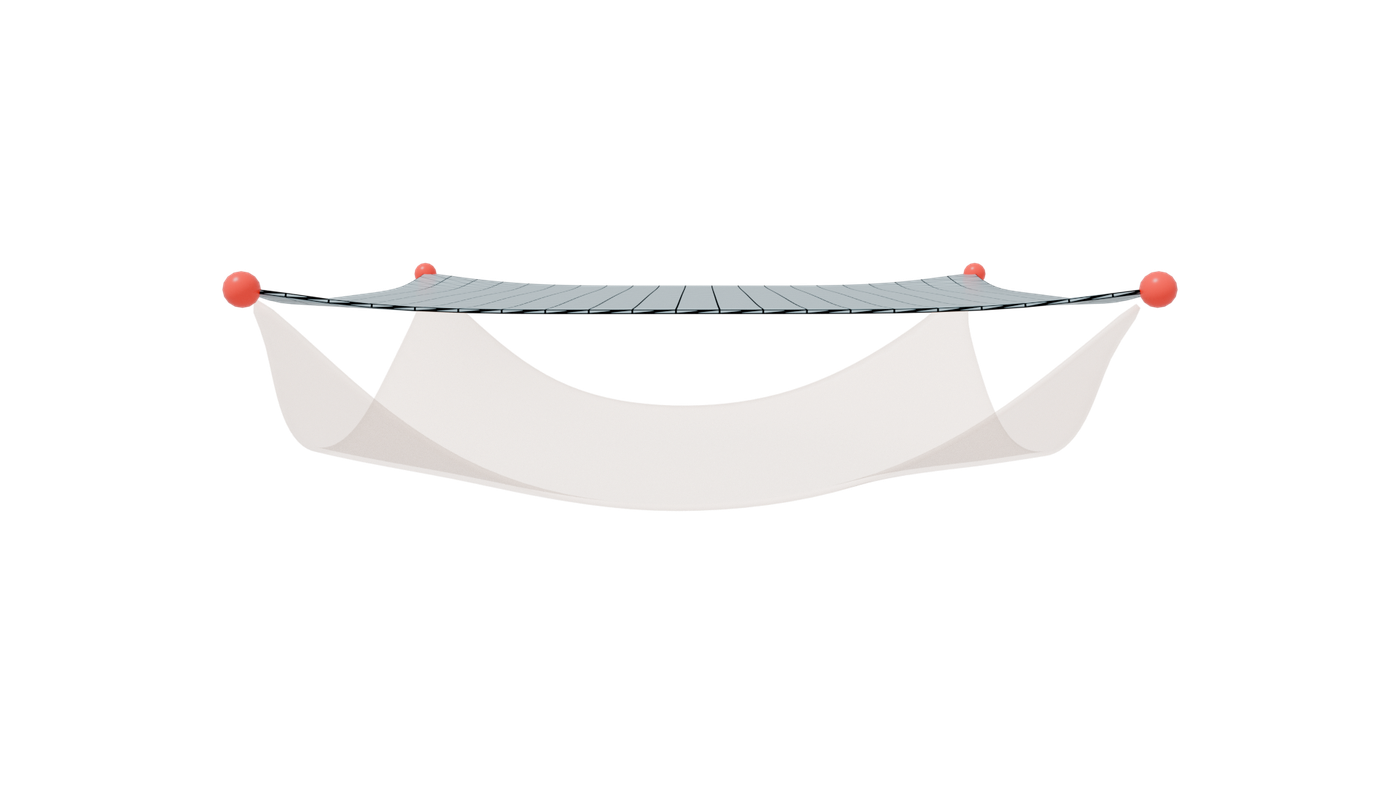}\hfill
  \includegraphics[width=0.163\linewidth]{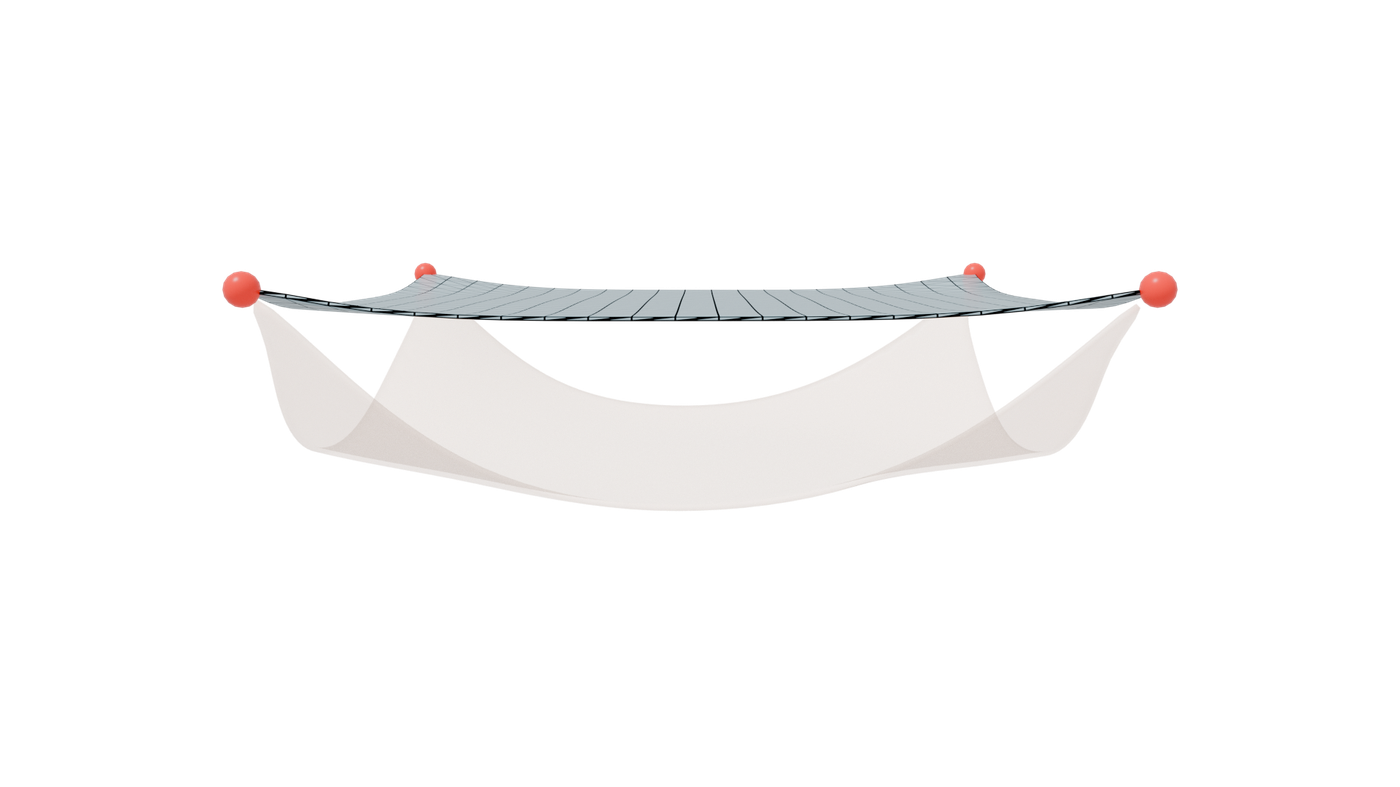}\hfill
  \includegraphics[width=0.163\linewidth]{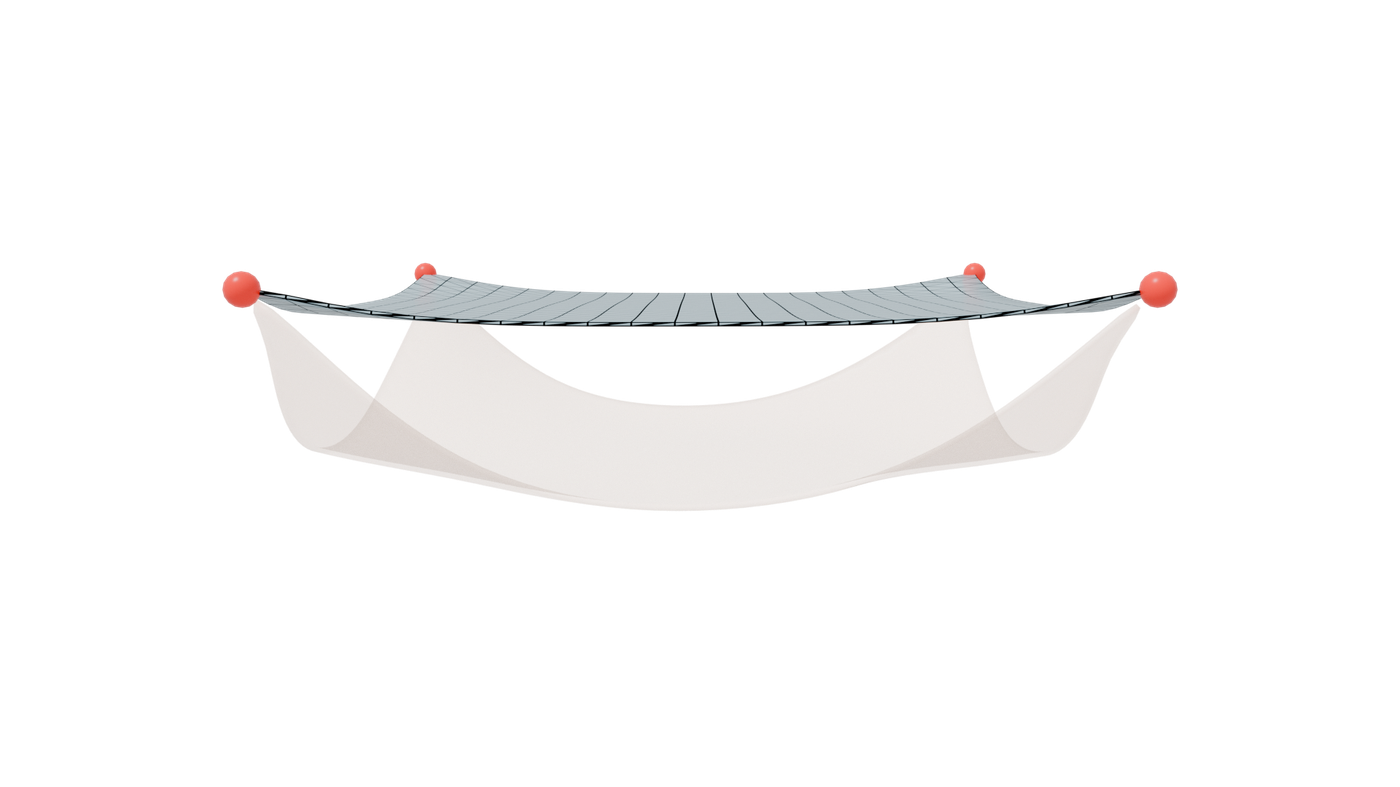}\par\vspace{2pt}
  \includegraphics[width=0.163\linewidth]{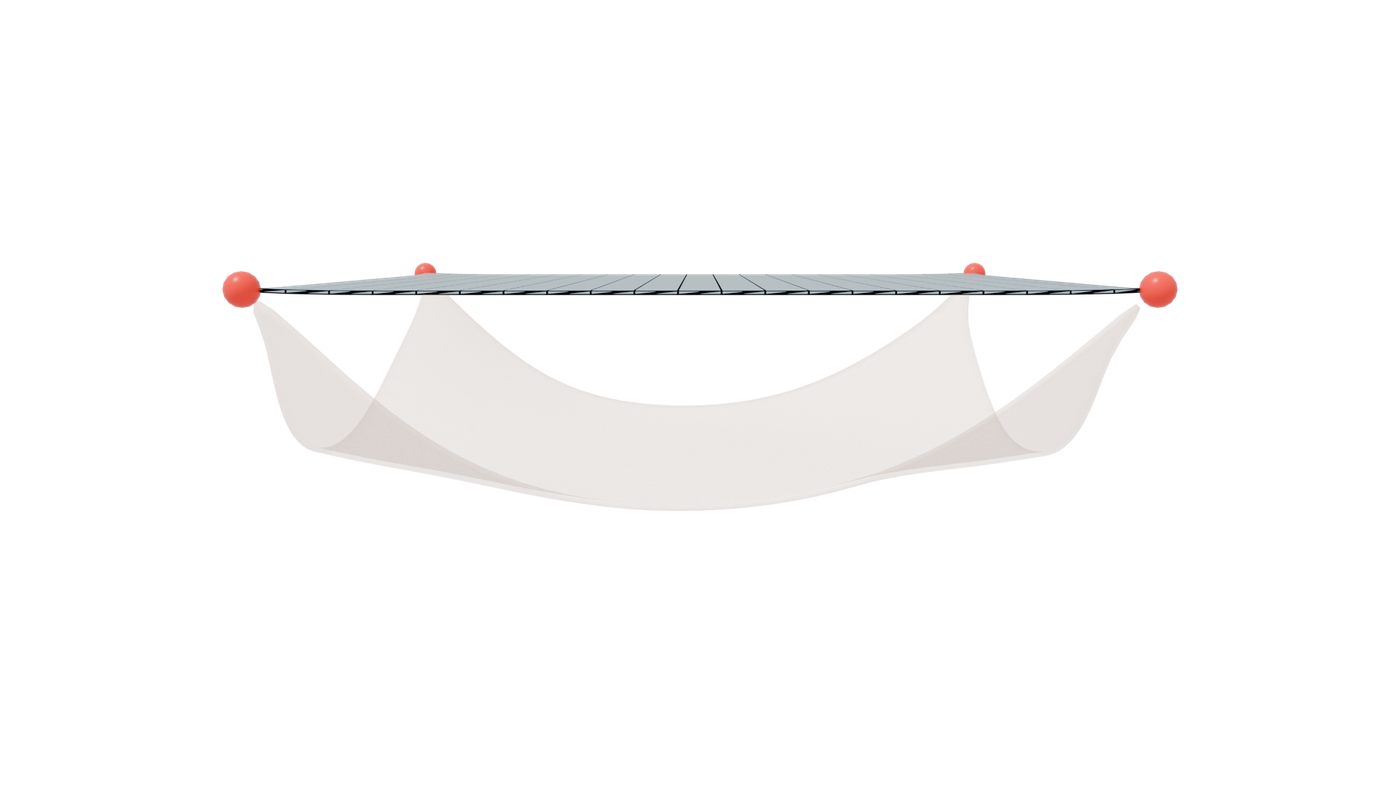}\hfill
  \includegraphics[width=0.163\linewidth]{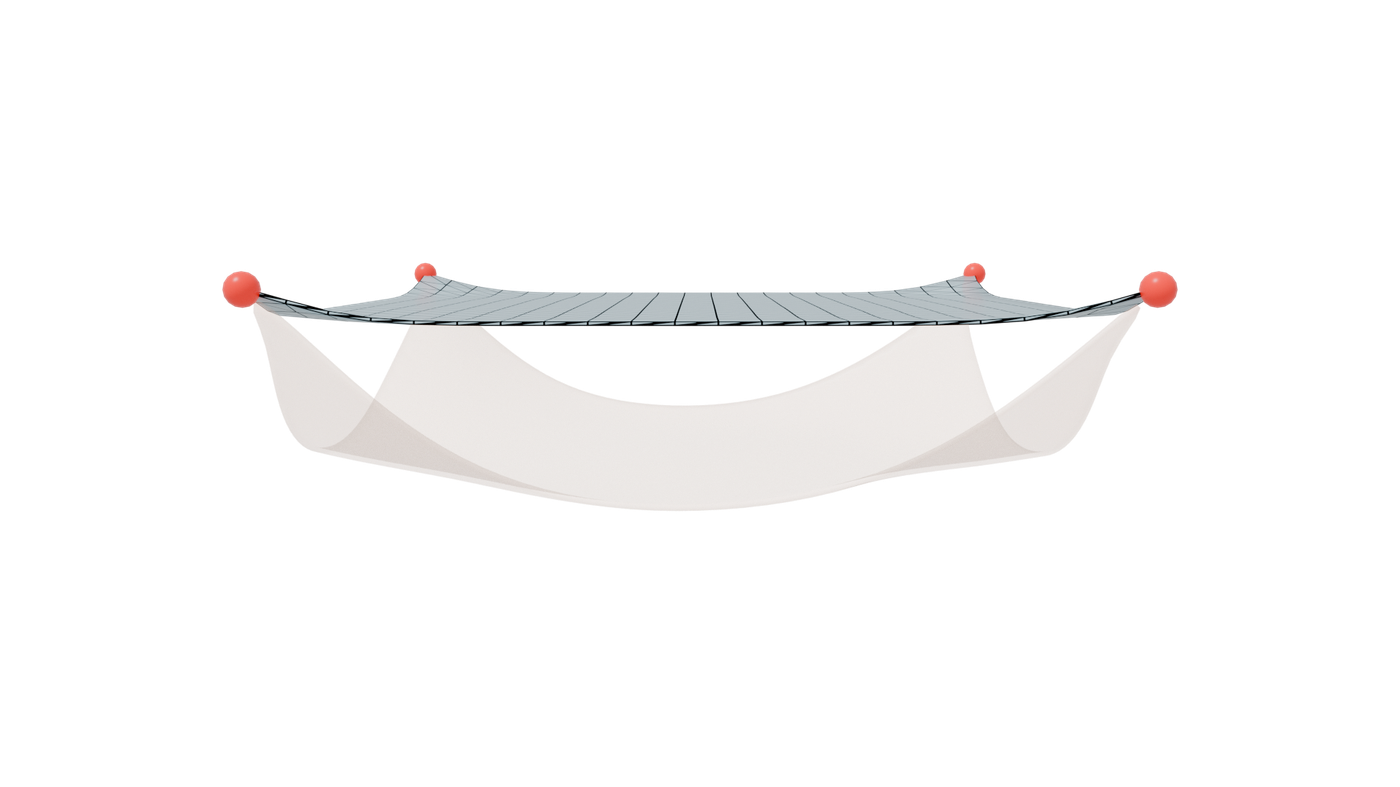}\hfill
  \includegraphics[width=0.163\linewidth]{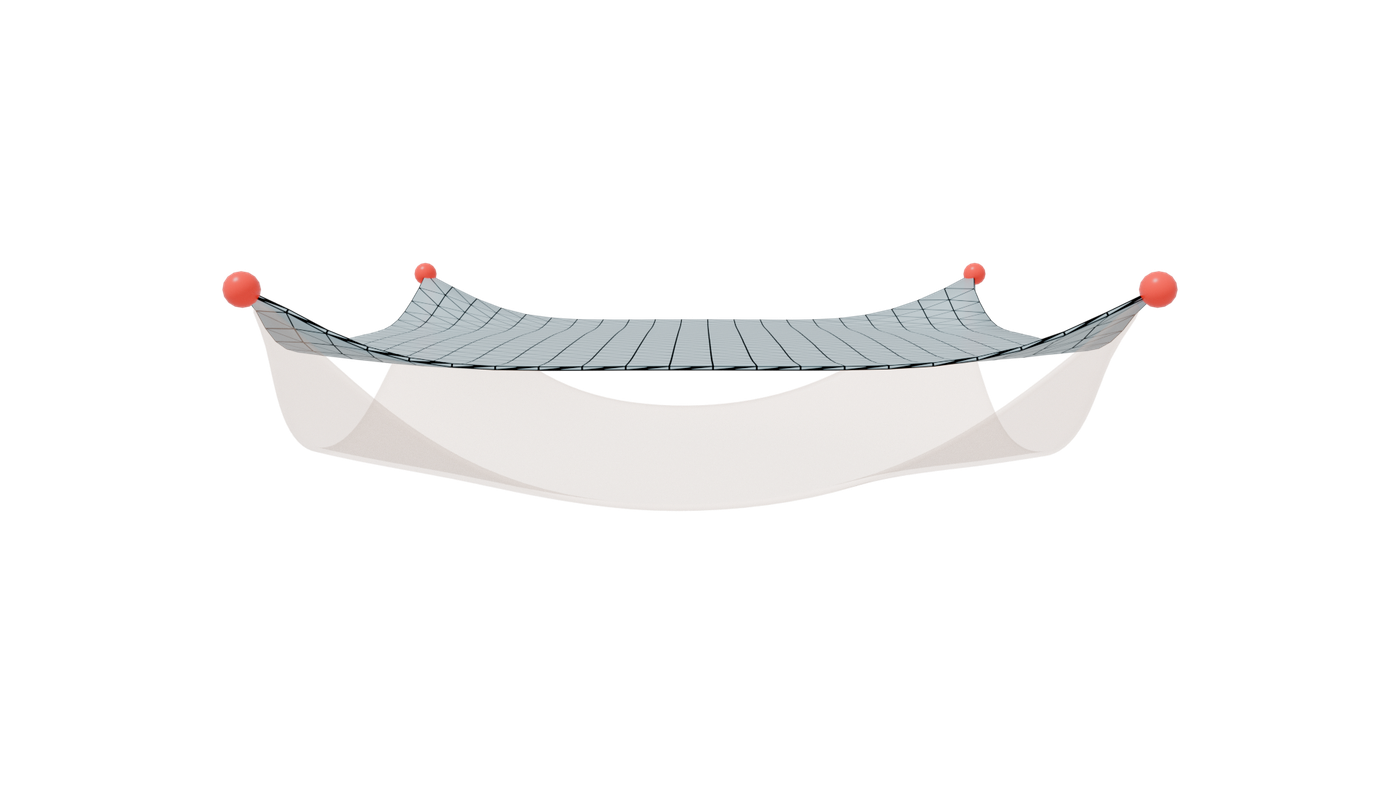}\hfill
  \includegraphics[width=0.163\linewidth]{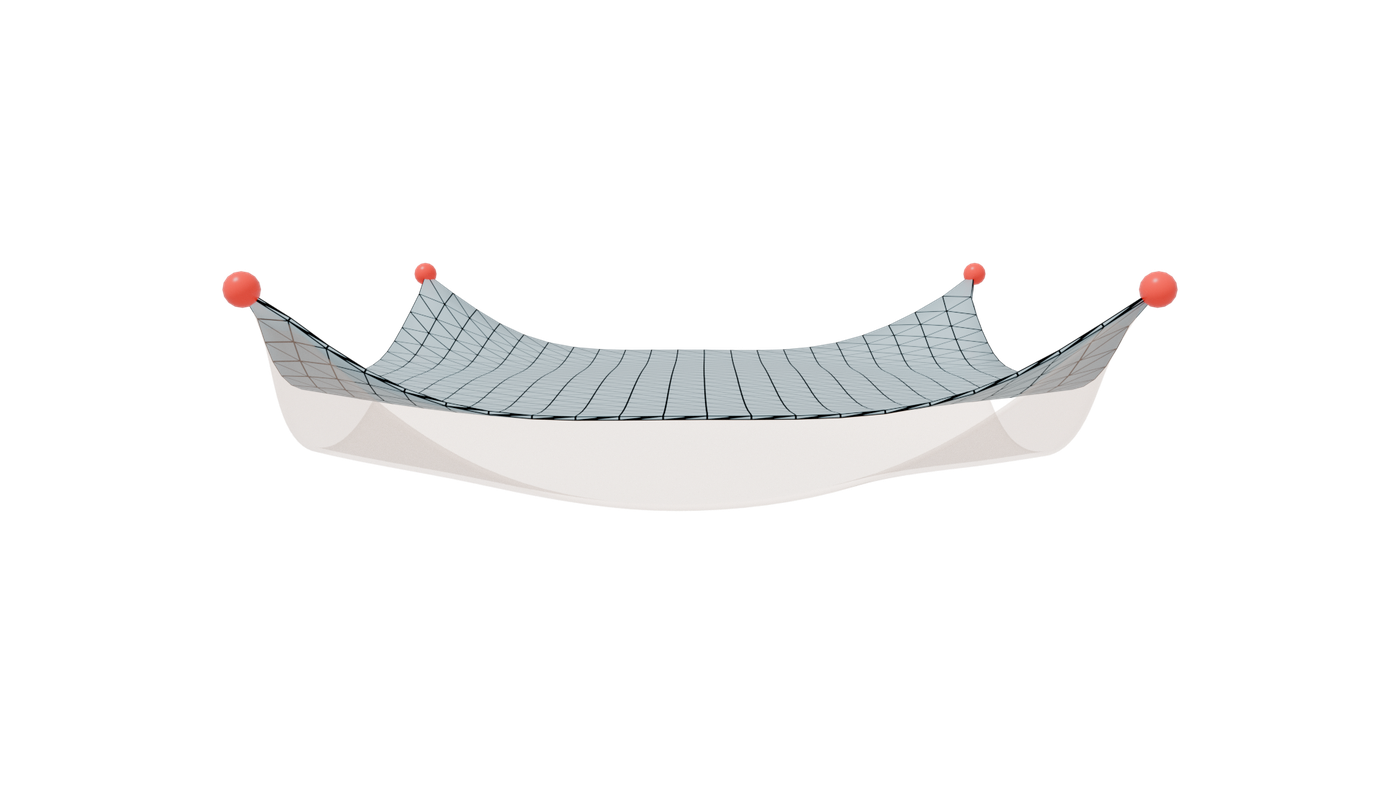}\hfill
  \includegraphics[width=0.163\linewidth]{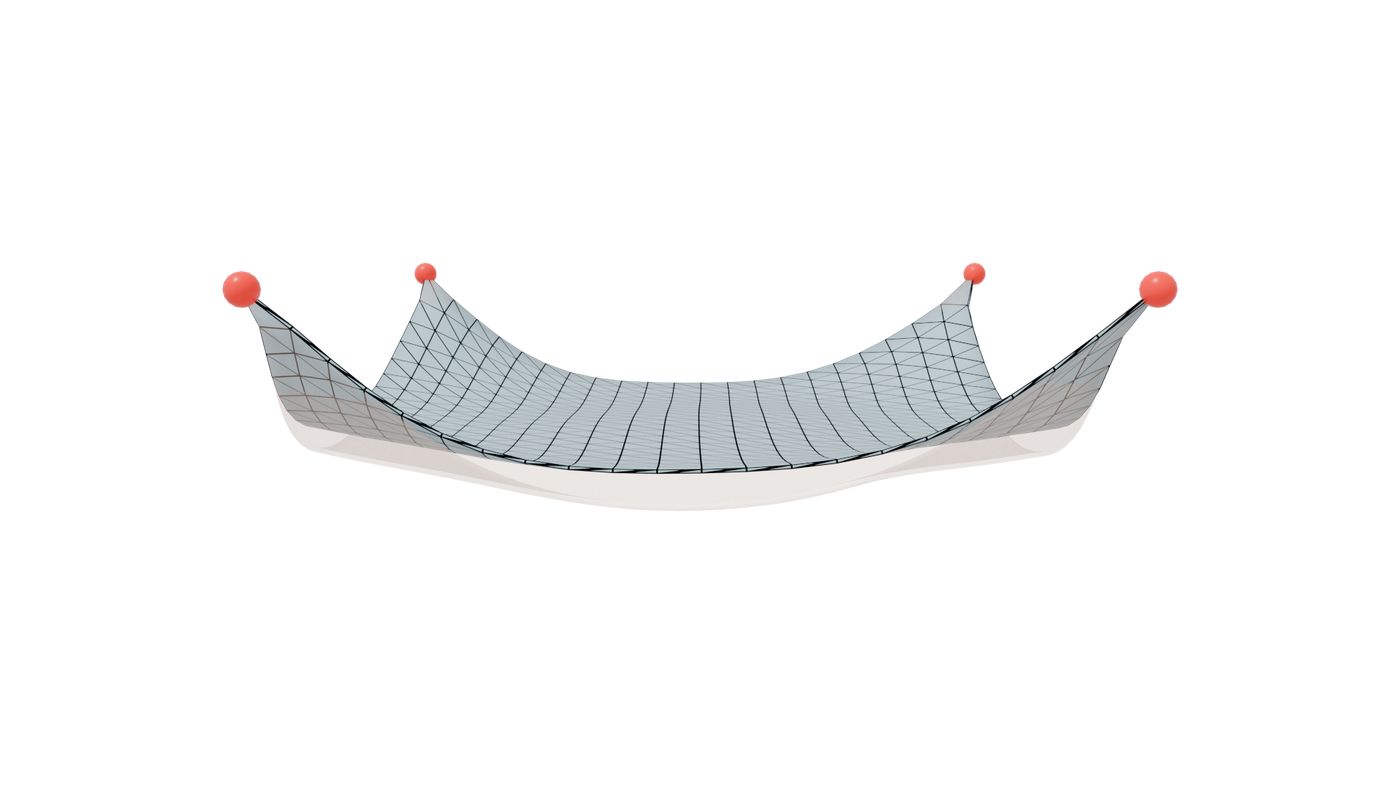}\hfill
  \includegraphics[width=0.163\linewidth]{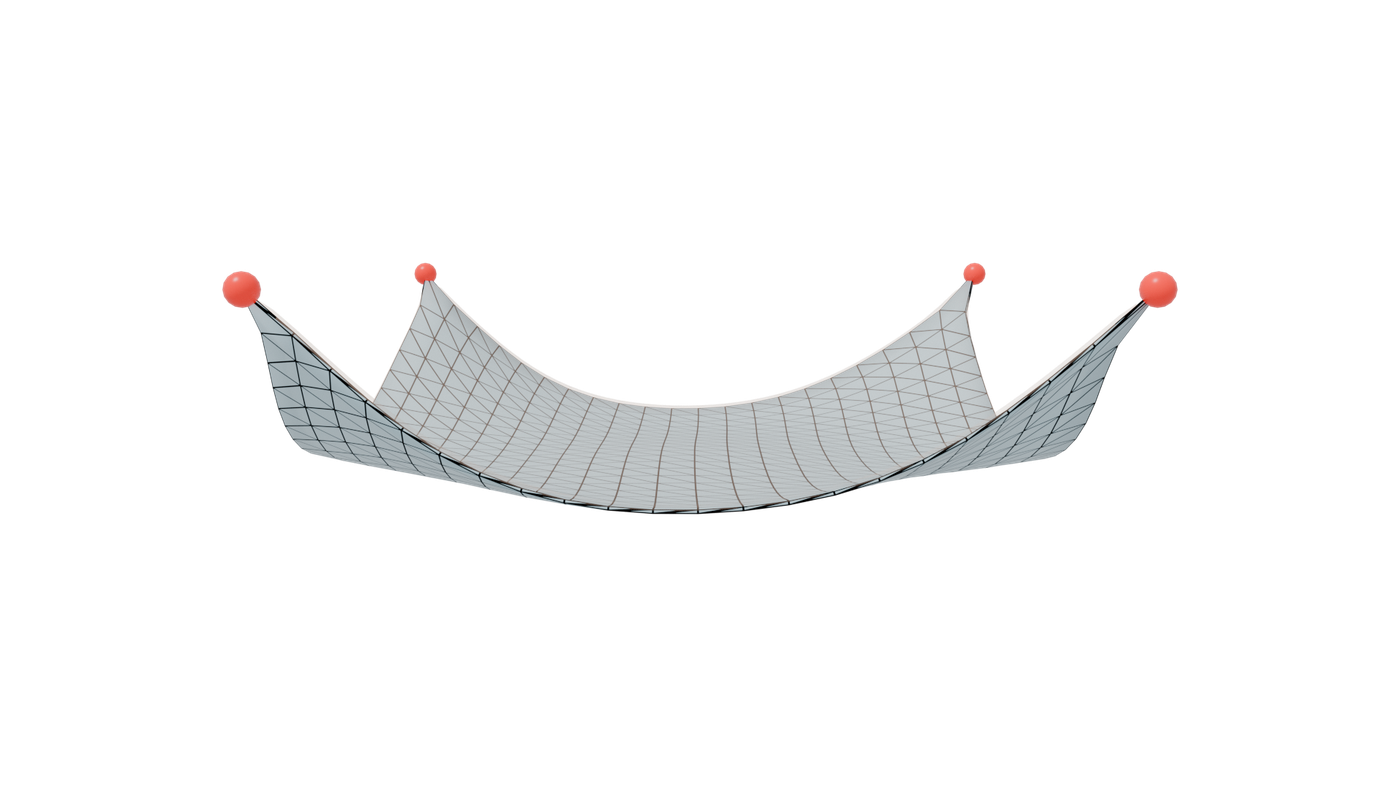}
  \caption{\textbf{Shear modulus recovery on a four-corner-pinned cape.} The
  translucent silhouette below the cloth is the optimization target (deeper
  sag). Frames at $t\in\{0.02,0.08,0.15,0.22,0.28,0.33\}$\,s.}
  \label{fig:mu-bptt}
\end{figure}

\paragraph{Shear modulus from a scalar sag target.}
A $484$-vertex square cloth (side $0.6$\,m, extruded to a thin three-tet
shell) is pinned at its four corners and sags into a hammock under gravity.
Stiffer $\mu$ gives shallower center sag, so the inverse problem
$\mathcal{L}(\mu){=}(y_{\text{ctr}}(\mu)-y_{\text{ctr}}^{\star})^{2}$ is
scalar and monotonic. From an over-stiff initialization $\mu_0{=}8000$\,Pa
against a target $\mu^{\star}{=}2000$\,Pa, Adam (learning rate $0.10$, cosine decay,
$80$ iterations) on $\partial\mathcal{L}/\partial\mu$ -- obtained from the
per-tetrahedron parameter-gradient outputs of the reverse sweep, summed over
tetrahedra -- recovers $\hat\mu{=}1996.7$\,Pa ($0.16\%$ error) and reduces the
loss by five orders of magnitude (Fig.~\ref{fig:mu-bptt}).

\begin{figure}[h]
  \centering
  \footnotesize\textit{(top) Init dress scale $s_0{=}0.45$ (too short, hem above waist);
                       (bottom) recovered $\hat{s}{=}0.910$ (hem at mid-thigh, err ${<}0.5\%$)}\par\vspace{2pt}
  \includegraphics[width=0.196\linewidth]{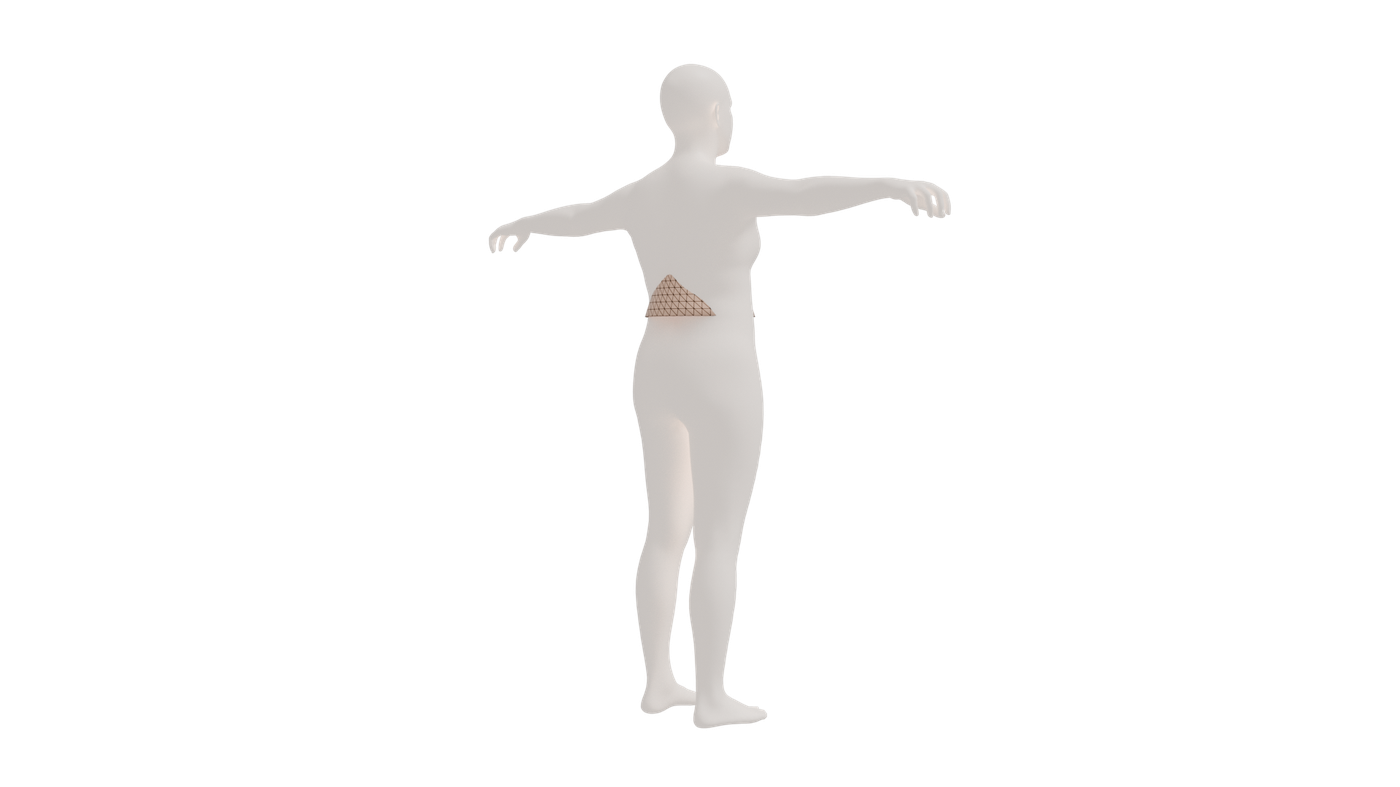}\hfill
  \includegraphics[width=0.196\linewidth]{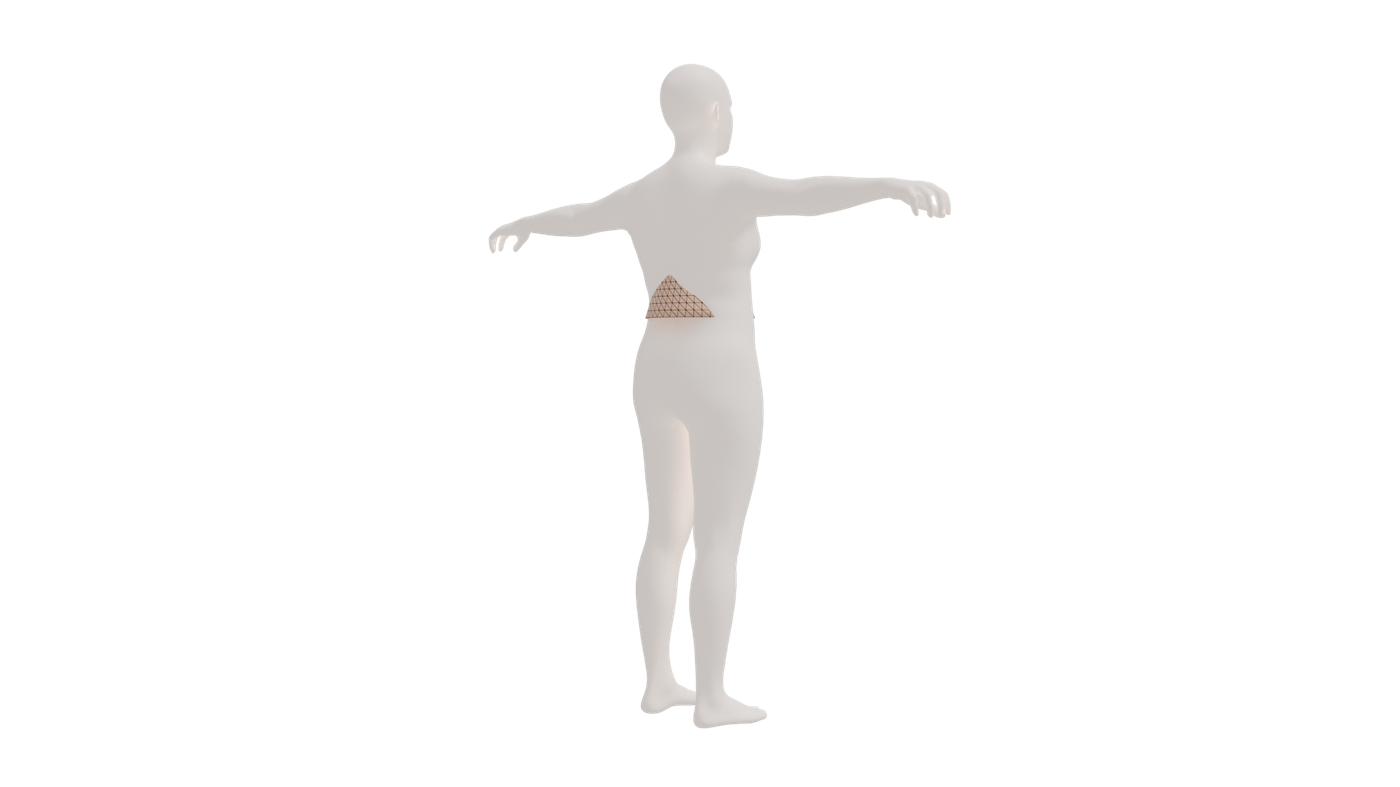}\hfill
  \includegraphics[width=0.196\linewidth]{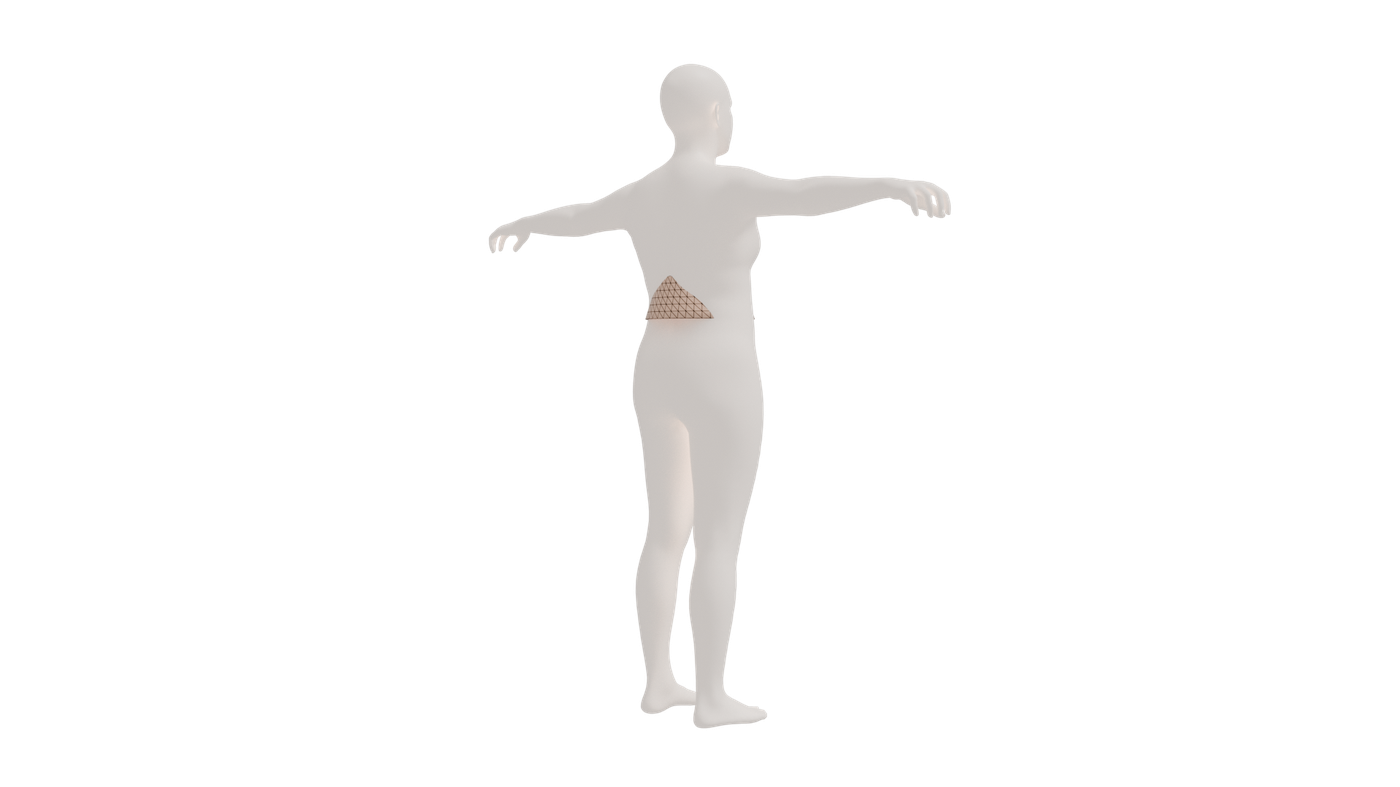}\hfill
  \includegraphics[width=0.196\linewidth]{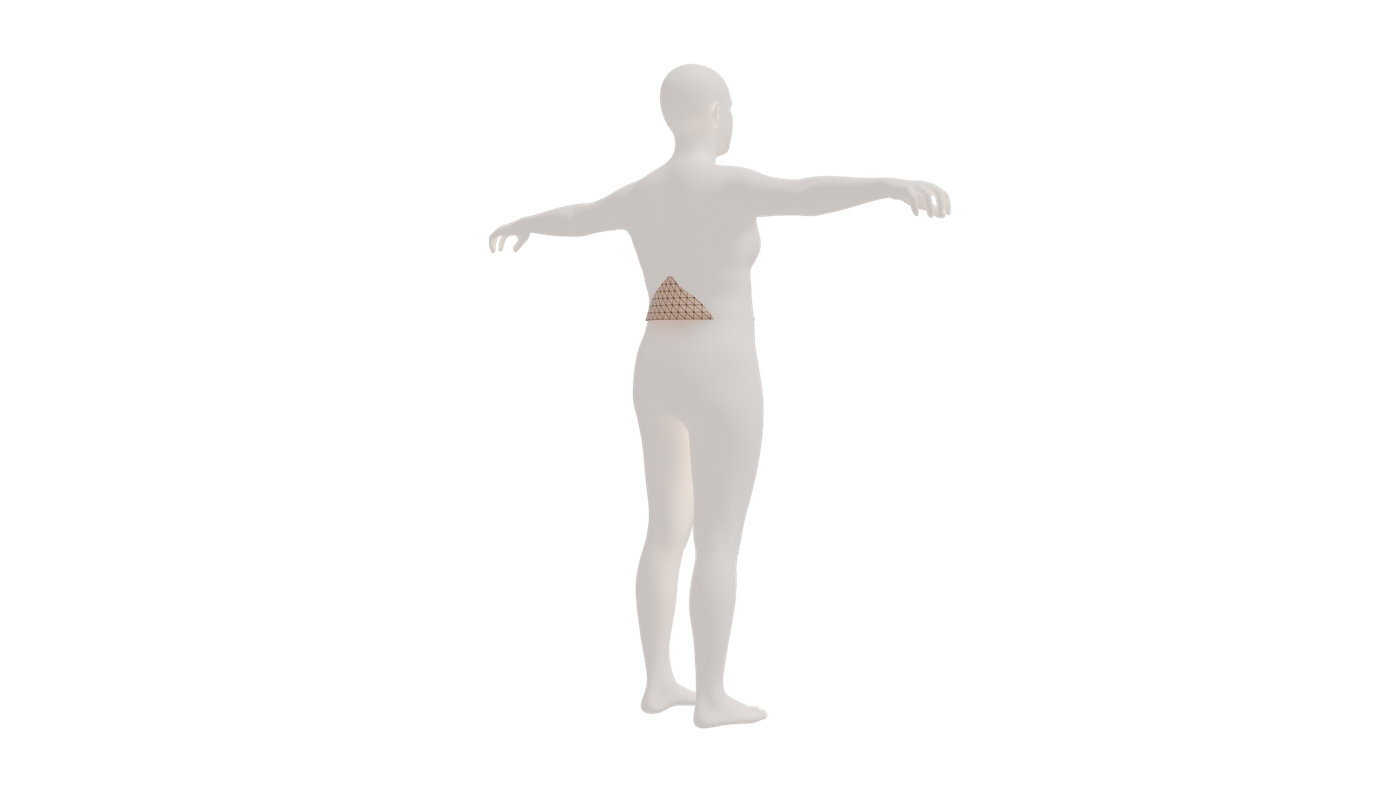}\hfill
  \includegraphics[width=0.196\linewidth]{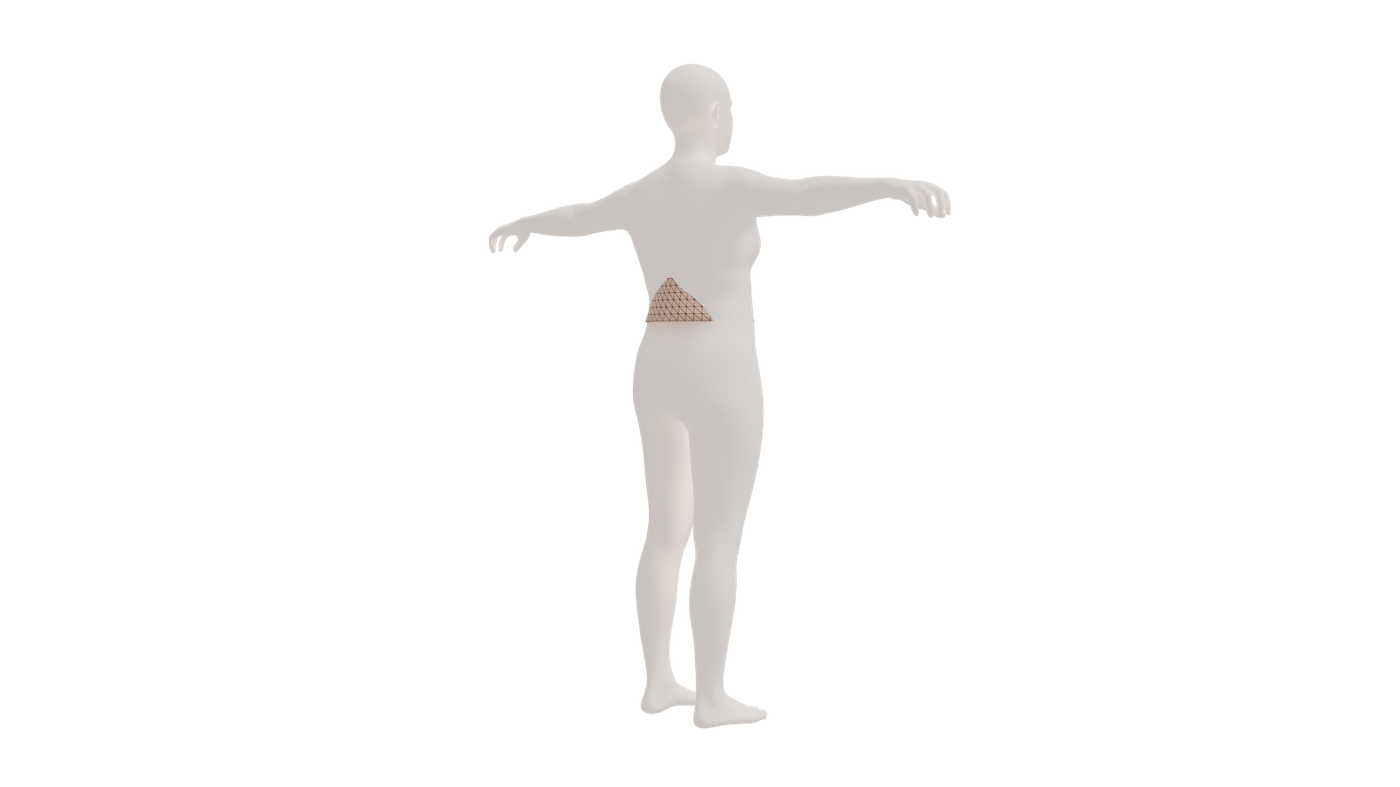}\par\vspace{2pt}
  \includegraphics[width=0.196\linewidth]{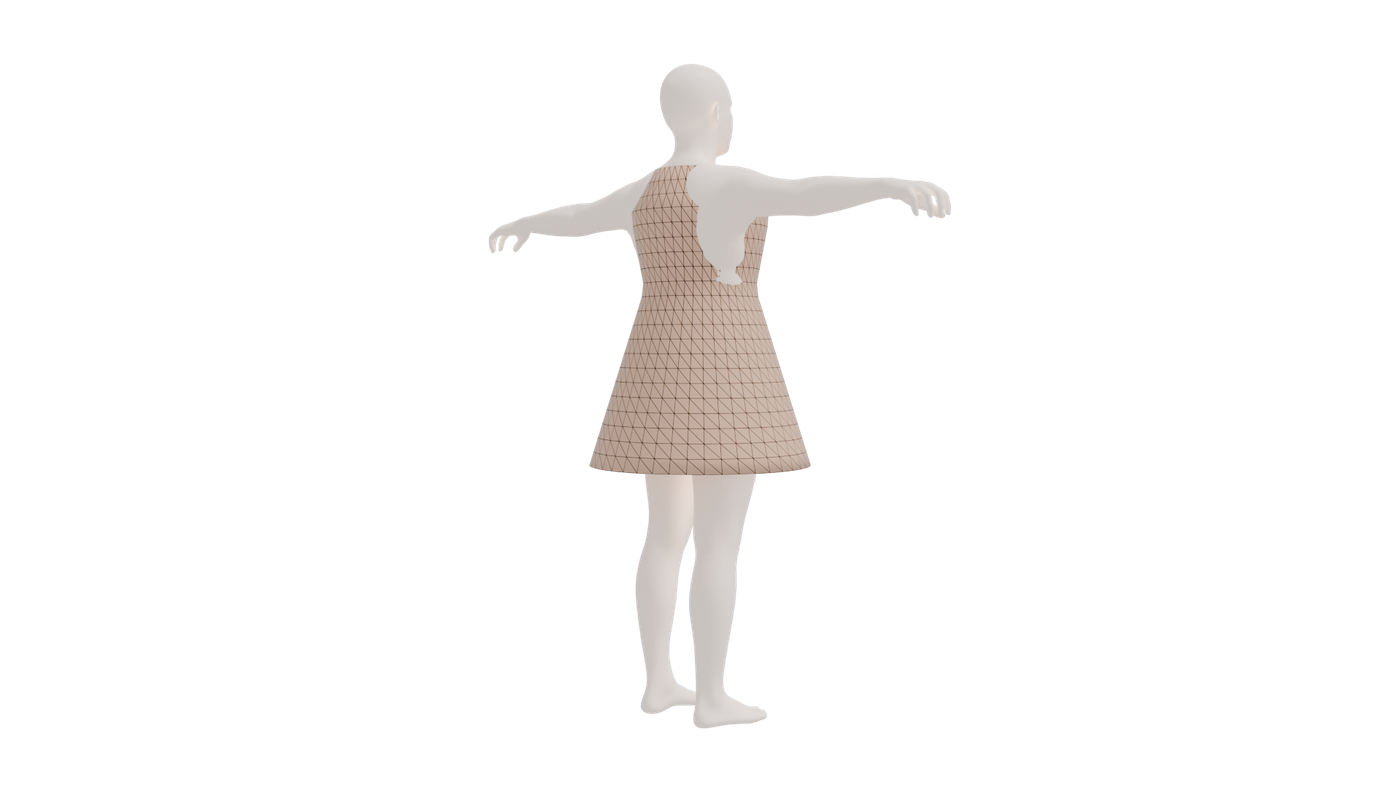}\hfill
  \includegraphics[width=0.196\linewidth]{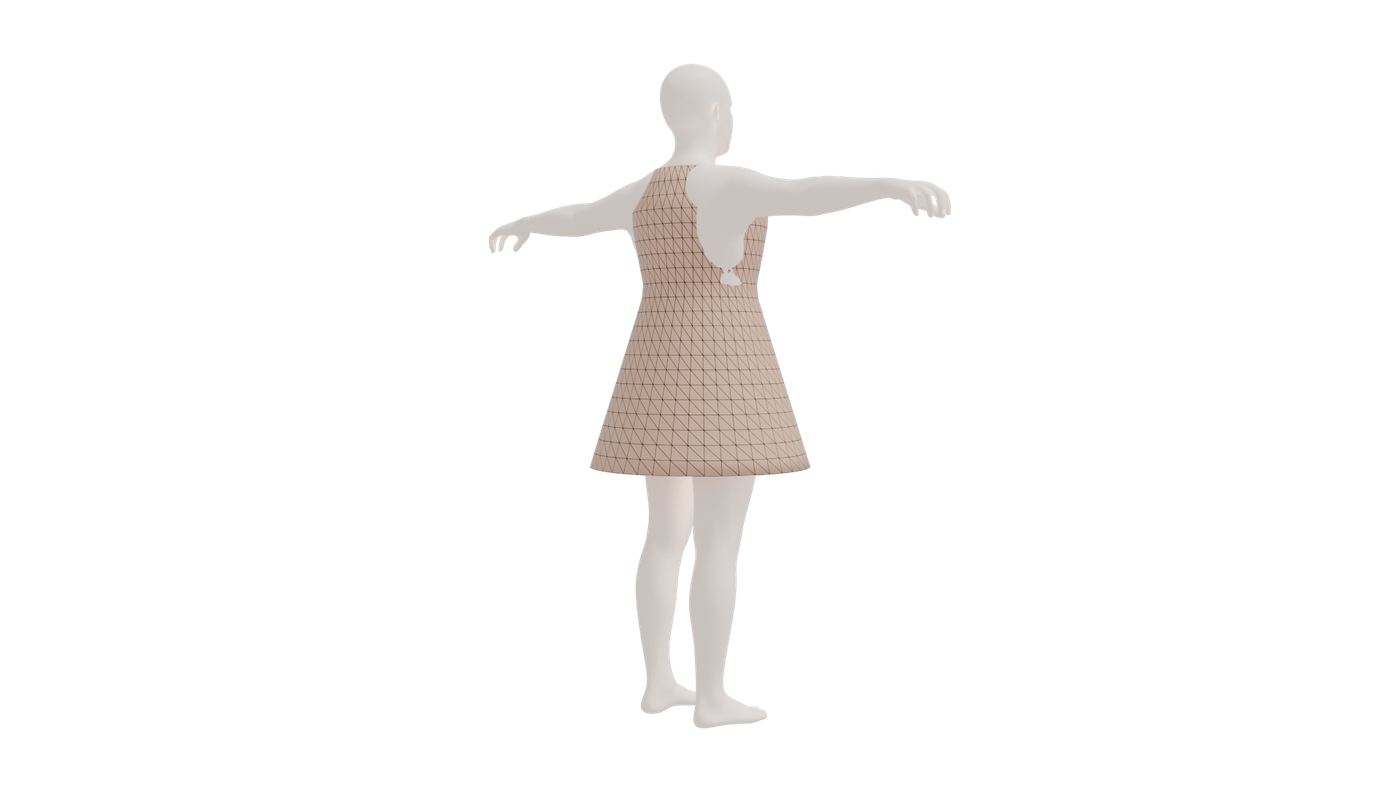}\hfill
  \includegraphics[width=0.196\linewidth]{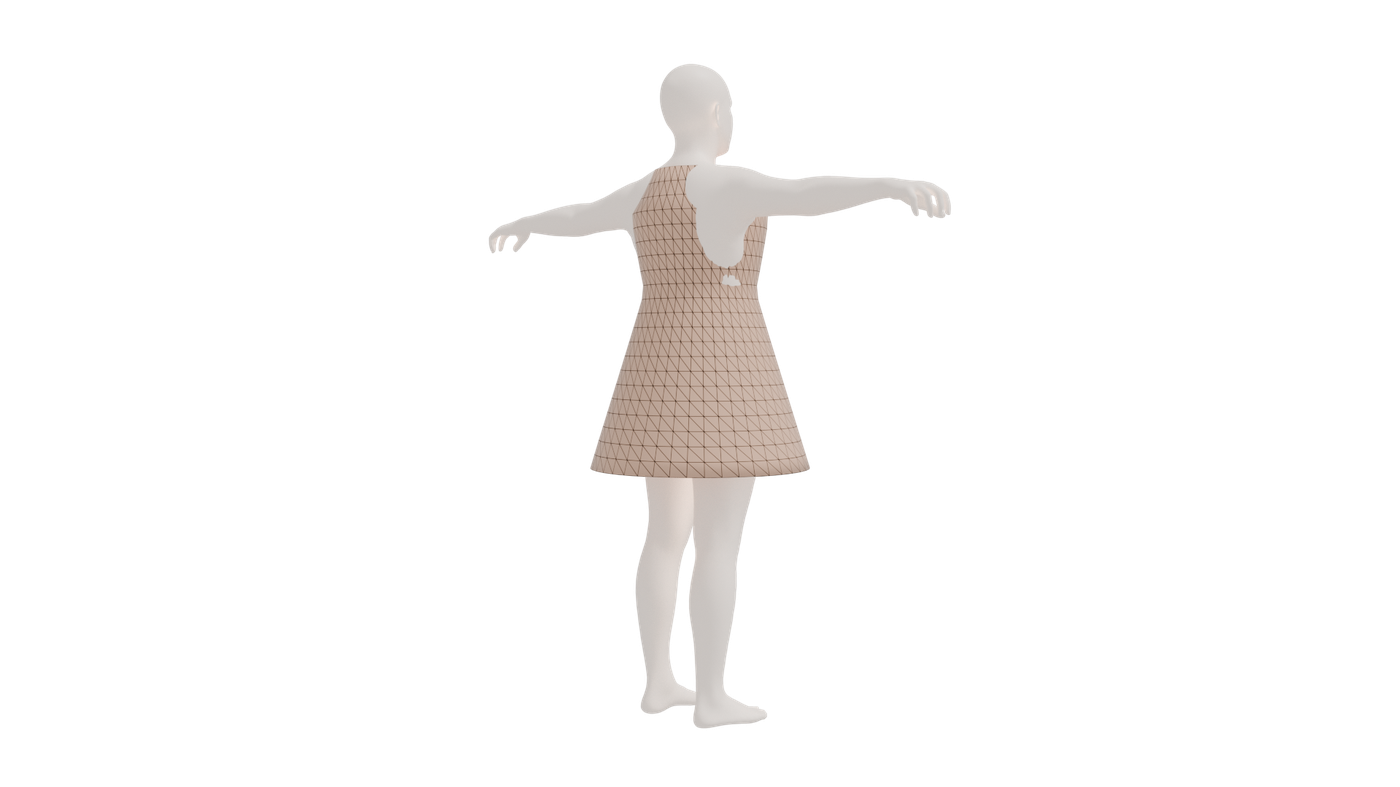}\hfill
  \includegraphics[width=0.196\linewidth]{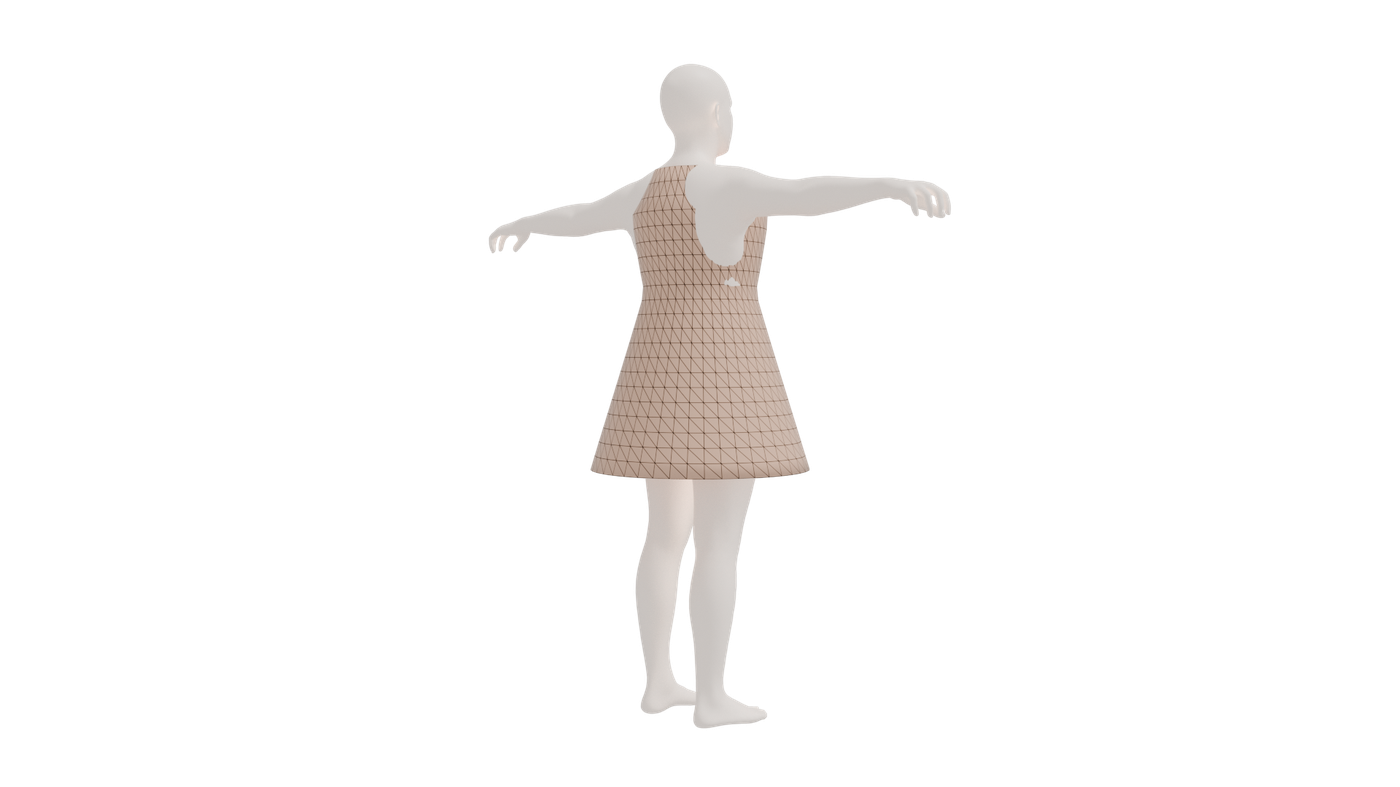}\hfill
  \includegraphics[width=0.196\linewidth]{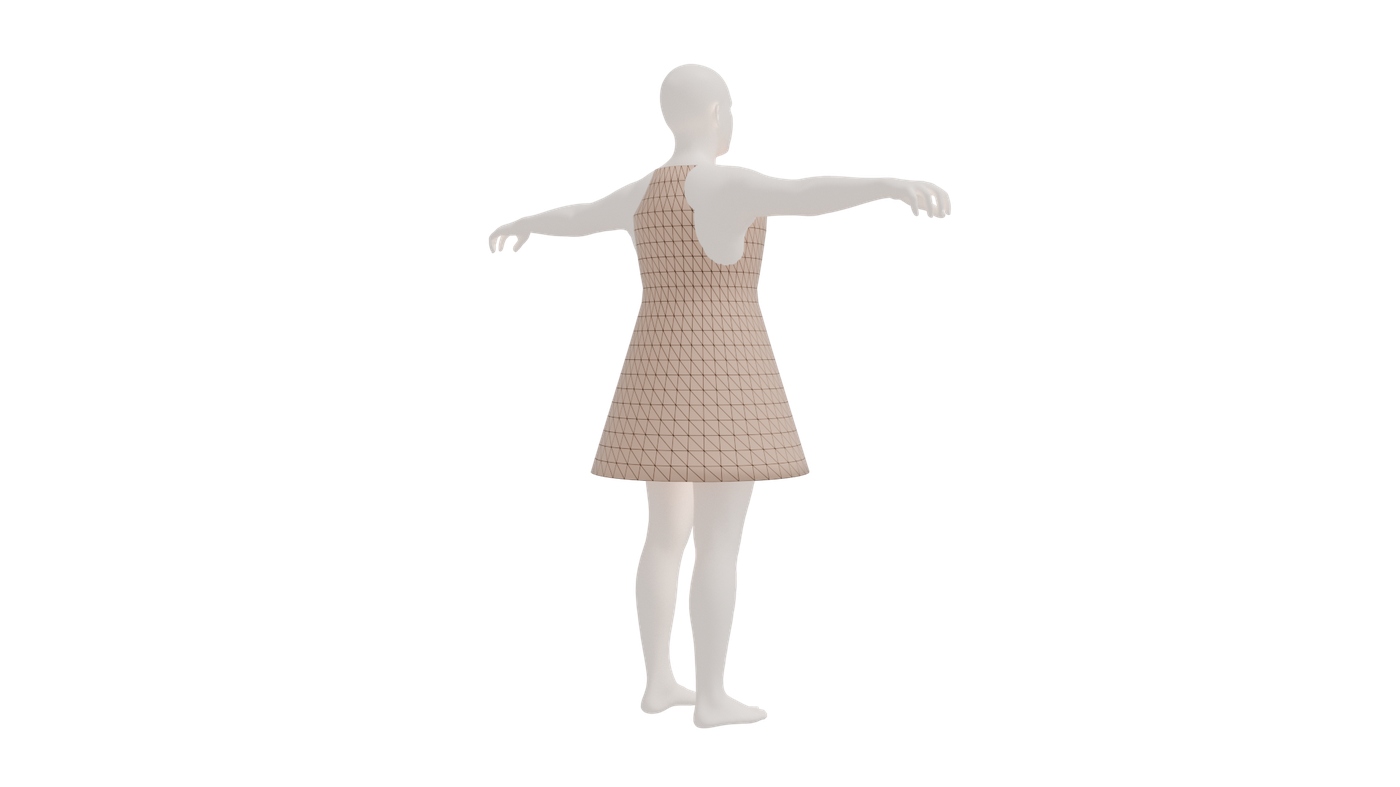}
  \caption{\textbf{Fitting a procedural dress to a body by optimizing a
  rest-shape scale.} Frames at $t\in\{0.02,0.13,0.25,0.37,0.58\}$\,s.}
  \label{fig:body-fit-bptt}
\end{figure}

\paragraph{Rest-shape size from a hem-height target.}
A $924$-vertex procedural A-line dress is pinned at an SMPL-X
\citep{Pavlakos2019SMPLX} neckline ($y{=}1.45$\,m) and drapes through $35$
DiffVBD steps. A single scalar \emph{scale} $s$ multiplies all rest-pose
radii, so the optimized quantity enters through the rest metric rather than
through the material or the initial state; the target is the hem height
$y_{\text{hem}}^{\star}{=}0.654$\,m under
$\mathcal{L}(s){=}(y_{\text{hem}}(s)-y_{\text{hem}}^{\star})^{2}$. Eighty Adam
iterations (learning rate $0.025$) drive $\mathcal{L}$ from $3.4\times10^{-2}$ to
$1.8\times10^{-7}$ and recover $\hat s{=}0.910$ to within $0.5\%$
(Fig.~\ref{fig:body-fit-bptt}).

\begin{figure}[h]
  \centering
  \footnotesize\textit{(top) Init $\bm{w}_0{=}\bm{0}$;
                       (bottom) recovered $\hat{\bm{w}}{=}(2.49,0,1.49)$}\par\vspace{2pt}
  \includegraphics[width=0.196\linewidth]{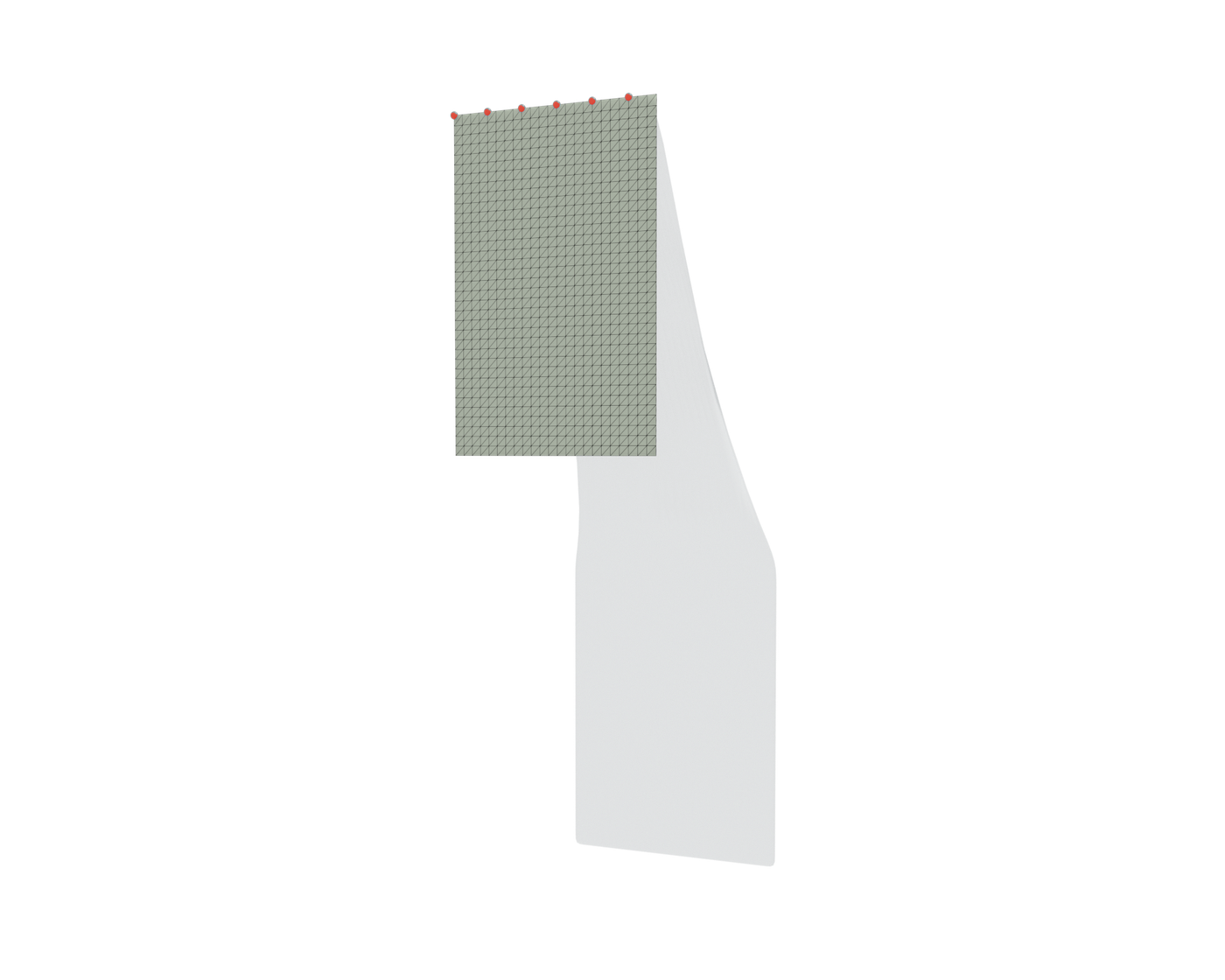}\hfill
  \includegraphics[width=0.196\linewidth]{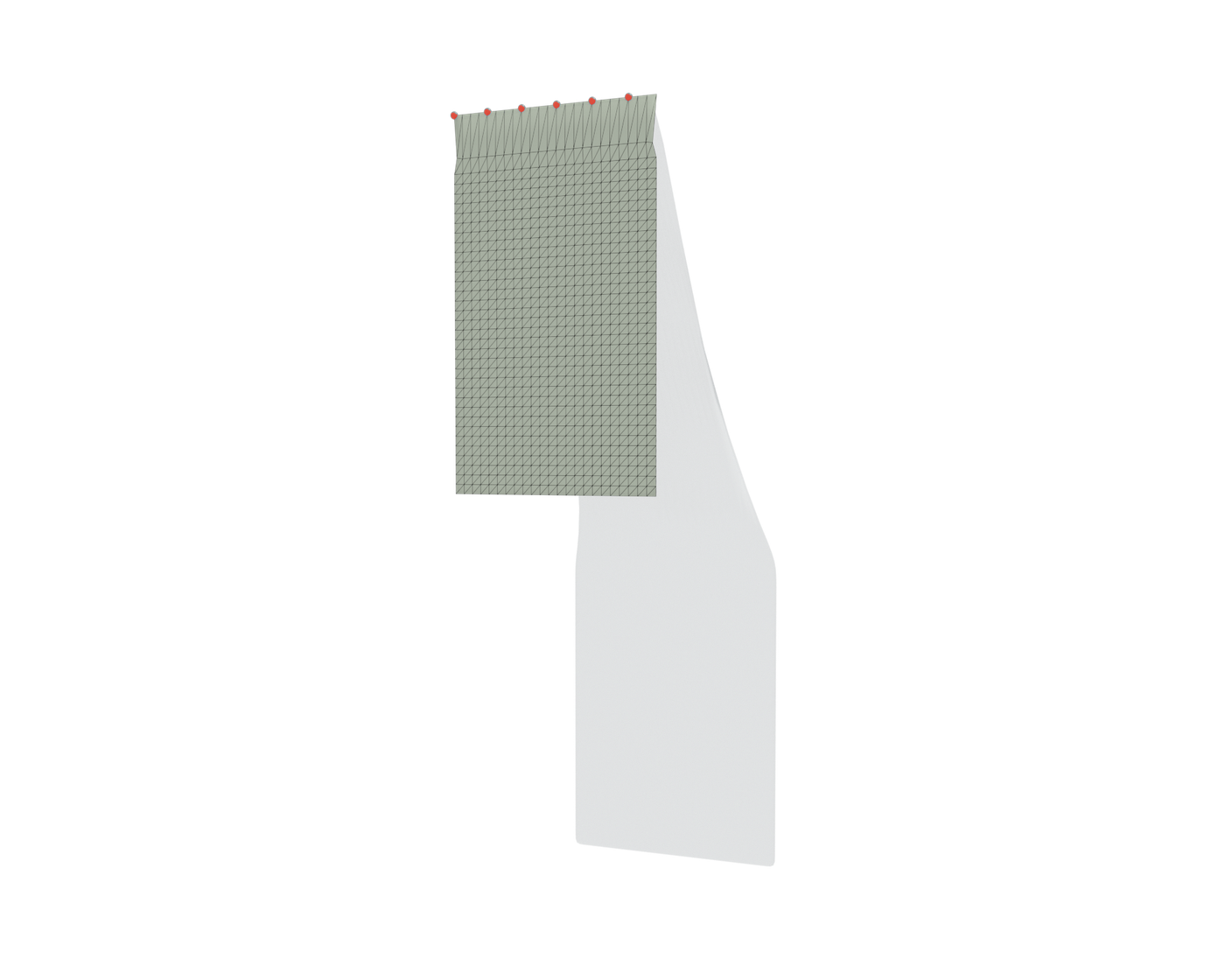}\hfill
  \includegraphics[width=0.196\linewidth]{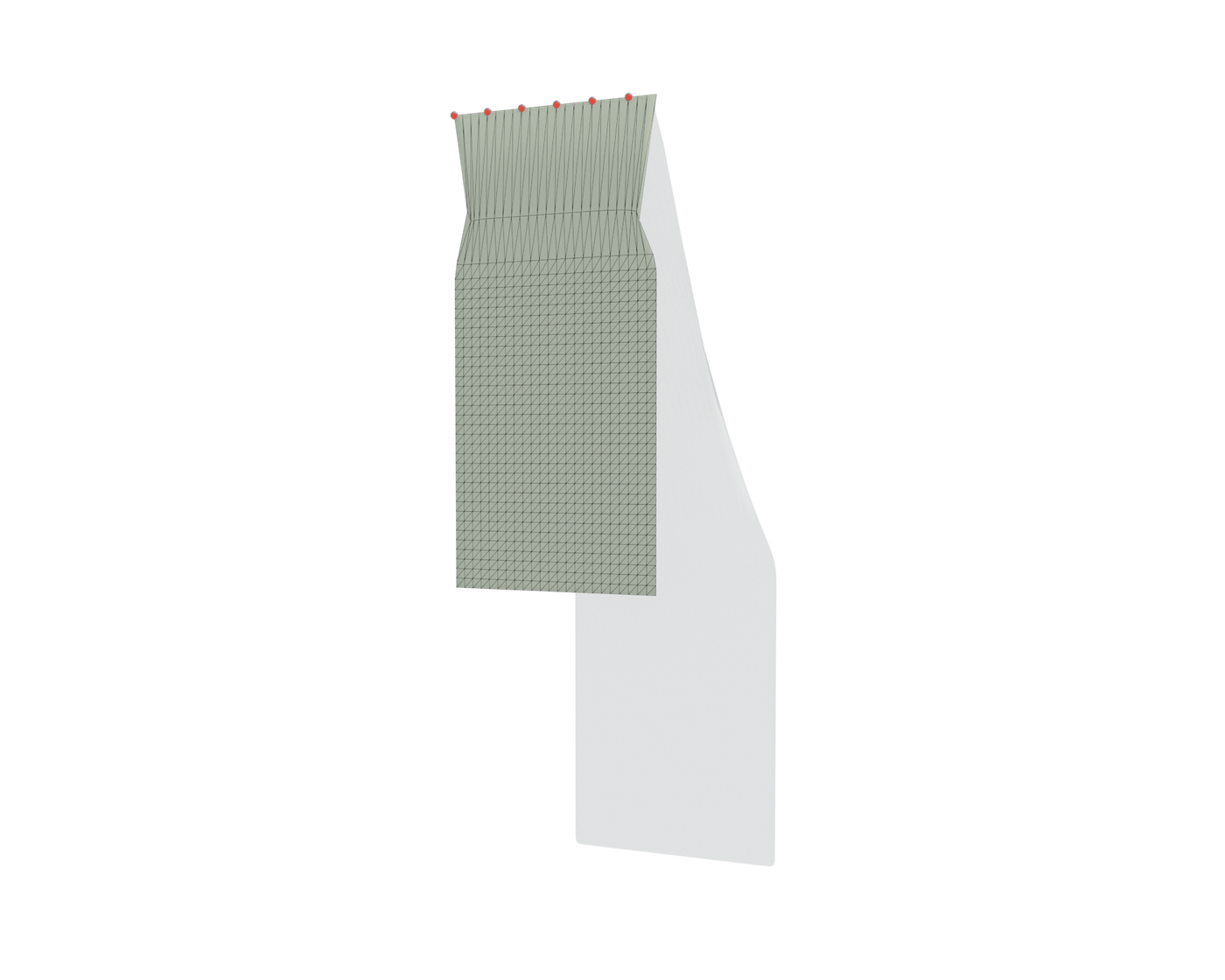}\hfill
  \includegraphics[width=0.196\linewidth]{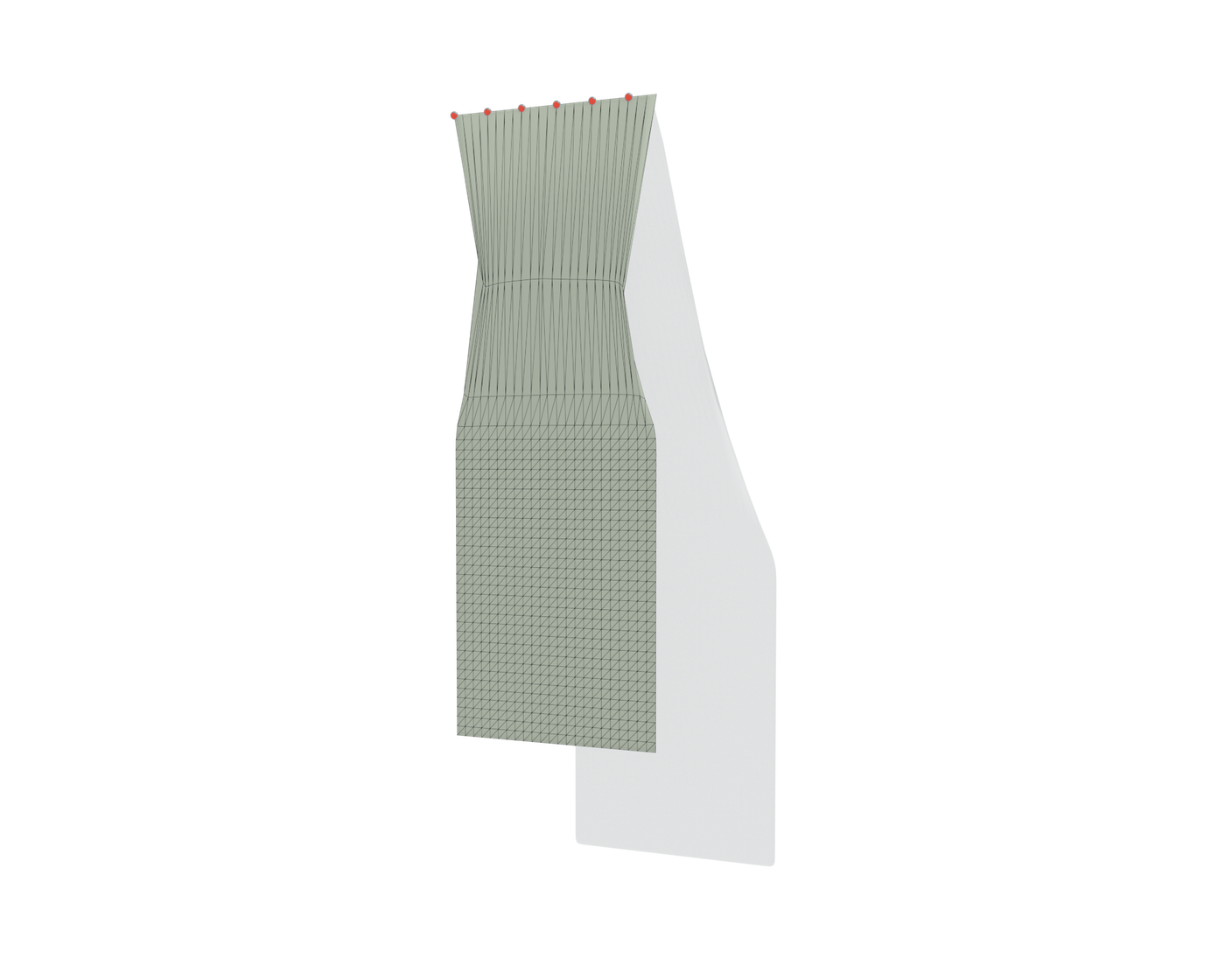}\hfill
  \includegraphics[width=0.196\linewidth]{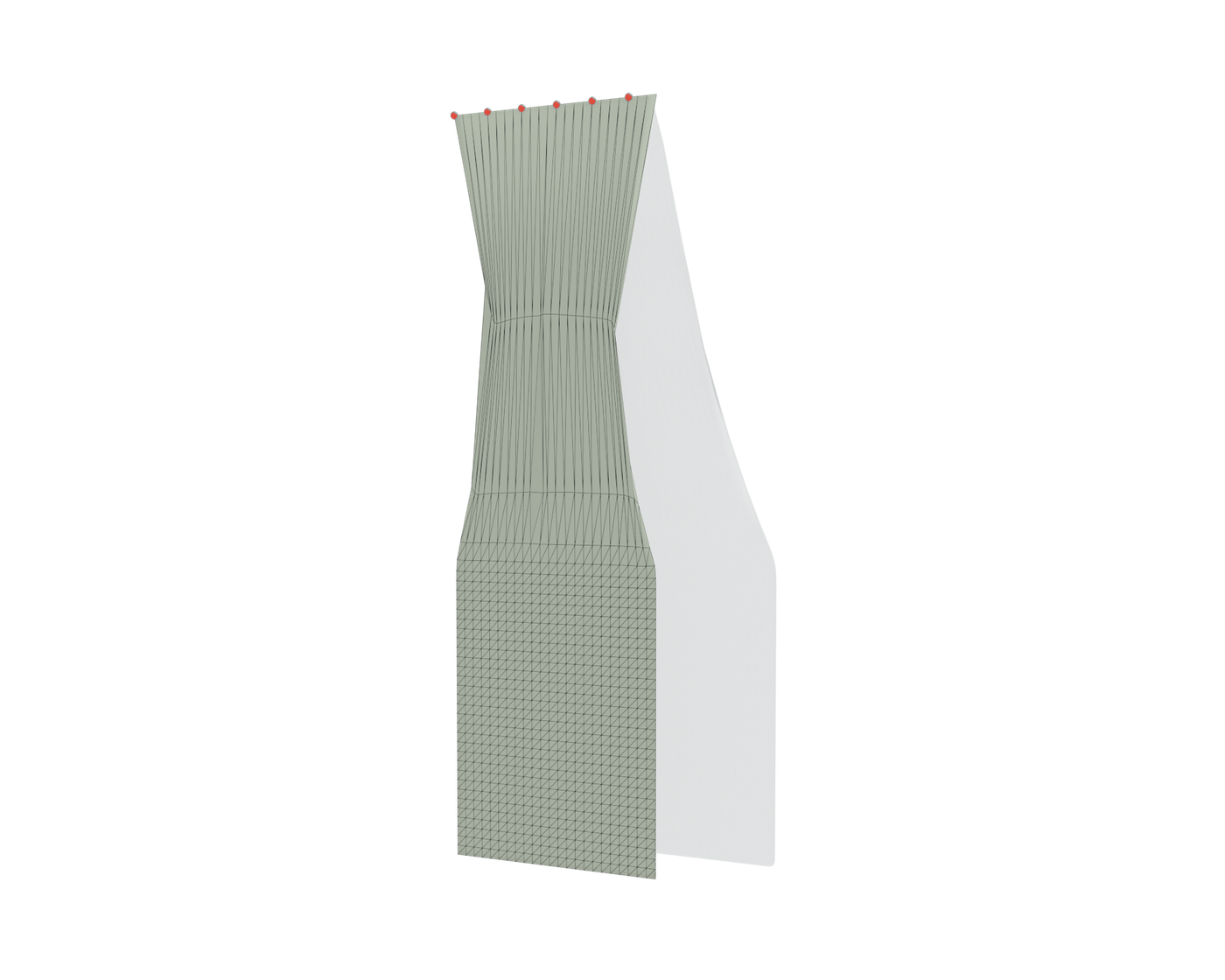}\par\vspace{2pt}
  \includegraphics[width=0.196\linewidth]{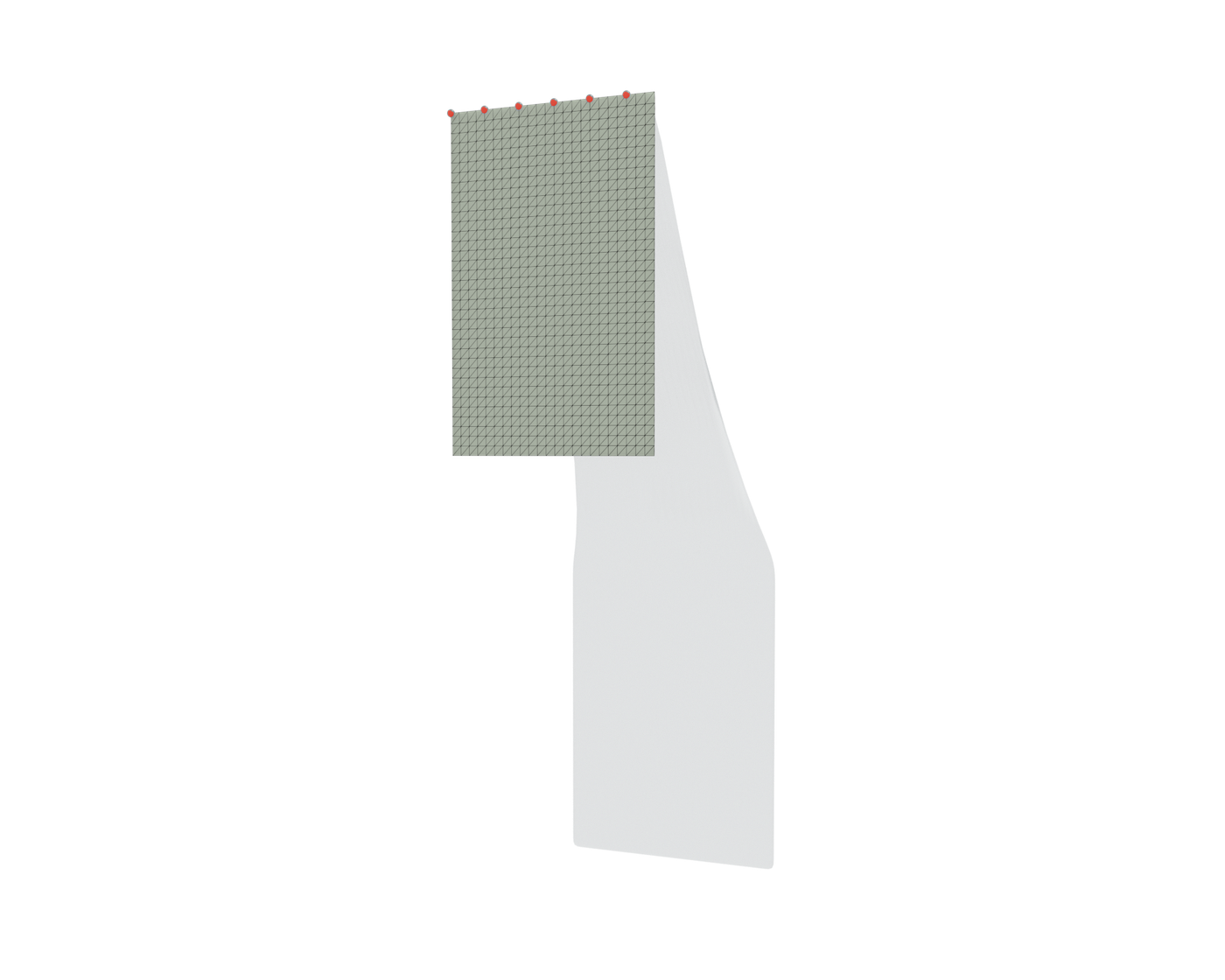}\hfill
  \includegraphics[width=0.196\linewidth]{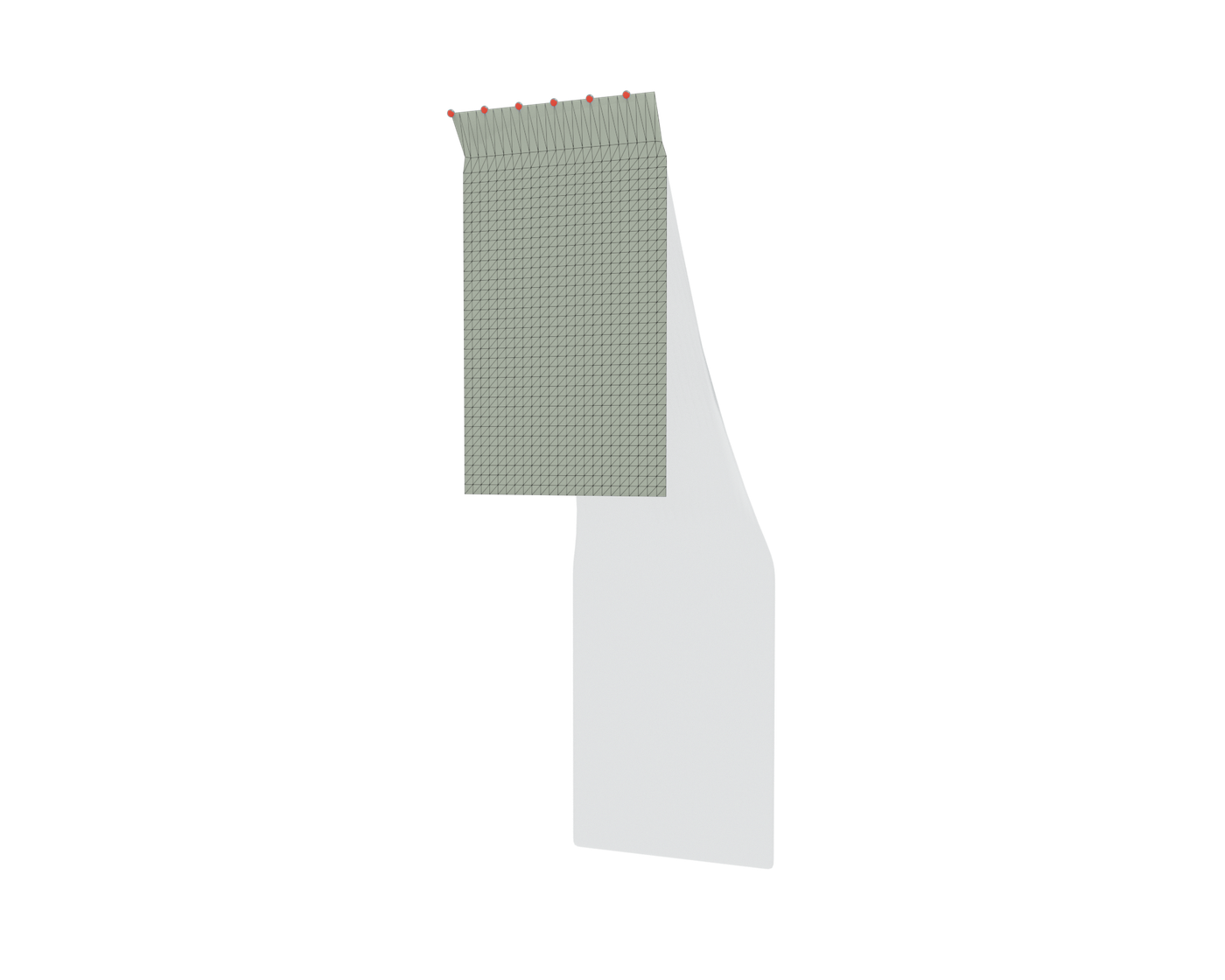}\hfill
  \includegraphics[width=0.196\linewidth]{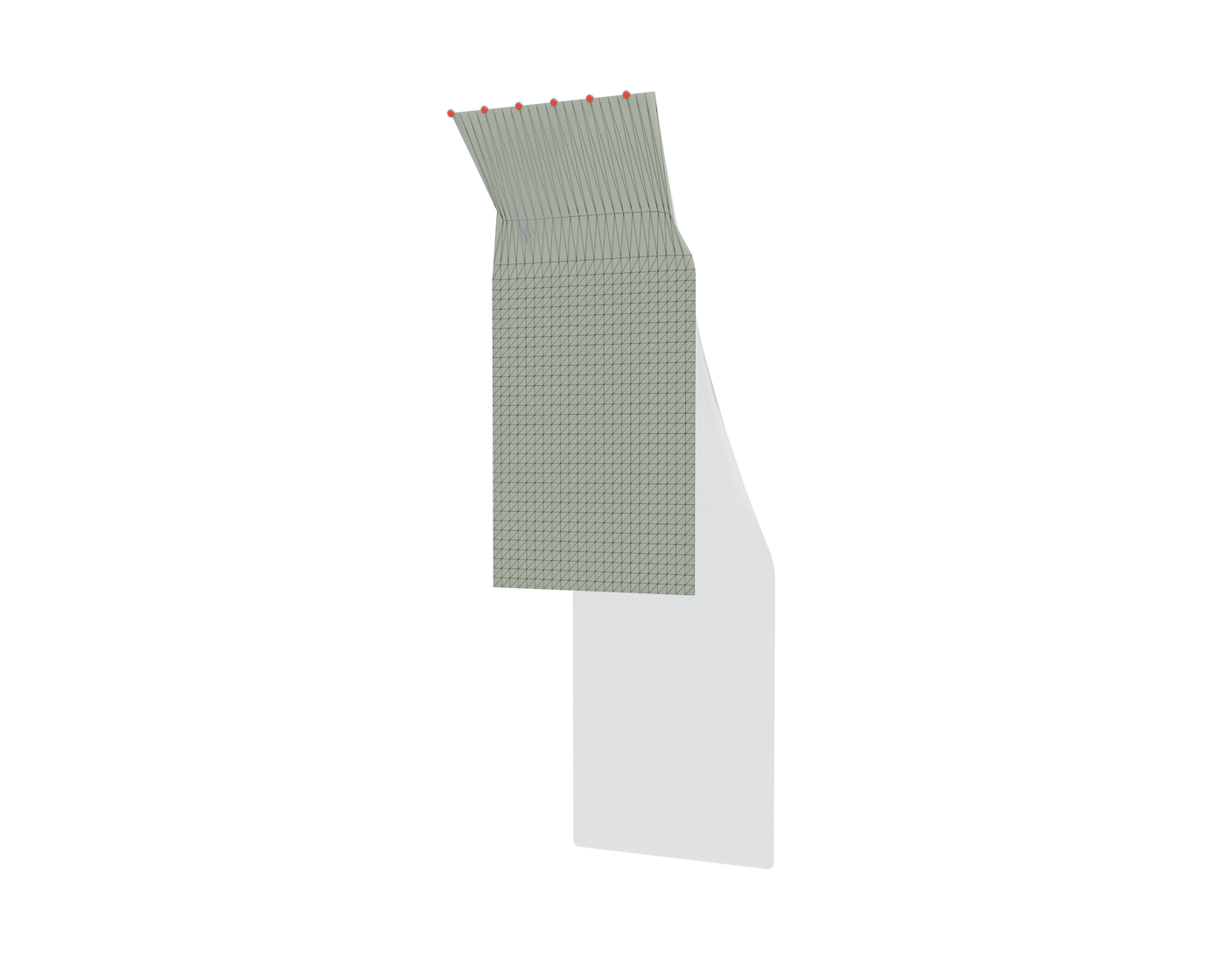}\hfill
  \includegraphics[width=0.196\linewidth]{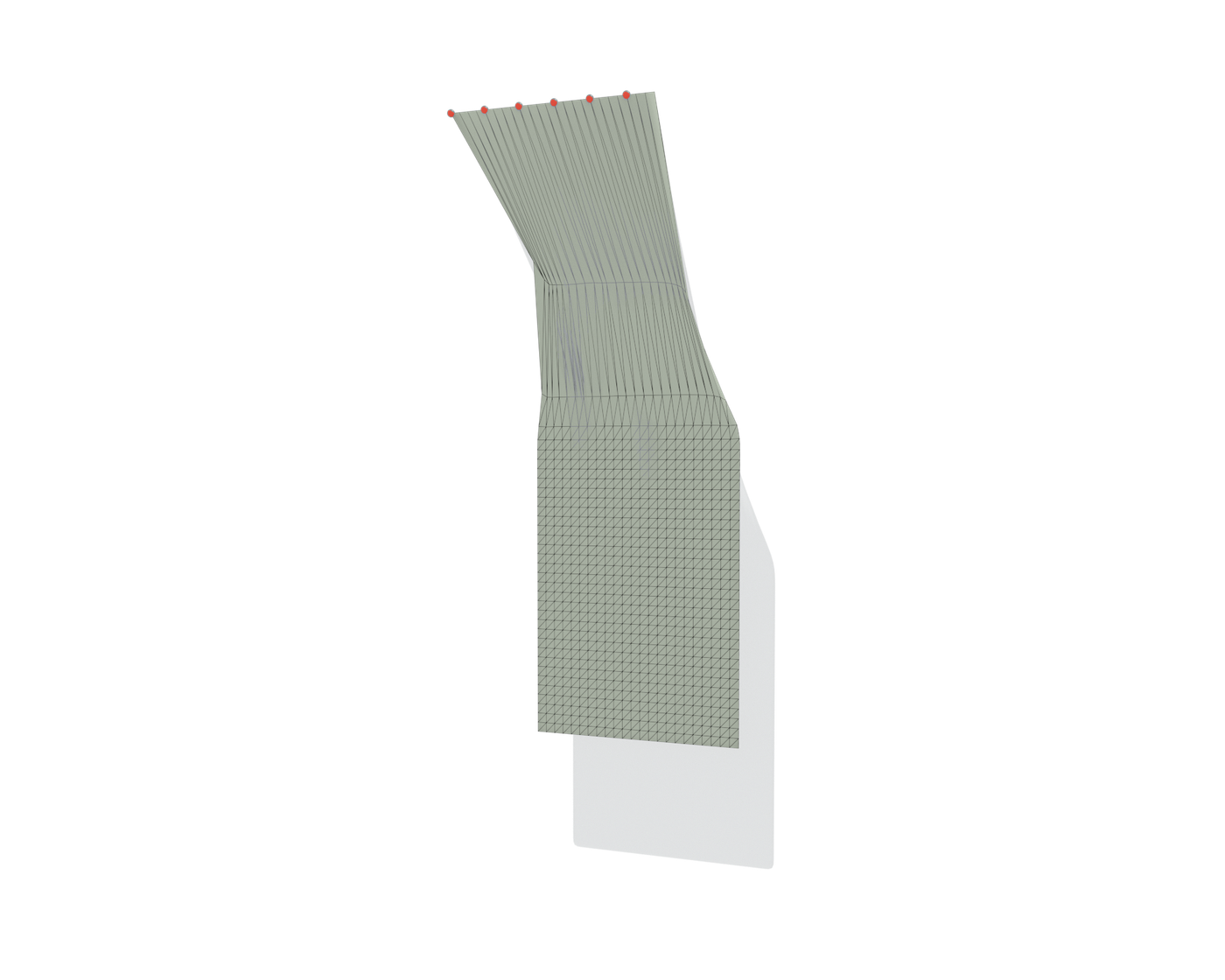}\hfill
  \includegraphics[width=0.196\linewidth]{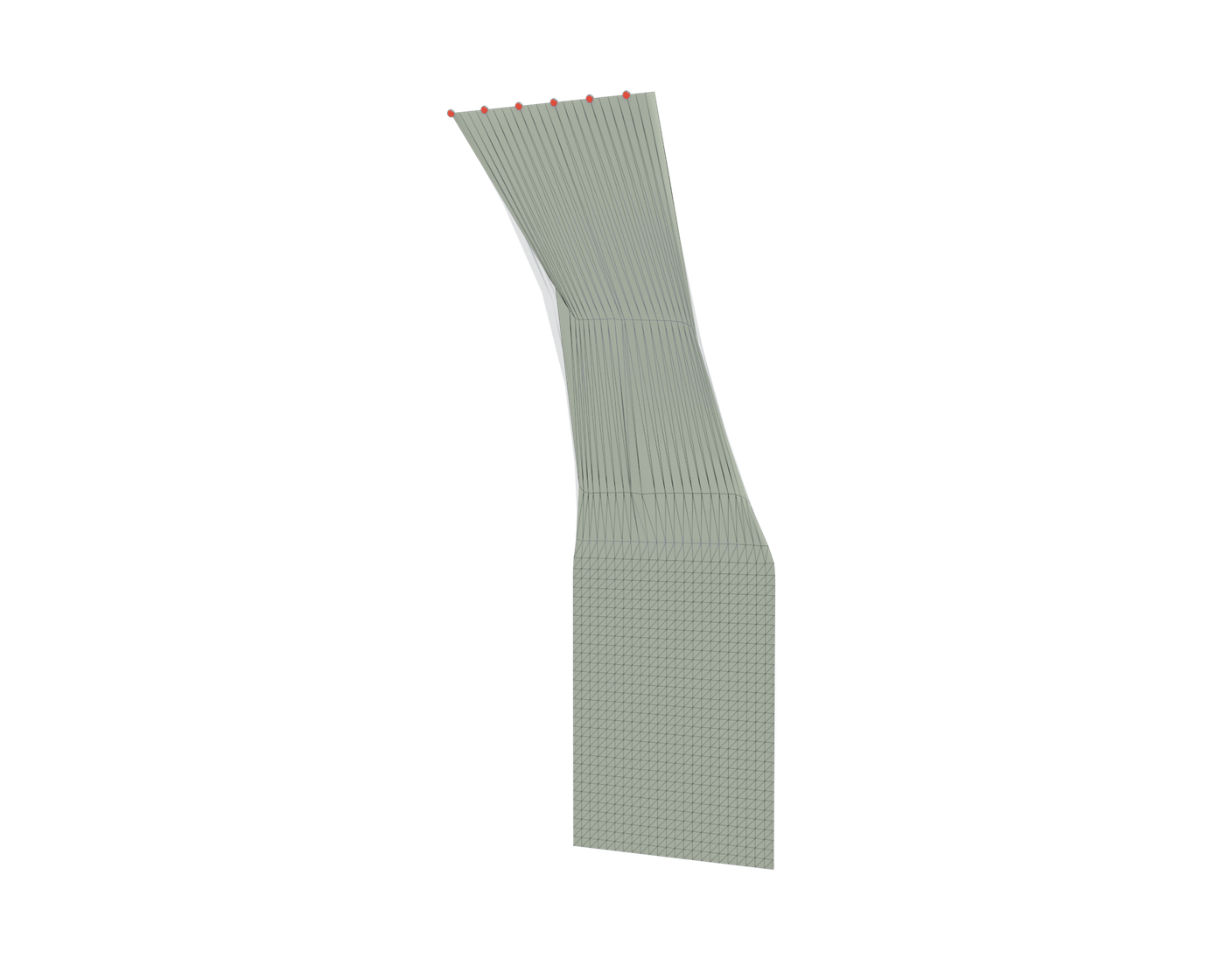}
  \caption{\textbf{Wind-force recovery on a top-pinned cape.} Frames at
  $t\in\{0.02,0.13,0.27,0.40,0.50\}$\,s.}
  \label{fig:cape-wind-bptt}
\end{figure}

\paragraph{Constant wind force from a hem-shape target.}
An $864$-vertex cape pinned along its top edge is exposed to an unknown
constant wind term $\bm{w}$ added to free-vertex velocities each step (a
per-step velocity increment, in m/s). Given
the target hem shape produced by $\bm{w}^{\star}{=}(2.5,0,1.5)$, we recover
$\bm{w}$ from a zero-wind start with $120$ Adam iterations on
$\mathcal{L}(\bm{w}){=}\lVert\bm{x}_T(\bm{w})-\bm{x}_T^{\star}\rVert^{2}$.
Each step's adjoint (\S\ref{sec:principle}) returns
$\partial\mathcal{L}/\partial\bm{v}_{\text{pre}}$, which chains to the wind by
$\bm{w}\leftarrow\bm{w}-\eta h\sum_i\bar{\bm{v}}_i$. Final relative error is
below $0.3\%$, with the loss falling from $4\times10^{-2}$ to
$2\times10^{-8}$ (Fig.~\ref{fig:cape-wind-bptt}). This task differs from the
flag-wind task of \S\ref{sec:exp-inverse}, which recovers a $24$-scalar
time-varying wind on a different mesh and is under-determined; here the wind
is constant and the recovery is unique.

\section{Gradient validation against finite differences}
\label{sec:exp-fd}

Independently of that reference, we validate against central finite differences on
a single-tetrahedron scene, sweeping the \emph{relative} FD step
$\epsilon\in[10^{-3},10^{-8}]$ for $\bm{x}_0$, $\bm{v}_0$, $\mu$ and $\lambda$.
The quantity reported is the maximum relative error between the analytic and FD
gradients; a configuration passes if that error is below $10^{-6}$ and stable
across the $\epsilon$ sweep (so that it reflects the gradient, not FD
truncation or round-off). All $11$ checks pass; Table~\ref{tab:fd} (App.~\ref{app:figs}) lists six
representative rows. CPU and GPU agree to machine precision.

\section{Solver depth and achieved outcome}\label{app:kdepth}

\begin{figure}[t]
\centering
\includegraphics[width=0.92\linewidth]{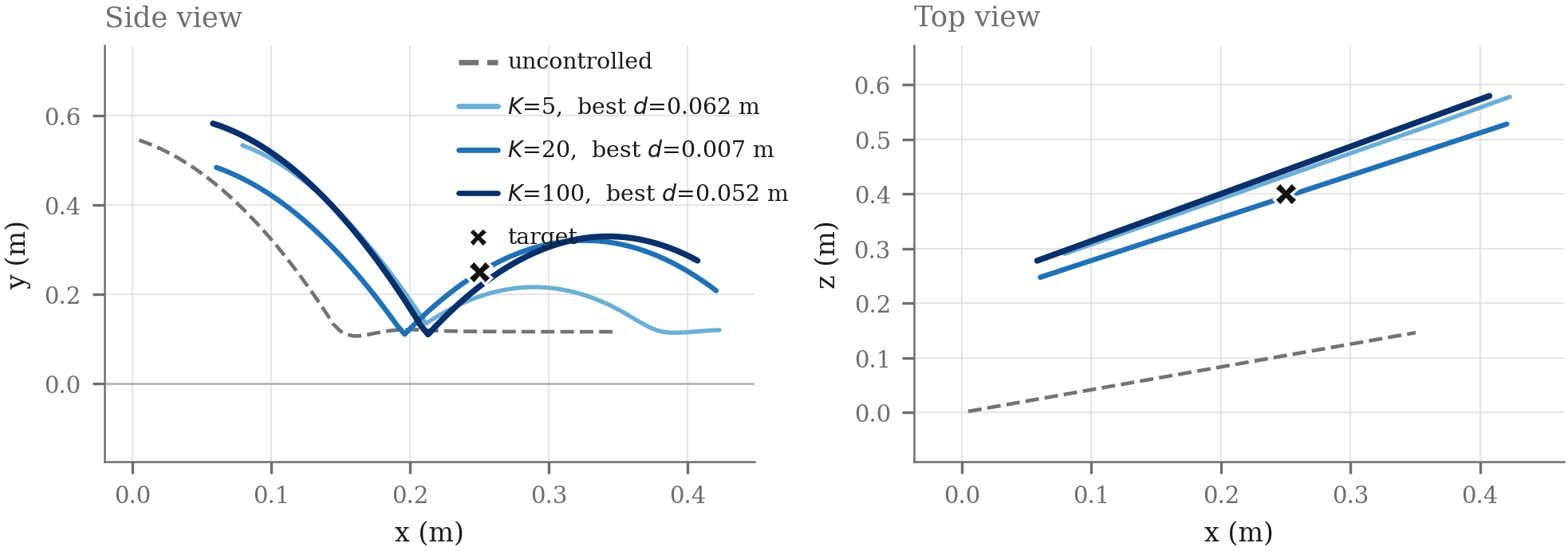}
\caption{Optimized cube center of mass trajectories at three solver depths under identical
$40$-iteration Adam budgets, with the uncontrolled trajectory under the initial
parameters overlaid (gray dashed). Every optimized trajectory passes close to
the target; the ordering between depths is within run-to-run spread on this
task (App.~\ref{app:kdepth}).}
\label{fig:Ksweep}
\end{figure}

Deeper solvers give more accurate gradients, but the operative question is
whether that changes what an optimizer achieves. Whether better gradients help the optimizer at all is
subtle in general \citep{Suh2022PolicyGradients}. Repeating the cube
initial-state task at $K\in\{5,20,100\}$ under identical $40$-iteration Adam
budgets from the default initialization gives best center-of-mass-to-target
distances of $0.039$, $0.058$ and $0.024$\,m
(Fig.~\ref{fig:Ksweep}). All three land far closer than the uncontrolled
trajectory, but the ordering is not monotone in $K$, which from a single
initialization is uninformative: with six optimization variables and a short
budget the gap between depths could easily be smaller than the spread induced
by where the optimizer starts.

We therefore measured that spread. The optimization is deterministic given an
initialization, so we re-ran the full sweep from $8$ initializations drawn
around the default with $15\%$ relative jitter, holding everything else fixed
(Table~\ref{tab:kspread}).

\begin{table}[h]
\caption{\textbf{Initialization spread of the cube initial-state task}, best
center-of-mass--target distance in metres over $8$ perturbed initializations per depth,
identical $40$-iteration budgets. Deeper solves are better on average and
roughly twice as consistent, but the mean gap is at the edge of what $8$
initializations resolve.}
\label{tab:kspread}
\centering\small
\renewcommand{\arraystretch}{1.12}
\begin{tabular}{lcccc}
\toprule
$K$ & mean & std & min & max \\
\midrule
$5$   & $0.052$ & $0.020$ & $0.023$ & $0.081$ \\
$20$  & $0.034$ & $0.018$ & $0.015$ & $0.065$ \\
$100$ & $0.035$ & $0.010$ & $0.021$ & $0.053$ \\
\bottomrule
\end{tabular}
\end{table}

Two things follow, and we state both. The mean does improve from $K{=}5$ to
$K{=}20$ ($0.052\to0.034$\,m) and then flattens, with $K{=}20$ and $K{=}100$
statistically indistinguishable (Welch $t{=}0.11$). But the improvement from
$K{=}5$ is only borderline resolved at this sample size (Welch $t{=}2.16$,
$\mathrm{df}{\approx}10.5$ against $K{=}100$; $t{=}1.87$ against $K{=}20$), so
we do \emph{not} claim a significant depth effect on the achieved objective.
The clearer effect is on variance: the standard deviation falls by $1.9\times$
from $K{=}5$ to $K{=}100$, i.e.\ deeper solves make the outcome more
repeatable across initializations even where they do not make it better on
average. The claim we do make remains the one in \S\ref{sec:exp-exact}, about
gradient accuracy, which is measured against a reference rather than inferred
from optimizer outcomes.

\section{Implementation details}\label{app:impl}

\paragraph{Backends.} Two GPU implementations share one algorithm. The first
is written in NVIDIA Warp \citep{Macklin2022Warp} kernels; the second is a
fused CUDA C++ extension compiled on first use through PyTorch's
\texttt{load\_inline} and cached thereafter. Both are \texttt{float64}
end to end. They agree to $10^{-16}$ on forward state and on gradients, which
is what lets us use either as the reference for the other; the fused path is
$7$--$22\times$ faster because one kernel launch per color performs assembly,
the $3{\times}3$ solve, the saturation and the write-back without a round trip
to global memory between stages.

\paragraph{Data layout.} Each vertex owns a \emph{star}: the list of
(tetrahedron, local vertex index) pairs it participates in, packed into flat
offset/element/local-index arrays so a thread reads a contiguous range. The
inverse rest metric of each tetrahedron is stored as nine separate scalar
arrays $\mathrm{IB}_{00}\dots\mathrm{IB}_{22}$ rather than as a $3{\times}3$
struct, so that a 32-thread hardware warp reading one column is coalesced. Vertices are
distance-1 greedily colored, which guarantees that two vertices of one color
never share a tetrahedron and can therefore be updated in parallel without
write conflicts.

\paragraph{Replay buffer.} The backward does not store an AD graph. Per sweep
$k$ and vertex $i$ it stores only the applied update $\Delta\bm{x}_i^{(k)}$
(three doubles) and the six independent entries of the symmetric block
$H_i^{(k)}$, that is nine \texttt{float64} per vertex per sweep, or $72$ bytes.
Everything else the reverse sweep needs is recomputed from the replayed state.
The ${\approx}4$\,KB per vertex per sweep quoted in
Table~\ref{tab:scaling} is the caching-allocator peak
(\texttt{torch.cuda.max\_memory\_allocated}), which also contains the mesh,
the star index and the material arrays; the tape proper is the $72$ bytes. The reverse pass walks sweeps, colors and vertices in
exactly reversed order, undoing $\Delta\bm{x}_i$ to recover the pre-update
state rather than storing it.

\paragraph{Safeguards.} Two safeguards in the forward are differentiated
rather than treated as constants, which is what
\S\ref{sec:exactness} requires. The per-vertex block is regularized as
$H_i+\tau I$ with $\tau{=}10^{-6}$ in the $h^{2}$-scaled system the kernels
assemble (equivalently $\tau/h^{2}$ in the normalization of
Eq.~\eqref{eq:local-gh}), and every update is saturated by
$\Delta\bm{x}\leftarrow\Delta\bm{x}\,c/(\lVert\Delta\bm{x}\rVert+c+\varepsilon)$
with $c{=}0.1$\,m and $\varepsilon{=}10^{-12}$. Both constants are defined once
as compile-time macros shared by the forward and backward kernels, so the two
cannot drift apart. Two details matter for a faithful re-implementation of
Algorithm~\ref{alg:diffvbd}: the saturation Jacobian of
Eq.~\eqref{eq:sat-jac} is applied only to the copy of $\bar{\bm{x}}_i^{+}$
that enters the local adjoint solve (the identity branch of
Eq.~\eqref{eq:identity-adjoint} uses the untransformed adjoint), and the
Hessian-tangent and parameter contractions use the \emph{raw} update
$\Delta\bm{x}_i^{\text{raw}}$, recovered exactly by inverting the saturation
of the recorded step. Writing them exactly matters: Warp through version $1.11$
rounds a literal inside \texttt{wp.float64(0.1)} through \texttt{float32}
first, which perturbs $c$ by $1.5\times10^{-8}$ relative and was large enough
to move the Warp backend off machine precision until the constants were
written exactly.

\paragraph{Contact.} Two primitives coexist. The built-in self and floor model
is an IPC \citep{Li2020IPC} log barrier with stiffness $\kappa$, activation
distance $\hat d$ and a lagged dissipation term $c_{\mathrm{damp}}$ that damps
only downward motion, so the damping does not fight separation. For many-body
scenes a differentiable center-of-mass penalty acts between bodies: a
uniform-grid fixed-radius search (cell size equal to the search radius,
linearized cell keys sorted once, candidates from the $27$ neighbor cells by
binary search) supplies candidate pairs, and a ReLU (optionally softplus,
sharpness $\alpha$) penalty on the overlap supplies the force, differentiable
in the centers of mass and therefore in the vertex positions that define
them. The pair list is index data with no gradient, so the broad phase adds
nothing to the adjoint.

\paragraph{Tensor interface and large runs.} Both backends expose forward and
backward entry points that consume and return GPU-resident tensors, so an outer
optimization loop never copies through the host; this is roughly $4\times$
faster than the array-based interface for backpropagation-through-time loops. At $10^{6}$ bodies the rollout is gradient-checkpointed
\citep{Chen2016SublinearMemory} every $5$ steps, bodies are packed into a
single mega-mesh processed in chunks of $65{,}536$, gradients are clipped at
norm $5$, and the material is optimized through a sigmoid parameterization that
confines $\mu$ to a bounded range instead of relying on projection.

\section{Inverse-problem tasks in detail}\label{app:inverse-detail}

These are the per-task descriptions for \S\ref{sec:exp-inverse}; the
headline numbers are Table~\ref{tab:inverse} in the main text, and the
reproduction protocols are Appendix~\ref{app:tasks}.

\begin{wrapfigure}[9]{r}{0.38\textwidth}
\vspace{-10pt}
\centering
\includegraphics[width=\linewidth]{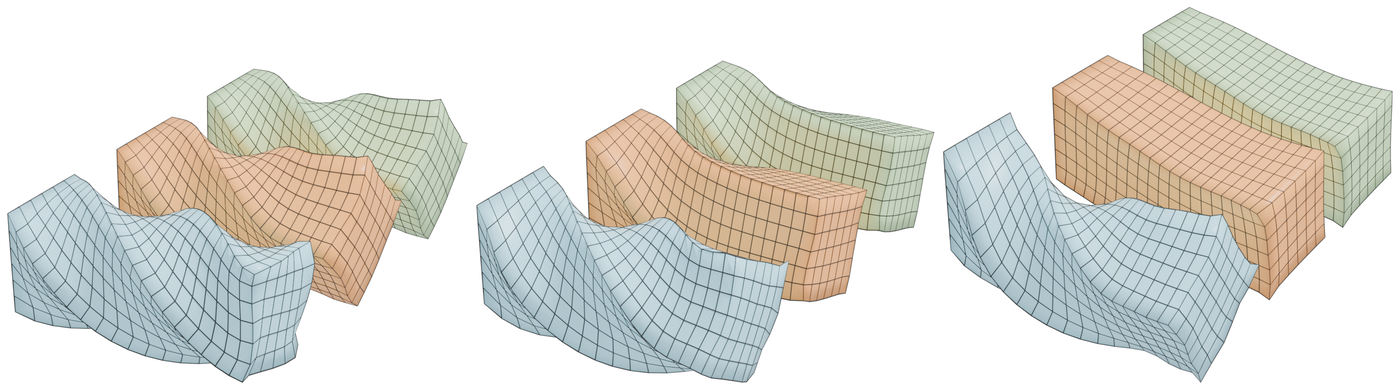}
\caption{Beam $\mu$ identification at three twist stages. Blue: initial ($\mu_{\text{init}}{=}10^{5}$). Coral: target ($\mu_{\text{true}}{=}10^{6}$). Green: recovered.}
\label{fig:beam-mu}
\vspace{-14pt}
\end{wrapfigure}

\textbf{Material identification.} Given a target rollout produced by unknown
Lam\'e parameters, recover them. On a $1029$-vertex cantilever twisted
$180^\circ$ and released, Adam with backpropagation through time recovers
$\mu_{\text{opt}}{=}1.001\times10^{6}$ from an initialization $10\times$ too
soft ($0.1\%$ error, $76$ iterations to the $10^{-4}$ stopping criterion). With
both parameters free, $(\mu,\lambda){=}(800,400)$ is recovered from $(300,150)$
to $<0.25\%$ (Figs.~\ref{fig:beam-mu},~\ref{fig:beam-mulam}). On a
$600$-vertex dress draping over a capsule mannequin with IPC ground contact,
single-stage log-space Adam recovers $(1471,880)$ from $(5000,3000)$
(Fig.~\ref{fig:applications}); replacing the vertex-position loss with a radial
silhouette profile gives $(1528,900)$.

The dress pair is recovered by a two-stage schedule whose structure mirrors
the solver's own block-coordinate principle. No single observation window
identifies both parameters: the time-averaged settled drape pins $\mu$ (to
$0.4\%$ under a 1-D Newton) but leaves $\lambda$ nearly free, while per-step
matching of the early frames -- before the contact-rich trajectory decoheres
\citep{Metz2021Gradients}
-- pins $\lambda$ (to ${\approx}1\%$) but constrains $\mu$ weakly from the
soft side. Stage A runs Adam on the sum of both objectives, which moves the
pair together into the correct basin; stage B then alternates damped 1-D
Newton solves, $\mu$ against the settled-shape objective and $\lambda$
against the early-dynamics objective. The alternation alone is insufficient:
started from $(5000,3000)$ it locks into a self-consistent wrong fixed point
$(1325,3310)$, which is why the basin-entering stage comes first. From the
same initialization the full schedule reaches $(1499.9, 800.2)$ against the
true $(1500, 800)$, bit-identical over three repeated runs. The same schedule
transfers to the silhouette variant with every observation restricted to
radial profiles (a silhouette \emph{sequence} rather than a single settled
silhouette): it recovers $(1500.0, 800.0)$ exactly, and although the stage-A
iterates differ slightly across repeated runs, stage~B contracts all of them
to the same fixed point, so the answer is run-to-run identical.

\begin{figure}[t]
\centering
\begin{minipage}[c]{0.44\linewidth}
\centering
\includegraphics[width=\linewidth]{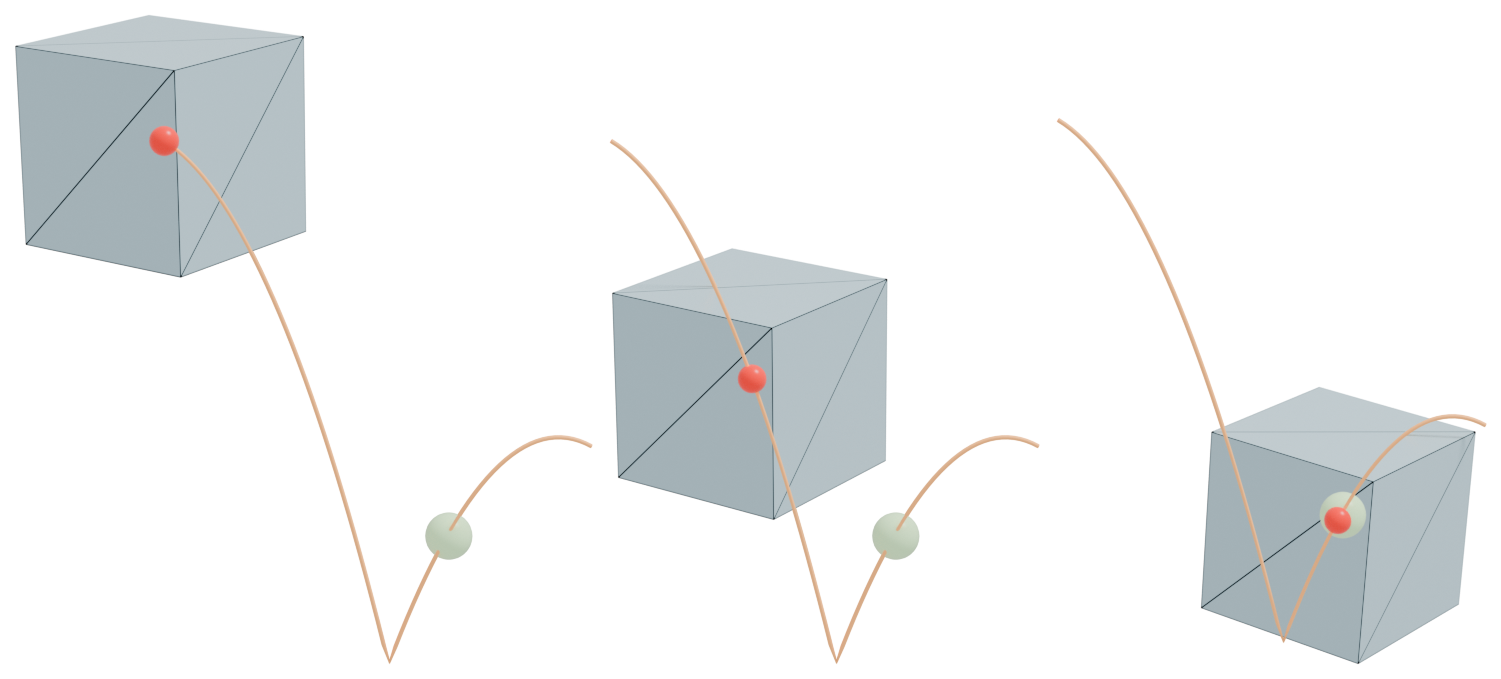}
\end{minipage}\hspace{0.03\linewidth}
\begin{minipage}[c]{0.48\linewidth}
\caption{Cube initial-state optimization: start, mid-bounce, and near target.
The center of mass trajectory comparison is Fig.~\ref{fig:cube-traj}.}
\label{fig:cube}
\end{minipage}
\vspace{-4pt}
\end{figure}

\textbf{Initial-state optimization.} Given a target voxel, optimize
$(\bm{x}_0,\bm{v}_0)$ so the center of mass reaches it after ground contact. The best
center-of-mass--target distance over the rollout drops from $0.304$\,m under the
initial parameters to $0.013$\,m for a cube ($8$ vertices, $6$ tetrahedra),
confirming that the optimization, not the initialization, drives the result.
The bunny starts much closer to its target ($0.035$\,m) and ends at
$0.034$\,m, so that task exercises the gradient without demonstrating a large
improvement
(Figs.~\ref{fig:cube},~\ref{fig:bunny}; Fig.~\ref{fig:Ksweep} overlays the
uncontrolled baseline).

\begin{wrapfigure}[9]{r}{0.38\textwidth}
\vspace{-10pt}
\centering
\includegraphics[width=\linewidth]{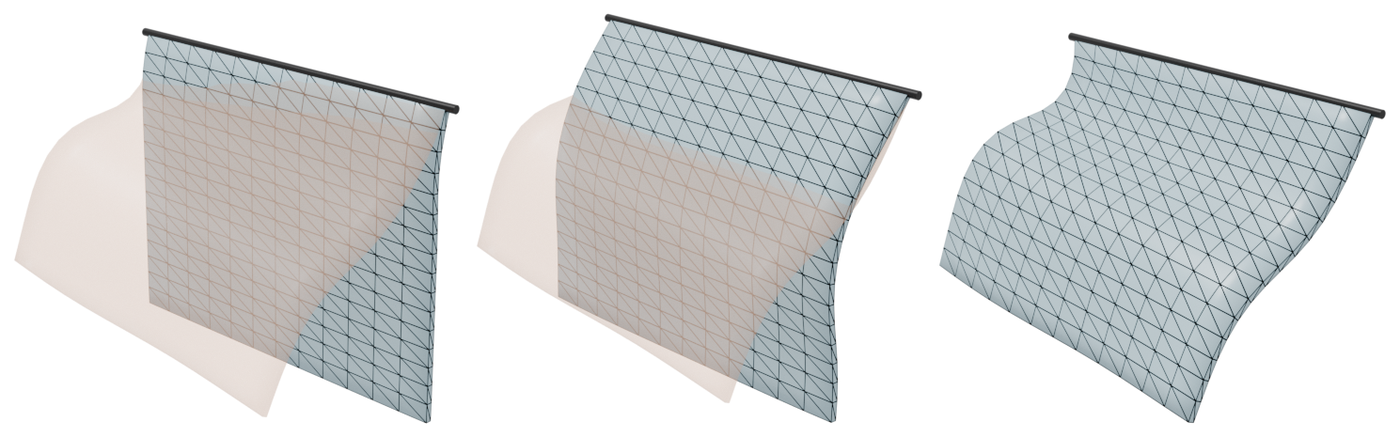}
\caption{Flag under the recovered time-varying wind, at three times in the rollout.}
\label{fig:flag}
\vspace{-14pt}
\end{wrapfigure}

\textbf{Wind control.} Given a target flag trajectory, recover a
spatially-uniform, time-varying wind on a $442$-vertex flag pinned along its
top edge, parameterized as piecewise-constant over $8$ equal time windows
($8\times3=24$ scalars, dimensionless force units relative to a unit reference
wind). Loss falls from $3.2\times10^{1}$ to $5.1\times10^{-3}$\,m$^2$ in $199$
iterations (Fig.~\ref{fig:flag}). The recovered wind differs from the reference by $\|\Delta w\|=15.3$
because the problem is under-determined: $24$ wind scalars act through
${\sim}12$ effective trajectory degrees of freedom, so trajectory matching does
not identify the wind uniquely.

\textbf{Contact-rich planning (Cover the Spot).} Three soft disks
($r{=}0.42$, $r/R{=}0.933$, just above the $R\sqrt{3}/2$ infeasibility bound)
drop from $2.5$\,m onto a target of radius $R{=}0.45$\,m. All three share one
mega-mesh in a single Vertex Block Descent context, so disk--disk and disk--floor
interactions
are resolved by the IPC log barrier ($\kappa{=}400$, $\hat d{=}2$\,cm, lagged
damping $c_d{=}8$; frictionless). We jointly optimize per-disk $\bm{v}_0$ and a
small $\bm{x}_0$ offset ($18$ scalars) under a smooth coverage objective with
sigmoid sharpness annealed $4\to25$; each forward rolls $160$ steps with full
backpropagation through time, so adjoints flow through every barrier
evaluation. $600$ Adam iterations
in $<40$\,s reduce the hard uncovered fraction from $89\%$ to $\mathbf{0.0\%}$
(App.~\ref{app:figs}, Fig.~\ref{fig:cover-the-spot}).

\section{Task protocols}\label{app:tasks}

Every task below is a script in the released harness and is run by a single
command; scenes are constructed procedurally, so there are no data
dependencies. We state what is given, what is optimized, and what counts as
success. Reported values are read out of the run logs by the harvesting script
rather than transcribed by hand.

\paragraph{Beam $\mu$ identification.} \emph{Given}: a target rollout of a
$13\times4\times4$ tetrahedralized cantilever ($0.5\times0.2\times0.2$\,m),
twisted $180^\circ$ and released, simulated at $h{=}2$\,ms for $100$ steps with
$K{=}15$ under a known $\lambda{=}10^{6}$. \emph{Optimized}: the scalar $\mu$,
from an initialization $10\times$ too soft ($10^{5}$ against a true $10^{6}$),
by Adam on $\partial\mathcal{L}/\partial\mu$ with the per-tetrahedron parameter
gradients summed. \emph{Criterion}: relative error in the recovered $\mu$.

\paragraph{Beam $\mu,\lambda$ identification.} As above but with both Lam\'e
parameters free and a $60^\circ$ angular kick as excitation; initialization
$(300,150)$ against a true $(800,400)$. \emph{Criterion}: relative error in
both parameters.

\paragraph{Dress $\mu,\lambda$ and dress silhouette.} \emph{Given}: a target
drape of a garment mesh over a capsule mannequin with IPC ground contact.
\emph{Optimized}: $(\mu,\lambda)$ in log space, so that the iterates stay
positive without clipping. The two variants differ only in the loss: a
per-vertex squared position error, or a radial silhouette profile that
discards the correspondence and compares only the outline.
\emph{Criterion}: relative error against the parameters that generated the
target.

\paragraph{Flag wind control.} \emph{Given}: a target trajectory of a
$442$-vertex flag pinned along its top edge, rolled out for $120$ steps at
$K{=}10$. \emph{Optimized}: a spatially uniform wind held piecewise constant
over $8$ equal time windows, i.e.\ $24$ scalars, by Adam at learning rate
$2.0$. \emph{Criterion}: squared trajectory error. We additionally report
$\lVert\Delta w\rVert$ against the generating wind, which stays large because
the map from $24$ wind scalars to the trajectory is not injective; the task is
trajectory matching, not wind identification, and we label it as such.

\paragraph{Cube and bunny initial state.} \emph{Given}: a target voxel.
\emph{Optimized}: the initial position offset and velocity ($6$ scalars for the
cube, $3$ for the bunny) by Adam with gradient clipping, over a rollout that
includes floor contact. \emph{Criterion}: the closest center-of-mass approach
to the target over the rollout, compared against the same quantity under the
initial parameters, so that the initialization cannot be credited with the
result.

\paragraph{Cover the Spot.} \emph{Given}: a circular target of radius $R$ and
three soft disks released from a fixed height. \emph{Optimized}: per-disk
initial velocity and a small initial position offset ($18$ scalars) under a
smooth coverage objective whose sigmoid sharpness is annealed during
optimization, with all three disks in one mega-mesh so that disk--disk and
disk--floor contact is resolved by the IPC barrier and differentiated through
every step. \emph{Criterion}: the hard (non-smoothed) uncovered fraction of the
target. The task is feasible only when the disk radius exceeds
$R\sqrt3/2$; below that bound no three disks can cover the target and the
script refuses to run, so the radius is part of the protocol and not a free
knob.

\section{Full $K\times|V|$ scaling grid}\label{app:scaling}

The backward-to-forward ratio is not a single number, and its dependence is
easy to misread, so we state it explicitly: at $339$
vertices the ratio is $4.2\times$ for the Warp backend and $1.7\times$ for the
fused backend, both dominated by fixed per-color kernel-launch overhead rather
than by adjoint arithmetic; the overhead amortizes as $|V|$ grows, and
Table~\ref{tab:scaling} reports the full grid.

Controlled scaling study on a held-out soft jelly-cube tetrahedral mesh (not
used elsewhere), on an NVIDIA RTX~A6000. 1-to-8 tetrahedral subdivision gives
four resolution levels L0--L3; we roll out $50$ implicit steps and
backpropagate through all of them
and sweep both solver depth $K$ and mesh resolution. Every row is measured.

\begin{table}[h]
\caption{\textbf{Measured $K\times|V|$ grid} (RTX~A6000, all rows measured).
Across the grid the backward/forward time ratio stays in $0.96$--$1.14$
(excluding the $K{=}5$ first-call warm-up, which includes kernel compilation),
and peak memory divided by $K|V|$ stays in $3.6$--$4.6$ kilobytes per vertex
per sweep, so linearity in $K|V|$ holds to $\pm13\%$. At fixed $K$, $360\times$
more vertices costs $370\times$ memory.}
\label{tab:scaling}
\centering\footnotesize
\renewcommand{\arraystretch}{1.12}
\begin{tabular}{llrrrr}
\toprule
Mesh & $K$ & $|V|$ & forward (ms) & backward (ms) & peak (MB) \\
\midrule
L0 & 5   & 558      &   1.4 &   4.3 &    12 \\
L0 & 10  & 558      &   1.5 &   1.6 &    22 \\
L0 & 20  & 558      &   2.7 &   2.7 &    42 \\
L0 & 50  & 558      &   6.4 &   6.2 &   100 \\
L0 & 100 & 558      &  12.5 &  12.2 &   201 \\
L0 & 200 & 558      &  24.8 &  24.0 &   393 \\
\midrule
L1 & 10  & 3{,}766  &   2.8 &   3.2 &   149 \\
L1 & 50  & 3{,}766  &  12.6 &  14.4 &   677 \\
L1 & 100 & 3{,}766  &  24.9 &  28.3 & 1{,}336 \\
\midrule
L2 & 10  & 27{,}169 &  17.0 &  17.1 & 1{,}074 \\
L2 & 20  & 27{,}169 &  31.7 &  32.9 & 2{,}025 \\
L2 & 50  & 27{,}169 &  76.0 &  80.5 & 4{,}926 \\
\midrule
L3 & 10  & 201{,}341 & 113.3 & 115.8 &  8{,}136 \\
L3 & 20  & 201{,}341 & 214.0 & 224.5 & 15{,}001 \\
\bottomrule
\end{tabular}
\end{table}

This grid measures our method alone; the controlled cross-method comparison
is Table~\ref{tab:cost} (one forward, one GPU, one mesh for all four
routes).

\section{Reproducibility details}\label{app:repro}

\paragraph{Hardware and software.} The controlled backward comparison
(\S\ref{sec:exp-exact}--\S\ref{sec:exp-cost}), the finite-difference checks and
the inverse problems were run on a single NVIDIA RTX~4090 Laptop GPU (16\,GB) with CUDA
12.1 and PyTorch 2.3, inside one process so that all four backward routes see
the same device state. The $K\times|V|$ grid of Table~\ref{tab:scaling} was run
on an RTX~A6000, and the $10^{6}$-body run on one H100 (80\,GB). All solver
arithmetic is \texttt{float64}; no result in this paper depends on
mixed-precision behavior.

\paragraph{Determinism and seeds.} Scene construction is deterministic:
meshes, colorings and rest metrics are generated procedurally from fixed
integer parameters, so a given task builds the identical problem on every
machine. The one deliberate exception is that the backward scatters
neighbor contributions with atomic adds, whose summation order is not fixed;
run-to-run gradient differences from this are at the level of
\texttt{float64} round-off and are far below every tolerance reported here,
but they mean bitwise reproducibility is not claimed. Random initializations are drawn from seeded generators recorded in the run
log.

\paragraph{What to run.} The released harness maps each numbered claim to a
single command; one driver re-runs every experiment and writes one log per
experiment, and a harvesting script reads the quoted numbers out of those logs
rather than transcribing them. Everything except the $10^{6}$-body run
(\S\ref{sec:exp-scale}) reproduces in minutes on one consumer GPU.

\end{document}